\documentclass[preprint]{aastex7}

\usepackage{graphicx}
\usepackage{verbatim}
\usepackage{listings}
\usepackage{tabularx}
\usepackage{booktabs}
\usepackage{amsmath} 
\usepackage{xspace}
  
\graphicspath{{./}{Figures/}}
\newcommand{\update}[1]{\textcolor{black}{#1}}

\newcommand{\package}[1]{\texttt{#1}}
\newcommand{\sorcha}{\package{Sorcha}\xspace}
\newcommand{\python}{\package{Python}\xspace}
\newcommand{\conda}{\package{conda}\xspace}
\newcommand{\mamba}{\package{mamba}\xspace}
\newcommand{\pip}{\package{pip}\xspace}
\newcommand{\condaforge}{\package{conda-forge}\xspace}

\definecolor{codegreen}{rgb}{0,0.6,0}
\definecolor{codegray}{rgb}{0.5,0.5,0.5}
\definecolor{codepurple}{rgb}{0.58,0,0.82}
\lstdefinestyle{mystyle}{
    commentstyle=\color{codegray},
    keywordstyle=\color{magenta},
    numberstyle=\tiny\color{codegray},
    stringstyle=\color{codegreen},
    basicstyle=\ttfamily\footnotesize,
    breakatwhitespace=false,
    breaklines=true,
    captionpos=b,
    keepspaces=true,
    showspaces=false,
    showstringspaces=false,
    showtabs=false,
    tabsize=2
}
\begin{document}

\title{Sorcha: A Solar System Survey Simulator for the Legacy Survey of Space and Time}

\author[0000-0001-5930-2829]{Stephanie R. Merritt}
\affiliation{Astrophysics Research Centre, School of Mathematics and Physics, Queen's University Belfast, Belfast BT7 1NN, UK}
\email{stephanierosemerritt@icloud.com}  

\author[0000-0002-8418-4809]{Grigori Fedorets}
\affiliation{Astrophysics Research Centre, School of Mathematics and Physics, Queen's University Belfast, Belfast BT7 1NN, UK}
\affiliation{Finnish Centre for Astronomy with ESO, University of Turku, FI-20014 Turku, Finland}
\affiliation{Department of Physics, University of Helsinki, P.O. Box 64, 00014
Helsinki, Finland}
\email{grigori.fedorets@helsinki.fi}  

\author[0000-0003-4365-1455]{Megan E. Schwamb}
\affiliation{Astrophysics Research Centre, School of Mathematics and Physics, Queen's University Belfast, Belfast BT7 1NN, UK}
\email[show]{m.schwamb@@qub.ac.uk}
\correspondingauthor{Megan E. Schwamb}

\author[0000-0002-0672-5104]{Samuel Cornwall} 
\affiliation{Department of Aerospace Engineering, Grainger College of Engineering, University of Illinois at Urbana-Champaign,
Urbana, IL 61801, USA}
\email{samuel41@illinois.edu}  

\author[0000-0003-0743-9422]{Pedro H. Bernardinelli} 
\affiliation{DiRAC Institute and the Department of Astronomy, University of Washington, 3910 15th Ave NE, Seattle, WA 98195, USA}
\email{phbern@uw.edu}  

\author[0000-0003-1996-9252]{Mario Juri\'c}
\affiliation{DiRAC Institute and the Department of Astronomy, University of Washington, 3910 15th Ave NE, Seattle, WA 98195, USA}
\email{mjuric@uw.edu}  

\author[0000-0002-1139-4880]{Matthew~J.~Holman}
\affil{Center for Astrophysics | Harvard \&Smithsonian, 60 Garden St., MS 51, Cambridge, MA 02138, USA}
\email{mholman@cfa.harvard.edu}  

\author[0009-0005-5452-0671]{Jacob A. Kurlander} 
\affiliation{DiRAC Institute and the Department of Astronomy, University of Washington, 3910 15th Ave NE, Seattle, WA 98195, USA} 
\email{jkurla@uw.edu}  

\author[0000-0002-1398-6302]{Siegfried Eggl}
\affiliation{Department of Aerospace Engineering,
Grainger College of Engineering,
University of Illinois at Urbana-Champaign,
Urbana, IL 61801, USA}
\affiliation{Department of Astronomy,
University of Illinois at Urbana-Champaign,
Urbana, IL 61801, USA}
\affiliation{National Center for Supercomputing Applications,
University of Illinois at Urbana-Champaign,
Urbana, IL 61801, USA}
\email{eggl@illinois.edu}  

\author[0000-0001-6984-8411]{Drew Oldag}
\affiliation{DiRAC Institute and the Department of Astronomy, University of Washington, 3910 15th Ave NE, Seattle, WA 98195, USA}
\affiliation{LSST Interdisciplinary Network for Collaboration and Computing Frameworks, 933 N. Cherry Avenue, Tucson AZ 85721}
\email{awoldag@uw.edu}  

\author[0009-0003-3171-3118]{Maxine West}
\affiliation{DiRAC Institute and the Department of Astronomy, University of Washington, 3910 15th Ave NE, Seattle, WA 98195, USA}
\affiliation{LSST Interdisciplinary Network for Collaboration and Computing Frameworks, 933 N. Cherry Avenue, Tucson AZ 85721}
\email{maxwest@uw.edu}  

\author[0009-0009-2281-7031]{Jeremy Kubica}
\affiliation{McWilliams Center for Cosmology, Department of Physics, Carnegie Mellon University, Pittsburgh, PA 15213, USA}
\affiliation{LSST Interdisciplinary Network for Collaboration and Computing Frameworks, 933 N. Cherry Avenue, Tucson AZ 85721}
\email{jkubica@andrew.cmu.edu}  

\author[0000-0001-9505-1131]{Joseph Murtagh}
\affiliation{Astrophysics Research Centre, School of Mathematics and Physics, Queen's University Belfast, Belfast BT7 1NN, UK}
\email{jmurtagh05@qub.ac.uk}  

\author[0000-0001-5916-0031]{R. Lynne Jones}
\affiliation{Rubin Observatory, 950 N. Cherry Ave., Tucson, AZ 85719, USA}
\affiliation{Aston Carter, Suite 150, 4321 Still Creek Drive, Burnaby, BC V5C6S, Canada}
\email{ljones.uw@gmail.com}  

\author[0000-0003-2874-6464]{Peter Yoachim}
\affiliation{DiRAC Institute and the Department of Astronomy, University of Washington, 3910 15th Ave NE, Seattle, WA 98195, USA}
\email{yoachim@uw.edu}  

\author[0009-0007-8602-2954]{Ryan R. Lyttle}
\affiliation{Astrophysics Research Centre, School of Mathematics and Physics, Queen's University Belfast, Belfast BT7 1NN, UK}
\email{rlyttle09@qub.ac.uk}  

\author[0000-0002-6702-7676]{Michael S. P. Kelley}
\affil{Department of Astronomy, University of Maryland, College Park, MD 20742-0001, USA}
\email{msk@astro.umd.edu}  

\author[0000-0001-5820-3925]{Joachim Moeyens}
\affiliation{Asteroid Institute, 20 Sunnyside Ave., Suite 427, Mill Valley, CA 94941, USA}
\affiliation{DiRAC Institute and the Department of Astronomy, University of Washington, 3910 15th Ave NE, Seattle, WA 98195, USA}
\email{moeyensj@uw.edu}  

\author{Kathleen Kiker}
\affiliation{Asteroid Institute, 20 Sunnyside Ave., Suite 427, Mill Valley, CA 94941, USA}
\email{kathleen@b612foundation.org}  

\author[0000-0003-4439-7014]{Shantanu P. Naidu}
\affiliation{Jet Propulsion Laboratory, California Institute of Technology, Pasadena, CA, USA}
\email{shantanu.p.naidu@jpl.nasa.gov}  

\author[0000-0001-9328-2905]{Colin Snodgrass}
\affiliation{Institute for Astronomy, University of Edinburgh, Royal Observatory, Edinburgh, EH9 3HJ, UK}
\email{colin.snodgrass@ed.ac.uk}  

\author[0000-0001-8633-9141]{Shannon M. Matthews}
\affiliation{Astrophysics Research Centre, School of Mathematics and Physics, Queen's University Belfast, Belfast BT7 1NN, UK}
\email{smatthews13@qub.ac.uk}  

\author[0000-0001-7335-1715]{Colin Orion Chandler}
\affiliation{DiRAC Institute and the Department of Astronomy, University of Washington, 3910 15th Ave NE, Seattle, WA 98195, USA}
\affiliation{LSST Interdisciplinary Network for Collaboration and Computing Frameworks, 933 N. Cherry Avenue, Tucson AZ 85721}
\email{coc123@uw.edu}



\begin{abstract}

The upcoming Legacy Survey of Space and Time (LSST) at the Vera C. Rubin Observatory is expected to revolutionize solar system astronomy. Unprecedented in scale, this ten-year wide-field survey will collect billions of observations and discover a predicted $\sim$5 million new solar system objects. Like all astronomical surveys, its results will be affected by a complex system of intertwined detection biases. Survey simulators have long been used  to forward-model the effects of these biases on a given population, allowing for a direct comparison to real discoveries. However, the scale and tremendous scope of the LSST requires the development of new tools. In this paper we present \sorcha, an open-source survey simulator written in \python. Designed with the scale of LSST in mind, \sorcha is a comprehensive survey simulator to cover all solar system small-body populations. Its flexible, modular design allows \sorcha to be easily adapted to other surveys by the user. The simulator is built to run both locally and on high-performance computing (HPC) clusters, allowing for repeated simulation of millions to billions of objects (both real and synthetic). 

\end{abstract}



\section{Introduction} \label{sec:intro}
The upcoming Legacy Survey of Space and Time \citep[LSST; ][]{lsst-sciencebook-ch5-2009,ivezic2019,bianco2022} using the Vera C. Rubin Observatory will revolutionize solar system astronomy. This ten-year synoptic ground-based survey boasts an unprecedented combination of six broad-band optical filters, a 9.6 deg$^2$ field-of-view (FOV), and a survey depth of $\sim$24.5 magnitude (in the $r$-filter). Moreover, the main Wide-Fast-Deep component of the survey will provide uniform sky coverage of $\sim$18,000 deg$^2$ of the southern sky \citep{lsst-SRD-2013}. With cataloging the solar system as one of its four main science goals, the LSST is expected to collect on the order of a billion individual observations of solar system objects, while discovering approximately five million new solar system objects across all populations \citep{lsst-sciencebook-ch5-2009}. This wealth of new information surpasses any solar system survey to date in its combination of depth, sky coverage and sheer number of observations, and this will allow us to probe the dynamics and formation history of the solar system on a scale never-before attempted. To highlight just some of the specific discoveries awaited: the LSST is expected to update the NEO (near-Earth object) inventory of objects with diameters larger than 140~meters to 66-86\%  from the 42\% estimated to have been discovered by 2022 \citep{veres2017a,jones2018}; the discovery and classification of many more active asteroids in the main asteroid belt will enable their population-level studies \citep{denneau2015,mcloughlin2015}; additional targets may be found for NASA's (National Aeronautics and Space Administration's) ongoing \emph{Lucy} mission \citep{schwamb2018a} and for ESA's (European Space Agency's) upcoming lay-in-wait \emph{Comet Interceptor} mission \citep{snodgrass2019}; and a handful of new interstellar objects (ISOs) are also predicted \citep{moro-martin2009, cook2016, engelhardt2017, trilling2017, seligman2018, levine2021, hoover2022}. Finally, it is expected that LSST will be paramount in settling the debate considering the existence of Planet Nine \citep[e.g.][]{trujillo2014,batygin2016,belyakov2022,brown2024}, by discovering a sufficient number of extreme trans-Neptunian objects (eTNOs) to definitely wield support for or against the clustering of eTNOs, or even discovering the putative planet itself \citep{trilling2018,belyakov2022,brown2024}.

While the size and scope of the LSST is unprecedented, astronomical surveys have a long history in the exploration and characterization of solar system small body populations in both the inner and outer solar system. All astronomical surveys are affected by a complex set of intertwined observational biases caused by a number of factors, including observational strategy and cadence, limiting magnitude, instrumentation effects and poor weather/seeing. The detections from an astronomical survey therefore provide a distorted view of the actual underlying population. Survey simulators have emerged as powerful tools for assessing the impact of observational biases and aiding in the study of the target population. As long as a survey is ``well-characterized,'' in terms of having a known  pointing history and calibrated image limiting magnitudes and detection efficiencies,  a survey simulator can be used to replicate the varied biases of the survey upon a small body model population. This allows forward-biased model populations (objects  from the input model that the simulator deemed detectable/discoverable by the survey) to be compared to survey's real discoveries.

Survey simulators have contributed to solar system science across many surveys, including the Canada–France Ecliptic Plane Survey (CFEPS) \citep{jones2006,kavelaars2009,petit2011}, the Palomar Distant Solar System Survey \citep{schwamb2010}, the Outer Solar System Origins Survey (OSSOS) \citep{bannister2016, bannister2018,lawler2018}, the Dark Energy Survey (DES) \citep{bernardinelli2022}, and the DECam Ecliptic Exploration Project (DEEP) \citep{trilling2024, bernardinelli2024}. In the outer solar system, survey simulators have been used to determine the structure of the Kuiper belt and Centaur region \citep{petit2011, kavelaars2021, gladman2012, bannister2017, bannister2018,kurlander2025}; explore size distributions of scattering TNOs (Trans-Neptunian objects) \citep{lawler2018b}; place constraints on formation models for distant Sedna-like objects \citep{schwamb2010}; probe the location of a color transition in the primordial TNO population \citep{buchanan2022}; test for clustering in the orbits of high-perihelion TNOs \citep{shankman2017, bernardinelli2020, napier2021}; and place limitations on the discoverability of Planet Nine \citep{sheppard2016,trilling2018, belyakov2022,brown2024}. In the inner solar system, survey simulators have found previously unreported detections of known NEOs \citep{masiero2023}, predicted the discovery yields of future planned space missions \citep{masiero2024}, estimated survey effectiveness at anticipating NEO impacts as a function of impact energy \citep{lay2024},  explored the size distribution , composition, and population size for  NEOs and potentially hazardous asteroids \citep{luu1989,bottke2002b,stuart2004}, and probed the population of asteroids interior to Earth and Venus \citep{pokorny2020,sheppard2022}. 

The LSST's impact on the investigation of all solar system small body populations will be immense: the potential utility of a survey simulator for use with the LSST is clear. However, previous survey simulators have been built with a specific survey and solar system small body population in mind, and are often tailored for a particular scientific objective. Additionally, they have universally been built to manage far fewer objects than the LSST is expected to discover -- OSSOS, for instance, tracks only about 800 objects \citep{bannister2018}. In comparison, a survey simulator for the LSST must be able to calculate the on-sky positions and magnitudes for input objects of all populations, and it must do this for approximately 2 million survey exposures/pointings, in six filters, taken over ten years. This presents a computational challenge of much greater scale than those faced by previous survey simulators. It is evident that previous survey simulators are inadequate for the LSST's needs and cannot simply be adapted. 

Considering the LSST's vast potential for advancing solar system research, there is an urgent need for a versatile and scalable survey simulator that can handle all solar system small body populations. For this purpose, we present \sorcha, our multipurpose, open-source solar system survey simulator for the LSST. \sorcha is built to be flexible, easy-to-use, and applicable to all solar system small body sub-populations, our survey simulator's modular design and configuration file allows each simulation to be highly customizable by the user for their specific needs, and while \sorcha was built with the LSST in mind, this inherent customizability also allows it to be easily adapted to other surveys. Additionally, the simulator was designed to work at the large scale demanded by the billions of upcoming LSST observations, and simulations can be run both locally and on high-performance computing clusters. We have designed \sorcha to be a key community tool for solar system science with the LSST.  \sorcha is \pip or \conda/\mamba installable, and the code is hosted in an open repository\footnote{\url{https://github.com/dirac-institute/sorcha}}. Online documentation is also available\footnote{https://sorcha.readthedocs.io}.

In Section~\ref{sec:lsst}, we discuss the particular challenges of the LSST pertaining to solar system science. Sections ~\ref{sec:software_design} and  \ref{sub:documentation}  give a general overview of \sorcha's software design and documentation. How to interface with \sorcha is discussed in Section~\ref{sec:inputs}. Section~\ref{sec:ephemeris} describes \sorcha's internal ephemeris generator, while Section~\ref{sec:post-processing} outlines the various calculations and ``filters''   within \sorcha's post-processing stage designed to model the effects of the observational biases of a survey. Section~\ref{sec:outputs} describes the outputs of the survey simulator. The validation tests performed to verify the output of \sorcha are described in Section~\ref{sec:validation}, and the benchmarking and information on \sorcha's performance are explained in Section~\ref{sec:benchmarking}. Utility scripts designed to perform various secondary functions useful to the user are explained in Section~\ref{sec:utilities}. Section~\ref{sec:caveats} discusses limitations and caveats of the software. Finally, Section~\ref{sec:conclusions} outlines our conclusions and plans for the future of \sorcha.  

\section{The LSST and the Solar System} \label{sec:lsst}
Building a survey simulator for the LSST comes with many unique challenges that were not faced by previous solar system survey simulators. Perhaps most obvious of these challenges is that the unprecedented number of LSST exposures (a total of approximately 2 million images), for example, will ultimately lead to an equally unprecedented number of discoveries. Consequently, the number of known members of each of the solar system small body sub-populations is expected to increase by an order of magnitude over the number currently reported in the Minor Planet Center (MPC) \citep{jones2009, lsst-sciencebook-ch5-2009, solontoi2010, shannon2015, silsbee2016, grav2016, veres2017b, jones2018, schwamb2018a, ivezic2019, fedorets2020}. The LSST will also integrate multiple surveys into its observational strategy. As of early-2025, the LSST observational cadence has been mostly finalized \citep{SCOC_Report_1,SCOC_Report_2,SCOC_Report_3}, but there are a few remaining decisions to be made in the next year. There is expected to be some final tweaks on the baseline observing strategy and rolling cadence plans, in addition to small changes necessary for optimizing science returns in the first year of operations. However, it has been decided that around 80-90\% of the observing time will be devoted to the Wide-Fast-Deep survey, a systematic survey covering 18,000 square degrees of the southern sky with a consistent, universal observing strategy. The remaining time will be dedicated to mapping Deep Drilling Fields (small pre-selected patches of the sky which will receive significantly more observations compared to the average number of Wide-Fast-Deep visits) as well as surveys of the southern polar cap, the galactic plane, and the northern ecliptic spur (which covers the sky ten degrees above the ecliptic in the northern celestial hemisphere \citep{lsst-sciencebook-ch5-2009,schwamb2018c,SCOC_Report_1,SCOC_Report_2,SCOC_Report_3}). Any survey simulator built for the LSST must thus be capable of processing tens of millions of synthetic input small bodies to identify which of the millions of exposures they each land in while tracking these orbits over ten years across multiple sub-surveys within the LSST. This is in itself a difficult task, and to the best of our knowledge no survey simulators built previously were required to work on such an immense computational scale required for the LSST's unprecedented combination of sky coverage and depth.

The LSST is not only focused on discovering solar system objects (SSOs), however. The survey will track five million new solar system objects over its ten-year duration, with each object receiving up to hundreds of observations across six broad-band \textit{ugrizy} filters; moreover, each pointing will be observed at least twice a night in two different filters. LSST's  built-in follow-up will enable the study of activity throughout the solar system, such as cometary outbursts, outgassing, asteroid collisions, and rotational breakup events \citep{jones2009, lsst-sciencebook-ch5-2009, schwamb2018a, schwamb2021}. Furthermore, the large volume of observations for each object will facilitate the determination of rotational light curves, phase curves, and color information, significantly enhancing our understanding of the shape, rotation rate, and composition of small solar system bodies \citep{jones2009, lsst-sciencebook-ch5-2009, schwamb2018a}. This built-in ability to monitor and follow up on discoveries is vital for LSST science, and thus, needs to be accurately modeled in a survey simulator. At its most basic level, a simulator should be capable of computing apparent magnitudes of solar system objects in all six LSST filters, accounting for phase effects on brightness to accurately simulate phase curves using the user's preferred phase function model; it would furthermore be useful to apply the effects of activity or rotation to the light curves generated in survey simulations. 

In addition to the expected instrumental and observational biases linked to the optics and observational strategy of the survey, the solar system object detection and discovery pipeline is also to be considered. The Rubin Solar System Processing (SSP) pipeline \citep{myers2013, juric2020}, currently under development, is a fully-automated discovery pipeline which works to link observations of transient sources to discover previously-unknown moving solar system objects. In this technique, two or more observations of point sources in a single night, suspected to be the same unknown object moving across the FOV, are linked into a ``tracklet''; these tracklets are then further linked into a ``track'' spanning multiple nights, and a preliminary orbit can thus be calculated \citep{kubica2007}. This process is described in more detail in Section~\ref{sub:linking_filter}. In order to mimic the operation of the LSST, software such as \sorcha must have the option to simulate the effects of this linking procedure, to determine which objects of the input population of interest are discovered by the SSP pipeline. In addition, the SSP pipeline is only expected to link objects out to a maximum of $\sim$200 au \citep{schwamb2023}. Users of \sorcha and the LSST interested in, for example, Planet Nine, may be using their own bespoke linking software to detect distant objects beyond the SSP's $\sim$200 au limit. Ideally, \sorcha should be able to simulate both the SSP pipeline and more user-specific linking software.

The nature of data release for the LSST also provides some interesting considerations for a survey simulator. The distribution of solar system data for LSST will occur through two main avenues, prompt data processing and annual releases, as described in the Rubin Data Products Definition Document \citep[DPDD;][]{juric2021}. The prompt data products will include all data observed on a given night, and will be released on a daily basis, with the Minor Planet Center being one of the distribution channels for solar system objects. During the night, the Rubin prompt processing pipelines will also send out alerts to the Rubin alert stream when known solar system objects are matched to transient sources identified in LSSTCam exposures. This is expected to occur within 60s of image readout.  Annual data releases, meanwhile, will provide an annual catalog of solar system discoveries, regenerated and updated using only LSST data. A fast, easy-to-use survey simulator will allow users to predict which prompt products will contain their objects of interest or compare the results of many simulations of many different models to the annual data release catalogs.

The LSST's ambitious goals, from tracking millions of new solar system objects to integrating diverse observational strategies, require a simulator that not only handles the vast computational load but also offers flexibility and precision in modeling the intricate dynamics of small body populations. \sorcha must therefore rise to meet these demands by providing detailed, customizable simulations that can account for the LSST’s unique capabilities, including its frequent observations, varied filters, and the sophisticated linking procedures of the SSP pipeline.

\section{Software Design Overview and Implementation Choices} \label{sec:software_design}
\begin{figure}
    \centering
    \includegraphics[width=0.96\columnwidth]{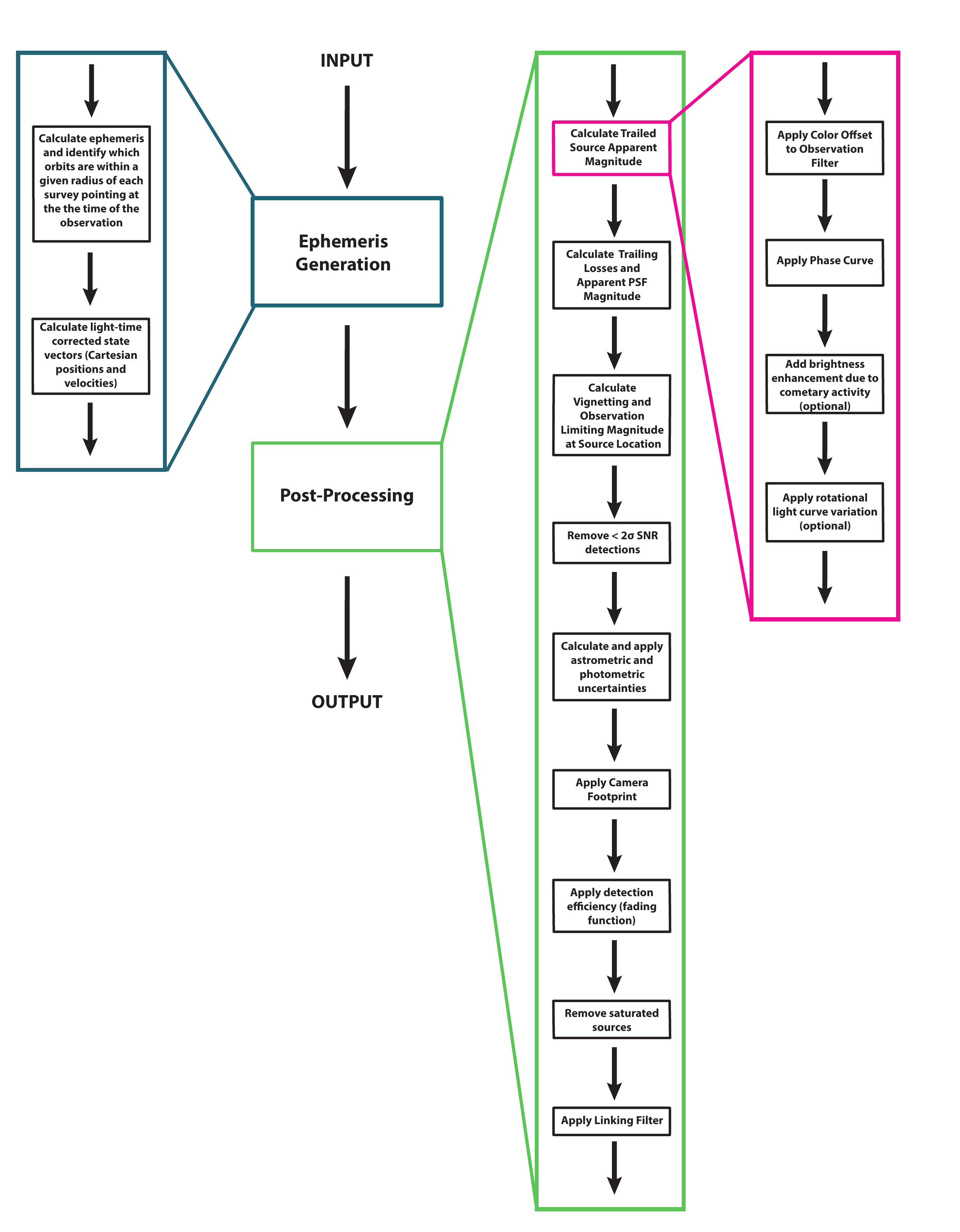}
    \caption{ The main \sorcha workflow. \sorcha is split into two stages. The ephemeris generation  stage (described in Section \ref{sec:ephemeris}) is in gray, and the post-processing stage (described in Section \ref{sec:post-processing}) is in green and magenta. We note for the reader that this workflow is for simulating what the LSST is expected to discover. We note there are a small number of  additional functions/filters available (described in Section \ref{sec:additional_advanced_filters}) that can be applied to produce different desired outputs of the simulated small body detections and discoveries based on the user's desired science case.}
    \label{fig:main_workflow}
\end{figure}

High level modules of \sorcha are written in \python making use of a C/C++ backend where possible. Our focus throughout development of \sorcha has been to make the survey simulator as flexible and easy-to-use as possible. \sorcha's basic structure consists of two main halves, illustrated in Figure~\ref{fig:main_workflow}: ephemeris generation and post-processing. Ephemeris generation, described further in Section~\ref{sec:ephemeris}, is the process by which \sorcha calculates the on-sky position of an input set of small bodies and matches them to the survey observations in which they appear. Post-processing, covered in Section~\ref{sec:post-processing}, applies a number of sequential steps designed to calculate the brightness of these input small bodies and apply the optical and instrumental effects of the survey to each individual observation. \sorcha allows either of these stages to be run in stand-alone mode. One might, for instance, use the built-in ephemeris generator by itself to simply calculate which survey observations a simulated small body population will appear in. Similarly, it is possible to only use the components of post-processing that will calculate the predicted brightness of those objects; if a user prefers pre-generated ephemerides from other software or a previous \sorcha run, only the post-processing steps may be desirable and can be performed independently. Together, these two steps of ephemeris generation and post-processing complete the full survey simulation for a target small body population.

Details about \sorcha's repository setup, software testing, and package architecture  and deployment are described in Appendix \ref{sub:testing_and_deployment}. \sorcha has been designed to facilitate easy customization both for new and advanced users. For the majority of users, \sorcha's configuration file offers everything needed to tailor the simulation to their particular science case. In this file, described further in Section~\ref{sub:config_file}, users can easily customize the parameters of every aspect of the simulation and switch off or on the available functions. This configuration file is also copied in its entirety to the log file of every \sorcha run, so that the user may easily locate and even copy out the specific configuration for that run. For users who wish to apply the effects of rotation or activity to their objects, we provide an additional package, \texttt{sorcha-addons}, designed to interface easily with \sorcha. This package, described in Section~\ref{sub:add-ons}, contains simple models to simulate activity or rotational light curves, and also allows the user to simply add their own custom classes modeling these effects. Finally, \sorcha was written to be highly modular and human-readable. The more advanced user should therefore find it simple to edit or replace any function in the simulation to suit their own needs.

Attention was also paid to the many ways in which a user might wish to run \sorcha. Getting started with \sorcha on a local machine or laptop is straightforward. \sorcha comes bundled with demonstration input files and a demo command, both easily accessible via the command line interface (CLI), and example configuration files which are designed for simulating the LSST are also available via the same method. A quick-start guide is presented in Appendix~\ref{app:running_sorcha}. However, it is understood that many users will be using HPC clusters to perform multiple runs of \sorcha on their simulated populations. \sorcha has thus been built and extensively tested for HPC, and comes with useful utility scripts designed to aid the user in collating and exploring the results of multiple runs. These are explained further in Section~\ref{sec:utilities}. Additionally, all inputs to \sorcha can be ``chunked'', where \sorcha reads in and iterates over segments of the input files in sequence, with a chunk size controlled by the configuration file to enable the user to tailor \sorcha runs for their own specific setup. This feature -- especially useful in the case of very large input files -- reduces the memory footprint of \sorcha simulations. 

\section{Documentation} \label{sub:documentation}

Extensive documentation is \href{https://sorcha.readthedocs.io}{available online}.
The documentation is automatically compiled in the \sorcha \href{https://github.com/dirac-institute/sorcha}{\texttt{Github} repository} and published on ReadTheDocs using the Sphinx\footnote{\url{https://www.sphinx-doc.org/en/master/index.html}} generator. 
The documentation includes guides on installation,  getting started with \sorcha, and examples on how to run the software. The docstrings\footnote{We use the \texttt{NumPy} \citep{harris2020} style for docstrings: see \url{https://numpydoc.readthedocs.io/en/latest/format.html}} within the \sorcha code base are also extracted automatically in the API Reference section\footnote{\url{https://sorcha.readthedocs.io/en/latest/autoapi/index.html}}. 

\section{Inputs} \label{sec:inputs}

\begin{figure}
\begin{center}
\includegraphics[width=1.0\columnwidth]{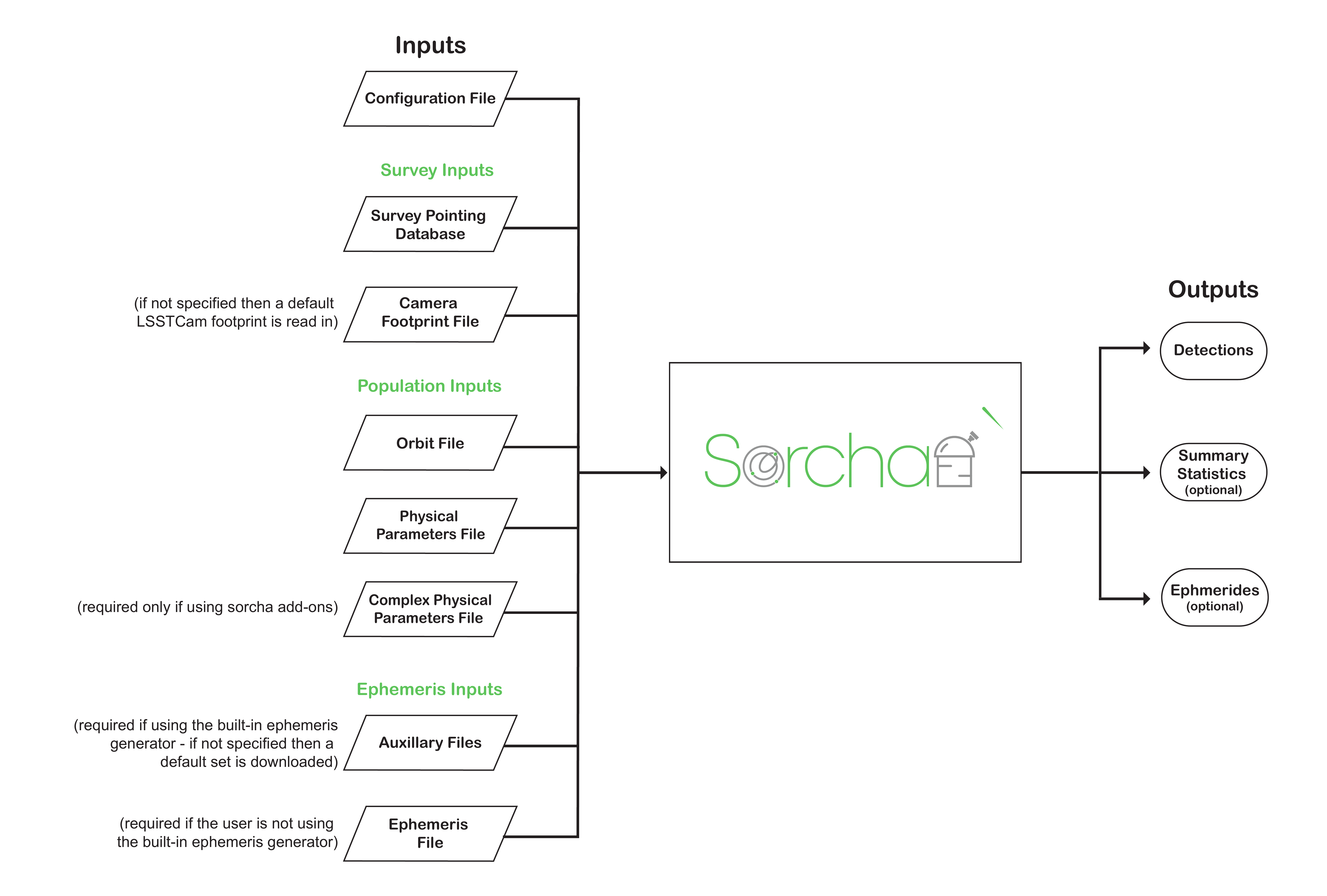}
\caption{\sorcha inputs and outputs. This figure provides an overview of the input files which are ingested by \sorcha and the possible resulting simulation outputs.}
\label{fig:overview}
\end{center}
\end{figure}

\sorcha is executed as a stand-alone \python package requiring a collection of mandatory and optional input files. Behavior specific to the execution and fine-tuning of the simulation is controlled through the configuration file, whereas more general parameters, such as paths to the input small body population files and output locations, are controlled through the CLI. This design allows for efficient simulations using HPC facilities: a large number of differing input small body populations can thus be run concurrently with only minor changes to the command line call, while the overall simulation-level behavior can be easily maintained using a single configuration file. 

\sorcha's input files are divided into three categories: the aforementioned configuration file, which sets aspects of \sorcha's behavior; the files defining the  input synthetic small body population to be run through \sorcha; and the files defining the survey, camera, and telescope. To enable ephemeris generation, \sorcha also requires auxiliary data and resource files to be downloaded once per installation: these are described further in Section~\ref{sub:auxiliary}. An overview of the input files can be seen in Figure~\ref{fig:overview}.

\subsection{Configuration File} \label{sub:config_file}
The configuration file acts as a main control file for \sorcha, allowing the user to customize how \sorcha runs. This configuration file is divided into sections, each containing a number of customizable parameters. Example configuration files can be found in Appendix~\ref{app:configs}. \update{Throughout Sections~\ref{sec:inputs}, \ref{sec:ephemeris}, \ref{sec:post-processing} and \ref{sec:outputs} of this paper,} the configuration parameters which are used to control the aspect of \sorcha under discussion will be noted under a ``relevant configuration file parameters'' heading as an easy reference for the user.

\subsection{Population Input Files} \label{sub:pop_files}
The population input files describe the synthetic solar system object population to be run through \sorcha, providing the information needed to compute the on-sky location/state vector and apparent magnitude of the objects at any given time. These should be supplied as text files, with columns separated by either whitespaces or commas, and with the expected file format given in the configuration file. The files should contain one row for each synthetic object, and the first column must be \texttt{ObjID}, which contains a unique string identifier for each object: the remaining columns can be in any order. 

\subsubsection{Orbit File} \label{subsub:orbits_file}
The orbit file defines the orbital parameters of each of the simulated objects and is used by \sorcha to compute their position on-sky at any given time. The required column headings and input units for each format can be seen in Table~\ref{tab:orbits}. These orbital parameters can be supplied as heliocentric or barycentric defined orbits. The six orbital elements can be defined in Keplerian, cometary or Cartesian format. An epoch column must also be included which defines the moment in time used as the reference point for the orbital elements.
\\
\\
\noindent \textbf{Relevant configuration file parameters:} \texttt{aux\_format}

\begin{deluxetable}{ll}
\tabletypesize{\footnotesize}
\tablecaption{Orbit File Column Headings \label{tab:orbits}}
\tablehead{
\colhead{Heading} & \colhead{Description}
}
\tablecolumns{2}
\startdata
\cutinhead{Cometary Format}
\texttt{ObjID} & Unique object identifier for each input object (string) \\
\texttt{FORMAT} & Orbit format string (COM for heliocentric or BCOM for barycentric) \\
\texttt{q} & Perihelion distance (au) \\ 
\texttt{e} & Eccentricity \\ 
\texttt{inc} & Inclination (degrees)\\ 
\texttt{node} & Longitude of the ascending node (degrees) \\
\texttt{argPeri} & Argument of perihelion (degrees) \\ 
\texttt{t\_p\_MJD\_TDB$^*$} & Time of periapsis passage (MJD) \\ 
\texttt{epochMJD\_TDB$^*$} & Epoch (MJD) \\
\cutinhead{Keplerian Format}
\texttt{ObjID} & Unique object identifier for each input object (string) \\
\texttt{FORMAT} & Orbit format string (KEP for heliocentric or BKEP for barycentric) \\ 
\texttt{a} & Semimajor axis (au) \\
\texttt{e} & Eccentricity \\
\texttt{inc} & Inclination (degrees) \\
\texttt{node} & Longitude of the ascending node (degrees)\\
\texttt{argPeri} & Argument of perihelion (degrees)\\
\texttt{ma} & Mean anomaly (degrees)\\
\texttt{epochMJD\_TDB$^*$} & Epoch (MJD)\\
\cutinhead{Cartesian Format}
\texttt{ObjID} & Unique object identifier for each input object (string)\\
\texttt{FORMAT} & Orbit format string (CART for heliocentric or BCART for barycentric)\\
\texttt{x}$^{\dagger}$ & The x-component of the heliocentric/barycentric position vector (au)\\
\texttt{y}$^{\dagger}$ & The y-component of the heliocentric/barycentric position vector (au)\\
\texttt{z}$^{\dagger}$ & The z-component of the heliocentric/barycentric position vector (au)\\
\texttt{xdot}$^{\dagger}$ & The x-component of the heliocentric/barycentric velocity vector (au/day)\\
\texttt{ydot}$^{\dagger}$ & The y-component of the heliocentric/barycentric velocity vector (au/day) \\
\texttt{zdot}$^{\dagger}$ & The z-component of the heliocentric/barycentric velocity vector (au/day)\\
\texttt{epochMJD\_TDB$^*$} & Epoch (MJD)\\
\enddata
\tablenotetext{^*}{All times/epochs specified in the orbit file are expected to be defined as MJD (Modified Julian Date) in TDB (Barycentric Dynamical Time).}
\tablenotetext{^{\dagger}}{All Cartesian coordinates and velocities are defined in the ecliptic frame.}
\tablecomments{Orbital elements can either be heliocentric or barycentric, independent of format}
\tablecomments{\sorcha requires that the input orbit file have the same format/orbit type throughout.}
\end{deluxetable}

\begin{deluxetable}{ll}
\tablewidth{0pt}
\tablecaption{Physical Parameters File Column Headings\label{tab:physical_parameters}}
\tablehead{\colhead{Heading(s)} & \colhead{Description}}
\startdata
\texttt{ObjID} & Unique object identifier for each input object (string)\\
\texttt{H\_x} & Absolute magnitude of the object in the main filter, where the main filter is \texttt{x} \\
\texttt{u-x}, \texttt{g-x}, etc & Photometric colors in the relevant survey filters\\
\texttt{GS} \textbf{OR} \texttt{GS\_u}, \texttt{GS\_g}, etc & Phase parameter for $HG$ model  \\
\texttt{G1} \textbf{OR} \texttt{G1\_u}, \texttt{G1\_g}, etc & First phase parameter for $HG_1G_2$ model \\
\texttt{G2} \textbf{OR} \texttt{G2\_u}, \texttt{G2\_g}, etc & Second phase parameter for $HG_1G_2$ model \\
\texttt{G12} \textbf{OR} \texttt{G12\_u}, \texttt{G12\_g}, etc & Phase parameter for $HG_{12}$ model \\
\texttt{S} \textbf{OR} \texttt{S\_u}, \texttt{S\_g,} etc & Phase curve parameter for linear model\\
\enddata
\tablecomments{The letter \texttt{x} here represents the (main) filter in which $H$ is defined. The main filter and the phase function are required to be specified in the \sorcha configuration file.  The user also must specify in the configuration file whether all survey observations or a subset of observations in a smaller set of observing filters should be used in the \sorcha simulation. The phase parameter columns required will depend both on the phase function model the user has selected, and whether they are using filter-dependent phase parameters.}
\end{deluxetable}

\subsubsection{Physical Parameters File}\label{subsub:params_file}

The physical parameters file is used to calculate the apparent magnitudes of the input small body population in the filter of each observation, and thus contains the absolute magnitude in a set filter (which we refer to as the \emph{main filter}), photometric colors (provided in the form of offsets with regards to main filter), and phase function parameters for each of the input objects.  \sorcha enables the calculation of the phase function using several different models, as described in Section~\ref{sub:magnitude_calc}, and additionally allows the user to specify filter-dependent phase function parameters. Due to this flexibility, the required columns for this file differ depending on the selected phase function. The column headings for the physical parameters file are detailed in Table~\ref{tab:physical_parameters}.
\\
\\
\noindent \textbf{Relevant configuration file parameters:} \texttt{aux\_format}, \texttt{phase\_function}, \texttt{observing\_filters}

\subsubsection{Complex Physical Parameters File}
\label{subsub:complex_params}
\sorcha has the ability to implement cometary activity and rotational light curves. Any additional parameters required for these models should be supplied in the optional complex physical parameters file. This behavior is described further in Section \ref{sub:add-ons}.
\\
\\
\noindent \textbf{Relevant configuration file parameters:} \texttt{aux\_format}

\begin{deluxetable}{ll}
\tablecaption{Ephemeris File Column Headings\label{tab:ephemeris}}
\tablehead{
\colhead{Heading} & \colhead{Description}
}
\tablecolumns{2}
\tablewidth{0pt}
\startdata
\texttt{ObjID} & Object identifier for each input small body simulated (string)\\
\texttt{FieldID} & Observation pointing field identifier\\
\texttt{fieldMJD\_TAI$^*$} & Observation Mean Julian Date (MJD) in TAI (International Atomic Time)\\
\texttt{fieldJD\_TDB$^*$} & Observation Julian Date in TDB (Barycentric Dynamical Time) \\
\texttt{Range\_LTC\_km} & \update{ Light-time-corrected object-observer distance (km) }\\ 
\texttt{RangeRate\_LTC\_km\_s} & \update{Light-time-corrected rate of change of the object-observer distance (km/s) }\\
\texttt{RA\_deg} & Object right ascension (degrees)\\
\texttt{RARateCosDec\_deg\_day} & Object right ascension rate of motion multiplied by cos(Dec) (deg/day)\\
\texttt{Dec\_deg} & Object declination (degrees)\\
\texttt{DecRate\_deg\_day} & Object declination rate of motion (deg/day)\\
\texttt{Obj\_Sun\_x\_LTC\_km}$^{\dagger}$ & Light-time-corrected Cartesian X-component of the object's heliocentric \update{position} (km)\\
\texttt{Obj\_Sun\_y\_LTC\_km}$^{\dagger}$ & Light-time-corrected Cartesian Y-component of the object's heliocentric \update{position} (km)\\
\texttt{Obj\_Sun\_z\_LTC\_km}$^{\dagger}$ &  Light-time-corrected Cartesian Z-component of the object's heliocentric \update{position} (km)\\
\texttt{Obj\_Sun\_vx\_LTC\_km\_s}$^{\dagger}$ & Light-time-corrected Cartesian X-component of the object's heliocentric velocity (km/s)\\
\texttt{Obj\_Sun\_vy\_LTC\_km\_s}$^{\dagger}$ & Light-time-corrected  Cartesian Y-component of the object's heliocentric velocity (km/s)\\
\texttt{Obj\_Sun\_vz\_LTC\_km\_s}$^{\dagger}$ & Light-time-corrected Cartesian Z-component of the object's heliocentric velocity (km/s)\\
\texttt{Obs\_Sun\_x\_km}$^{\dagger}$ & Cartesian X-component of \update{the} observer's heliocentric \update{position} (km)\\
\texttt{Obs\_Sun\_y\_km}$^{\dagger}$ & Cartesian Y-component of the observer's heliocentric \update{position} (km)\\
\texttt{Obs\_Sun\_z\_km}$^{\dagger}$ & Cartesian Z-component of the observer's heliocentric \update{position} (km)\\
\texttt{Obs\_Sun\_vx\_km\_s}$^{\dagger}$ & Cartesian X-component of the observer's heliocentric velocity (km/s)\\
\texttt{Obs\_Sun\_vy\_km\_s}$^{\dagger}$ & Cartesian Y-component of the observer's heliocentric velocity (km/s)\\
\texttt{Obs\_Sun\_vz\_km\_s}$^{\dagger}$ & Cartesian Z-component of the observer's heliocentric velocity (km/s)\\
\texttt{phase\_deg} & Phase angle between the Sun, object, and observer (degrees)\\
\enddata
\tablenotetext{^*}{All times are reported for the midpoint of the simulated survey observation}
\tablenotetext{^{\dagger}}{Defined in the ecliptic frame.}
\tablenotetext{}{\update{Note: Light-time corrected distances, velocities,  and vectors account for the finite speed of light between the light being reflected from the object's surface and observed by the telescope.}}
\end{deluxetable}

\subsubsection{Ephemeris File}\label{subsub:ephem_file}

The first step in a \sorcha simulation is the determination of which of the input small body population will appear in or near the camera FOV for any given exposure. This process of ephemeris generation, as described in Section~\ref{sec:ephemeris}, requires the numerical N-body integration of each object's orbit over the course of the survey, and is thus the most computationally intensive part of \sorcha. For this reason, the user is given the option to supply their own ephemeris file, either from a previous \sorcha run or from an external ephemeris generator. This is useful in situations in which the user wishes to re-run \sorcha using the same object ephemerides but with a differing configuration or new physical parameters for the synthetic objects. The columns required in the ephemeris file are given in Table~\ref{tab:ephemeris}. Given the very large sizes of the ephemeris output, \sorcha has the capability of ingesting ephemerides in HDF5 (Hierarchical Data Format version 5) format. 
\\
\\
\noindent \textbf{Relevant configuration file parameters:} \texttt{ephemerides\_type, eph\_format}

\subsection{Survey and Telescope Input Files} \label{sub:survey_files}
The survey and telescope input files describe the observational schedule and camera architecture of the survey to be simulated, independent of the input small body population. While \sorcha was built with the LSST in mind, these files allow any survey to be simulated if its pointing history is known or simulated and its camera architecture can be described.

\subsubsection{The Pointing Database} \label{subsub:pointing_database}
The pointing database is a SQLite database containing metadata about the survey pointing history, image depth, and observing conditions. \sorcha uses the observation mid-time, center-of-field right ascension, center-of-field declination, rotation angle of the camera, observation $5\sigma$ limiting magnitude at the center of the FOV, filter, and seeing information to evaluate the effects of various observational biases. 

\sorcha is currently designed to work with the simulated pointing databases generated by the \package{rubin\_sim} package \citep{Connolly2014,yoachim2022} and the Rubin Observatory scheduler, \package{rubin\_scheduler} \citep{delgado2014,yoachim2024}\footnote{The most up-to-date versions of the \package{rubin\_sim}/\package{rubin\_scheduler} cadence simulations can be found at \href{https://s3df.slac.stanford.edu/data/rubin/sim-data/}{https://s3df.slac.stanford.edu/data/rubin/sim-data/}.}. This software is being used to simulate the cadence and on-sky coverage of differing survey strategies. A description of an early version of this \python software can be found in  \cite{Connolly2014}, \citet{delgado2014}, and \cite{jones2018}; further details can be found in \citet{schwamb2023}. Once the LSST has begun operations, future versions of \sorcha will be modified to read the actual pointing history.
\\
\\
\noindent \textbf{Relevant configuration file parameters:} \texttt{pointing\_sql\_query}

\subsubsection{The Camera Footprint File} \label{subsub:telescope_file}
The architecture of the survey camera, including the shape of its footprint and any gaps between its detectors, is relevant for object detectability. By ingesting a map of the camera detector layout and taking into account field rotation from the pointing database, \sorcha can track where the input objects land on the detectors and remove those which have fallen into gaps between the detectors and those fall on the very edges of the CCD, where they may not be detected (see Section~\ref{sub:footprint} for more details). \sorcha comes with a representation of the LSST Camera (LSSTCam) architecture already installed: this is used if the full camera footprint is requested in the configuration file and no alternative camera footprint file location is supplied. However, the user can also supply their own comma-separated text file describing the survey camera footprint in the configuration file. The expected columns for this file are given in Table~\ref{tab:footprint}.

\begin{deluxetable}{ll}
\tablecaption{Camera Footprint File Column Headings\label{tab:footprint}}
\tablehead{
\colhead{Heading} & \colhead{Description}
}
\tablecolumns{2}
\tablewidth{0pt}
\startdata
\texttt{detector} & Detector identifier (integer)\\
\texttt{x} & x position of the detector corner on the focal plane (float) \\
\texttt{y} & y position of the detector corner on the focal plane (float)\\
\enddata
\tablecomments{The x and y values are unitless and are respectively equal to \update{$\tan(\textrm{ra})$ and $\tan(\textrm{dec})$}, where ra and dec are the vertical and horizontal angles of the points from the center of the sphere tangent to the origin in the focal plane, \update{that is, whose origin is at the nominal boresight angle of the focal plane}. For each detector, all four corners must be specified in the camera footprint file.}
\end{deluxetable}

\noindent \textbf{Relevant configuration file parameters:} \texttt{camera\_model}, \texttt{footprint\_path}

\section{Ephemeris Generation} \label{sec:ephemeris}
After ingestion and validation checks of the input files, \sorcha proceeds to ephemeris generation (if requested by the user). The generation of ephemerides was one of the most challenging steps within \sorcha to optimize as the time span of the survey, limiting magnitude, and sky coverage of the LSST are unprecedented.  Traditional brute force methods of calculating the on-sky positions for every input orbit at every survey observation are too slow when scaled to LSST discovery rates (even when deploying on high performance computing clusters). The LSST is estimated to generate approximately 1 billion photometric/astrometric detections of approximately 5 million solar system objects \citep{lsst-sciencebook-ch5-2009}. To simulate the LSST survey detections (especially the main-belt asteroids) requires simulating a significantly larger population of synthetic objects than will be detected, with all of those orbits needing to be run through \sorcha with ephemerides predicted. In \sorcha, this barrier has been overcome by a combination of predicting positions at set times per night, on-sky grids, and interpolation. Full details about the implemented algorithm are described in Holman et al (in press); we provide a brief summary below. Additional details on how the auxiliary data files required by the ephemeris generator are downloaded and stored can be found in Section \ref{sub:auxiliary}. 

\sorcha's internal ephemeris generator iterates through each survey observation ingested from the pointing database, determining the heliocentric Cartesian positions for every model orbit at the time of the observation, converting these positions to right ascension and declination on-sky at the observatory's location, and identifying which of the \update{input} synthetic small bodies are within a given radius of the observation's pointing center. This radius is specified by the user in the configuration file (the \texttt{ar\_ang\_fov} + \texttt{ar\_fov\_buffer}) and should be larger than the survey's camera FOV, as later on in the simulation \sorcha can perform a more refined filtering using a model or approximation for the camera footprint (see Section \ref{sub:footprint}). The user also provides the relevant survey's MPC observatory code in the configuration file.  The right ascension, declination, and heliocentric state vector (Cartesian position and velocity) are returned for each synthetic small body within the search radius of the observation pointing. Using the location of the observatory, the phase angle and the light-time-corrected state vector are also computed and returned, as these will be later used when calculating the apparent magnitudes as described in Section \ref{sub:magnitude_calc}. All the values that are calculated by the ephemeris generator and passed on to the post-processing stage can be found in Table \ref{tab:ephemeris}. 

 \sorcha's ephemeris generator is powered by ASSIST \citep{holman2023}, an open-source \python and C99 software package for producing ephemeris-quality integrations of solar system test particles. 
 ASSIST is an extension of the REBOUND N-body package \citep{rein2012} and uses its IAS15 \citep[Gauss-Radau integrator with Adaptive Step-size control, 15th order][]{rein2015} integrator to integrate test particle trajectories in the field of the Sun, Moon, planets, and 16 massive asteroids and dwarf planets (listed in Table \ref{tab:perturber}).  The positions of the masses come from the JPL (Jet Propulsion Laboratory) DE440/441 ephemeris and its associated asteroid perturber file (see Section \ref{sub:auxiliary}). ASSIST includes the most significant gravitational harmonics and general relativistic corrections, and it also accounts for position- and velocity-dependent non-gravitational effects.  With this level of detail, ASSIST closely matches the results of JPL's small bodies integrator, a current standard of reference. Note: including any of the 16 bodies in a \sorcha simulation will yield inaccurate results due to gravitational self-perturbation.   A user calculated external ephemeris (see Section \ref{subsub:ephem_file})  could be used as input into a \sorcha simulation if these bodies do need to be included. Straightforward modifications to  \sorcha's internal ephemeris generator would be needed to initialize ASSIST with an additional massive perturber such as some of the recently proposed theories of a planetary-mass beyond Neptune \citep{trujillo2014, batygin2016, sheppard2016, brown2019, batygin2019, brown2021, oldroyd2021,lykawa2023}.  We note that we have designed this part of the software such that it would be straightforward in future versions of \sorcha to potentially incorporate other N-body integrators as additional engines for the generator. 
 \\
\\
\noindent \textbf{Relevant configuration file parameters:} \texttt{ephemerides\_type, ar\_ang\_fov, ar\_fov\_buffer, ar\_picket, ar\_obs\_code, ar\_healpix\_order, ar\_n\_sub\_intervals   }

\begin{deluxetable}{l||l}
\tablecaption{List of MBA and Dwarf Planet Perturbers Included Within \sorcha's Ephemeris Generation Stage \label{tab:perturber}}
\tablehead{
\colhead{Asteroid MPC Designation } &  \colhead{Asteroid MPC Designation }
}
\tablecolumns{2}
\tablewidth{0pt}
\startdata
(107) Camilla & \update{(704) Interamnia} \\
(1) Ceres  & (7)  Iris \\ 
(65) Cybele  & (3) Juno \\ 
 (511) Davida   & (2) Pallas \\
 (15) Eunomia  &  (16) Psyche \\
(31) Euphrosyne  &  (87) Sylvia \\
  (52) Europa  &  \update{(88) Thisbe}   \\ 
 (10) Hygiea &  (4) Vesta 
\enddata
\tablecomments{As these objects and their masses are included with the ASSIST integrations performed duing the ephemeris generation stage. Poor predictions will result if these specific MBAs and dwarf planets are simulated within \sorcha when using the simulator's internal ephemeris engine.}
\end{deluxetable}


\section{Post-Processing} \label{sec:post-processing}
Once the ephemerides have been generated or read in from an external file, \sorcha moves on to the second phase, which we call post-processing. For each of the input objects, \sorcha goes through the potential detections identified in the ephemeris generation step and performs a series of calculations and assessments in the post-processing stage to determine whether the objects would have been detectable as a source in the survey images and would have later been identified as a moving solar system object. \sorcha combines the calculated ephemerides with the physical parameters of the input objects and the effects originating from the telescope optics and automated data processing pipelines.  It then uses that to predict the apparent magnitude and eventual discoverability of the input small body population. This is performed by applying a sequential series of modular steps to every potential detection of the input objects, each of which calculates a particular attribute (such as apparent brightness) or simulates a specific optical/structural/software phenomenon/characteristic of the astronomical survey. Each of these steps can be controlled or modified to suit the user's needs via the configuration file, facilitating a high level of flexibility. An overview of the post-processing workflow for simulating what the LSST should find is shown in  Figure \ref{fig:main_workflow} in green and magenta; the sections below describe each stage of post-processing in detail, including the relevant configuration file parameters and code functions. Each of these functions has been separately validated to ensure correct operation: validation notebooks for each function can be found in the \href{https://sorcha.readthedocs.io}{online documentation} and within the main \sorcha \href{https://github.com/dirac-institute/sorcha}{repository on \texttt{GitHub}}.

\subsection{Rubin Apparent Magnitudes} 
\label{sec:rubin_apparent_mags}

\begin{figure}
    \centering
    \includegraphics[width=0.9\columnwidth]{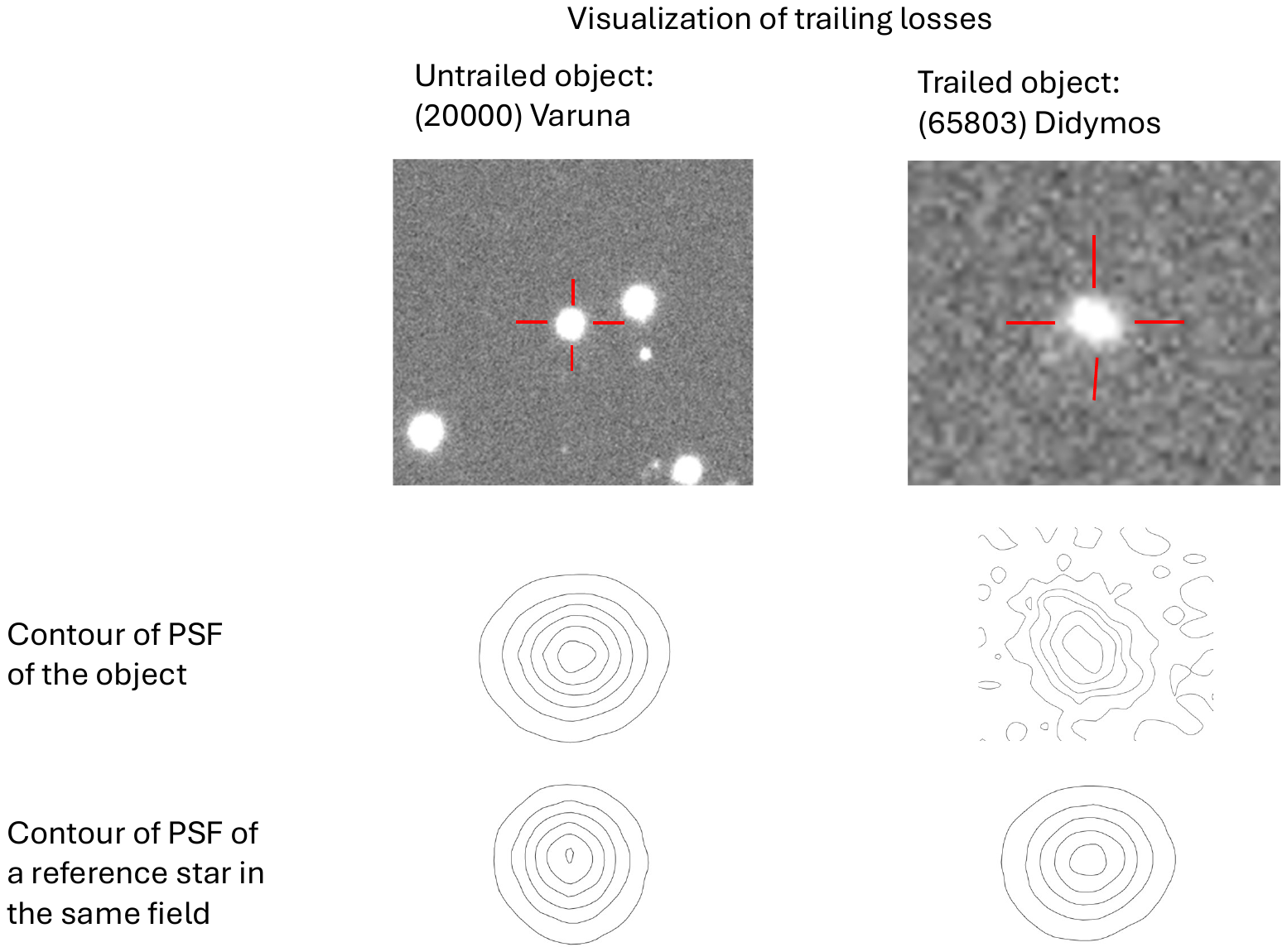}
    \caption{The trailing of moving solar system objects visualized with Dark Energy Survey \citep[DES;][]{abbott2021} observations. The top two rows show the image cutouts and PSF (point spread function) contours of example solar system objects (identified by the red crosshairs in the cutouts). The last row shows the PSF contours for representative reference stars within the sane images. Slow moving TNO (20000) Varuna has a stellar-like PSF (shown on the \textbf{left}) and fast moving NEO (65803) Didymos has an extended significantly trailed PSF as shown on the \textbf{right}. \update{Dark Energy Camera \cite[DECam,;][]{flaugher2015b} data was selected because of the expected similarity to the LSSTCam images} \citep{ivezic2019,bianco2022}.}
    \label{fig:trailing_loss}
\end{figure}

\begin{figure}
\begin{center}
\includegraphics[width=0.60\columnwidth]{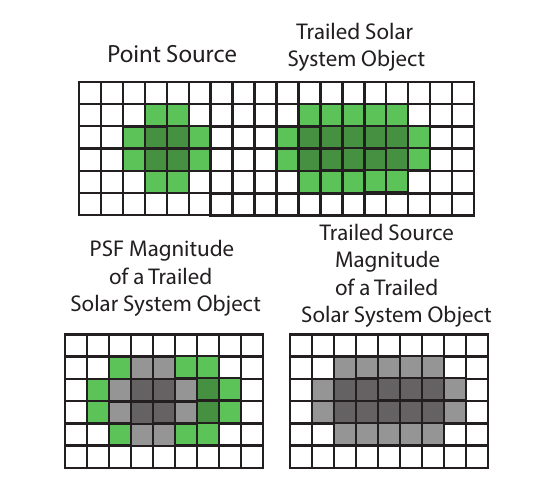}
\caption{A cartoon schematic depicting the difference between how the trailed source magnitude and the PSF (point spread function) magnitude for a moving solar system object observed on an LSST image are estimated. The Rubin Data Management source detection pipeline, the Difference Image Analysis (DIA) pipeline, uses PSF filter matching to find sources on the image. This pipeline will measure the PSF magnitude. Only transient sources identified by the DIA pipeline will be sent on to the  Rubin Solar System Processing (SSP) pipelines to search for moving objects. The SSP pipelines will report the trailed source magnitude. }
\label{fig:PSF_trailed_source_mags}
\end{center}
\end{figure}

In order \update{to} assess the detectability of the input synthetic small body population in LSST observations, \sorcha calculates two apparent magnitudes (which we refer to as the \emph{trailed source magnitude} and the \emph{PSF [point spread function] magnitude}). Unlike background stars, a moving solar system object may leave a streak on the detector, depending on the object's on-sky velocity, the image exposure time, the camera's pixel resolution, and the image quality, resulting in an extended PSF (``trail'') spanning more pixels than a point source would. This effect is shown in Figure~\ref{fig:trailing_loss}. The \emph{trailed source magnitude} (described in Section \ref{sub:magnitude_calc}) is the true apparent magnitude of the object (and the apparent magnitude that will be eventually calculated by the Rubin SSP pipelines), whereas the \emph{PSF magnitude} (discussed in Section \ref{sub:trailing_losses}) is the effective brightness of the solar system object measured by the Rubin source detection algorithm (Difference Image Analysis; DIA). The PSF magnitude accounts for the loss in flux from the Rubin source detection algorithms which use  stellar PSF-like matched filter to identify transient sources in the survey's difference images. The difference between these  two  magnitudes is illustrated in Figure \ref{fig:PSF_trailed_source_mags}. For the LSST and its expected $\sim$30 s exposures, objects whose rates of motion are $0.16 \, \mathrm{deg}/\mathrm{day}$ will trail $0.2''$, the size of a LSSTCam pixel. With a median expected seeing of $0.7'' \approx 3.5 \, \mathrm{pixels}$ \citep{ivezic2019}, we then expect that objects with rates of motion $\gtrsim 0.4 \, \mathrm{deg}/\mathrm{day}$, such as the closest MBAs and NEOs, will have trails roughly comparable to \update{their PSF full width at half maximum (FWHM)}, whereas the most distant objects (\update{e.g.} TNOs) will mostly appear as point sources (see Figure \ref{fig:trailing_loss}). The longer the trail,  the larger the difference between the trailed source magnitude and the PSF magnitude. For TNOs and more distant objects, the PSF magnitude and  the trailed source magnitude will be nearly identical because of their slow rates.

\begin{figure}
\begin{center}
\includegraphics[width=0.60\columnwidth]{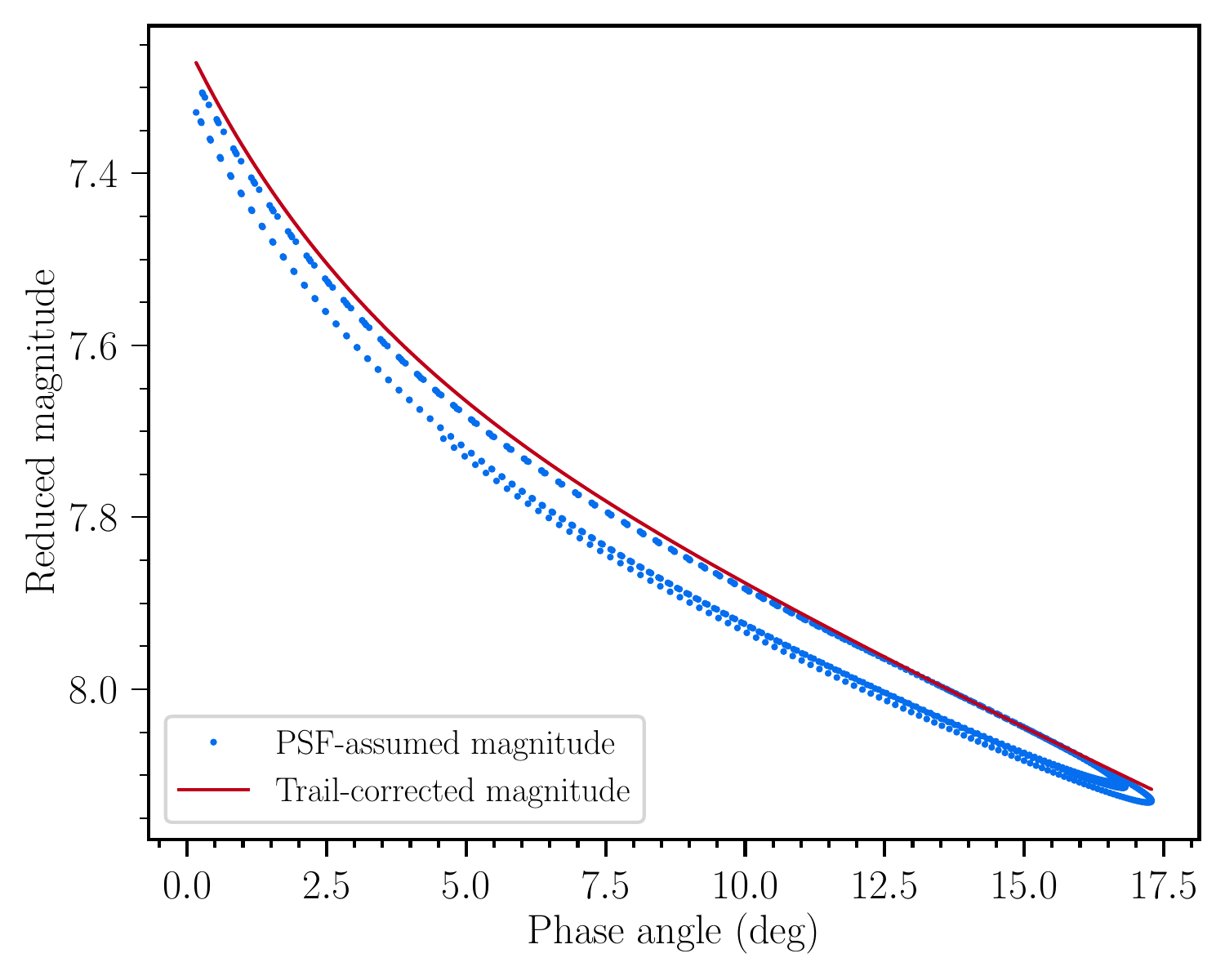}
\caption{Demonstration of the difference between the trailed source magnitude and the PSF magnitude. We plot the reduced magnitude derived from the PSF magnitude (blue dots) and the trailed source magnitude (red curve), simulated for asteroid (24) Themis. The PSF magnitude measures less flux because it is not sampling using the full image trail from the simulated detection, and therefore a clear flux loss can be seen in the reduced  PSF magnitude, that varies with the on-sky velocity of Themis, compared to the reduced magnitude calculated from the trailed source magnitude. For any photometric analyses of simulated detections/discoveries output by \sorcha, the user should use the reported trailed source magnitude. }
\label{fig:psf_vs_trail_recoverymag}
\end{center}
\end{figure}

When analyzing the detections and discoveries output from a \sorcha simulation, we caution the reader to only use the trailed source magnitude. Using the PSF magnitude will give incorrect results because it is missing some of the object's flux on the LSST image. To demonstrate this effect, we have simulated a toy scenario with two years of daily (between 2021 and 2023), single band, noiseless observations of asteroid (24) Themis, from the LSST site.  We use the $HG$ photometric model \citep[][as described in Section \ref{sub:magnitude_calc}]{bowell1989}, assuming $H = 7.25$ and $G = 0.19$. We then estimate the PSF magnitude from the prescription in Section \ref{sub:trailing_losses}. The reduced magnitude (distance-corrected magnitude) is calculated for each scenario, as shown in Figure \ref{fig:psf_vs_trail_recoverymag}. The reduced magnitudes estimated from the PSF magnitudes do not follow the shape of the absolute magnitude as a function of phase angle expected from the $HG$ model. To quantify this trend, we have performed a simple least-squares fit to this model, using an independent (i.e. not the one used by \sorcha) implementation of the $HG$ model. We recover, in the case of the PSF magnitude, $H = 7.29$, $G = 0.22$, demonstrating a bias toward fainter magnitudes, as expected. The same procedure on the reduced magnitudes estimated from the trailed source magnitude leads to the expected values of $H$ and $G$ to high precision. 

 \subsection{Calculating the Trailed Source Magnitude} \label{sub:magnitude_calc}

\sorcha first computes the trailed source magnitude and then later calculates the discovery trailing losses and the resulting PSF magnitude, described later in Section \ref{sub:trailing_losses}. The trailed source magnitude is the measured apparent magnitude when all pixels in the trail are included. When simulating a non-active body (e.g. no cometary activity or production of a dust tail/coma), we calculate the trailed source magnitude $m_{\textrm{trailed source,}i}$ in a given filter $i$ for a small body at a time $t$ as: 
\begin{equation} \label{eq:perfect_brightness}
m_{\textrm{trailed source,}i} = H_x + 5 \log_{10} (r \Delta) + \Phi(\alpha,i) + C(i-x) + \delta(i,t) 
\end{equation}

where: 

\begin{itemize}
\item $r$ and $\Delta$, respectively, are the heliocentric and geocentric distance to the object; 
\item $\alpha$ is the phase angle (the angle between the Sun, the object, and the observer); 
\item $H_x$ is the absolute magnitude of the object in photometric filter $x$ (referred to in \sorcha as the \emph{main filter}), defined as the trailed source magnitude at  opposition at the distance of 1 au both from the Sun and the observer at $\alpha=0$;
\item $\Phi(\alpha, i)$ is the disk-integrated phase function evaluated at phase angle $\alpha$ for filter $i$; 
\item $C(i-x)$ is the ($i-x$) color of the object to convert the apparent magnitude from the main filter ($x$) to the observed filter ($i$);  
\item $\delta(i,t)$ is an optional contribution from the rotational light-curve in filter $i$ at time $t$.
\end{itemize}
The phase function ($\Phi(\alpha,i)$) model applied by \sorcha is specified by the user in the configuration file, and we use the implementations available in the small body \python package \texttt{sbpy} \citep{mommert2019}. The models currently available for estimating the phase curve contribution are:
\begin{itemize}
    \item none (no phase function applied) 
    \item $HG$ \citep{bowell1989}
    \item $HG_1 G_2$ \citep{muinonen2010}
    \item Modified $HG_{12}$ \citep{penttila2016}
    \item linear 
\end{itemize}
The relevant phase curve parameters are specified per object in the physical parameters file. The phase function is filter-dependent \citep[e.g. ][]{alvarez-candal2022,dobson2023,robinson2024}. The user can either provide filter-specific phase curve parameters or single values applied to all survey filters evaluated (see Section \ref{subsub:params_file}).
We allow the user to optionally supply their own self-developed models for rotation effects, $\delta(i,t)$. This functionality is described further in Section \ref{sub:add-ons}. 

In the case of simulating active objects, \sorcha performs one more step. It assumes that the trailed source magnitude calculated in Equation \ref{eq:perfect_brightness} is the apparent magnitude of the nucleus ($m_{\textrm{nucleus trailed source,}i}$) and the trailed source magnitude of the nucleus and coma/tail together is estimated as

\begin{equation} \label{eq:activity}
m_{\textrm{trailed source,}i} = A(r, \Delta, \alpha,i, t, m_{\textrm{nucleus trailed source,}i}) 
\end{equation}
where $A(r, \Delta, \alpha, i,t,  m_{\textrm{nucleus trailed source,}i}$) is provided by the user. The user's self-developed models for cometary activity are expected to compute the flux contribution from any coma/tail and combine this flux with the contribution from the nucleus. This functionality is described further in the next section (Section \ref{sub:add-ons})
\\
\\
\noindent \textbf{Relevant configuration file parameters:}  \texttt{phase\_function},  \texttt{observing\_filters}
\\ 
\\
\textbf{Code function:}  \texttt{PPCalculateApparentMagnitude}

\subsection{Optional Activity and Light Curve Models}\label{sub:add-ons}
To support future extensions without opening the original source code to external contribution, \sorcha makes use of a technique that allows users to use their own models for comet activity or rotational light curve effects by providing abstract base classes (\texttt{AbstractCometaryActivity} and \texttt{AbstractLightCurve}) from which custom implementations can inherit. At run time, \sorcha automatically registers all classes found in the working environment which inherit from these abstract classes and makes those available for use within the code. Thus, any user-defined subclass can be used within \sorcha, without having to modify \sorcha's source code. To make it easier for new users to develop their own models,  we provide example subclasses for both comet activity and light curve models (\texttt{LSSTCometActivity} and \texttt{SinusoidalLightCurve}) that demonstrate the bare minimum required implementation: these can be found in the \sorcha \texttt{add-ons} package\footnote{\url{https://github.com/dirac-institute/sorcha-addons}}. 
To facilitate the use of these add-ons, \sorcha allows the ingestion of an optional complex physical parameters file, in which the user may include any necessary parameters for their models. These depend on the chosen model of variability and activity, and must reflect the necessary columns for the user-specified model. 

\subsubsection{Comet Activity} \label{subsub:comet_activity}
At Earth, most of what is observed from an active solar system object is due to mass loss. The measured flux from a comet will have a contribution from the ejected dust. The dust particles injected into a coma around the nucleus or shaped into a tail will scatter and reflect sunlight towards the observer (see \cite{agarwal2023}, \cite{jewitt2022}, and references therein). In \sorcha, the input absolute magnitude is assumed to be the value of the nucleus alone, but we have developed the framework to allow the calculated apparent magnitude to be modified to mimic the effect of cometary activity or a collision using the cometary activity class within the \sorcha add-ons package. \sorcha assumes that the active object is still a point source with its apparent brightness modified by the activity. Given the complexities in tail/comae shapes, \sorcha does not adjust the PSF of the simulated small body, but the detectability of the active object will change due to the flux enhancement as a result of the activity. Adapting from the base class, a user can develop or implement their own function to take the nucleus' measured flux (which can include a contribution from a rotational light curve as described in Section \ref{subsub:light_curves}) and add on an additional contribution for an unresolved coma/tail. This may be more representative of weak cometary activity \citep[e.g. ][]{cochran1989} or where the coma is bound to the object like in case of the recent epoch of cometary activity on Centaur Chiron \citep{dobson2021,dobson2023, pinillaalonso2024}. The input parameters needed to calculate the contribution from the cometary activity should be \update{provided in} the complex physical parameters file (see Section \ref{subsub:ephem_file}). 

\begin{figure}
    \centering
    \includegraphics[width=0.9\columnwidth]{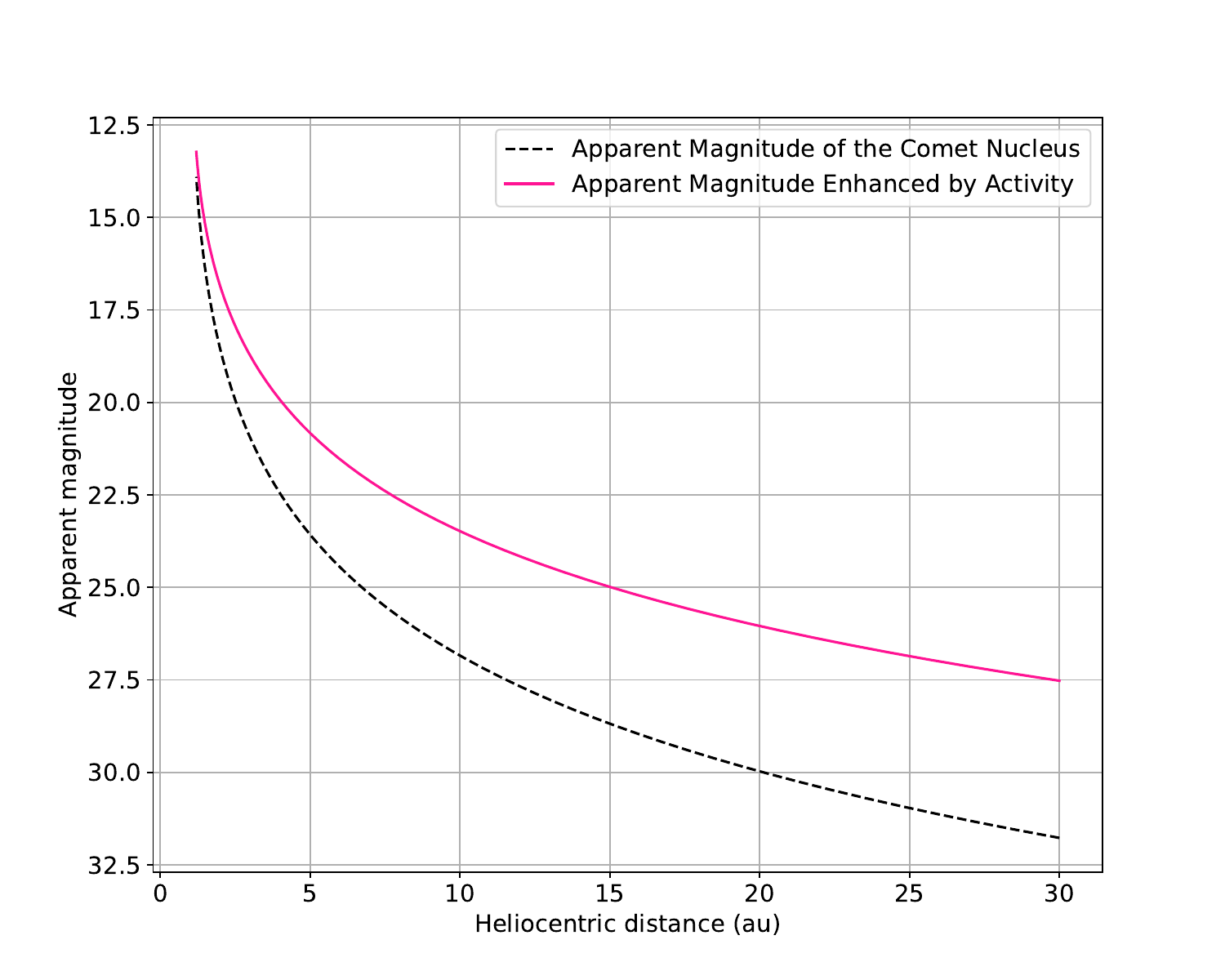}
    \caption{The \texttt{LSSTCometActivity} class's cometary activity model included in the \sorcha \texttt{add-ons} package is demonstrated in this figure. The apparent magnitude (trailed source magnitude) in $r$-band as a function of heliocentric distance is plotted for an inactive comet nucleus (black) and an active comet (Af$\rho_1$=150 cm and $k$=-0.3.). Both simulated objects have a nucleus of $H_r$=17. No phase function effects are included in order to highlight the impact from cometary activity($\alpha$=0 is assumed at all distances). }
    \label{fig:simple_comet}
\end{figure}

In \sorcha \texttt{add-ons}, we have implemented  the \texttt{LSSTCometActivity} cometary activity subclass. Dust production is typically parameterized by a comet's $Af\rho$ value \citep{ahearn1984}, and can be calculated from observed quantities within a fixed physical photometric aperture:
\begin{equation}
Af\rho = \frac{4r^2\Delta^2 F_{\textrm{comet}}}{\rho F_{\odot}}
\end{equation}
where $r$ is the heliocentric distance of the comet measured in au, $\Delta$ is the geocentric distance of the comet in centimeters, $F_{\textrm{comet}}$ is the measured flux of the comet in the observing band, and $F_{\odot}$ is the flux of the Sun at 1 au in the observing band, and $\rho$ is the physical size of the photometric aperture in centimeters (typically selected to be 10,000 km). The activity/dust production varies as a function of heliocentric distance. The user provides the $V$-band $Af\rho$ value of the comet at 1 au ( $Af\rho_1$)and the power-law slope that describes how the activity varies with heliocentric distance ($k$). The $Af\rho$ at the given observation is estimated as:
\begin{equation}
Af\rho(r) = Af\rho_1 r^k \textrm{f}(\alpha)
\end{equation}
where $Af\rho_1$ is the $V$-band $Af\rho$ value at 1 au, $k$ is the activity power law slope, $r$ is the heliocentric distance in au,  and f$(\alpha)$ is the Halley-Marcus phase function  at phase angle ($\alpha$) from Schleicher\footnote{\url{https://asteroid.lowell.edu/comet/dustphase/}}. The estimated $Af\rho(r)$ is calculated and transformed into an apparent magnitude in the observing filter ($i$) for the coma assuming a 1$^{\prime\prime}$ aperture and solar color for the dust, from \citet{willmer2018}.  Because of the uncertainty in what the profile of a coma should look like as a function of heliocentric distance and dust/gas production, the \texttt{LSSTCometActivity} class only applies an enhancement to the photometry. The coma/tail is treated as unresolved, meaning the active body has the same PSF as a non-active object of the same apparent brightness. This approximation will  be valid for very distant active objects and those bodies exhibiting low level activity where the contribution to the apparent magnitude from the coma will dominate the apparent magnitude (see Figure \ref{fig:simple_comet}). 
\\
\\
\noindent \textbf{Relevant configuration file parameters: }  \texttt{comet$\_$activity}

\subsubsection{Light Curves} \label{subsub:light_curves}
\begin{figure}
    \centering
    \includegraphics[width=0.9\columnwidth]{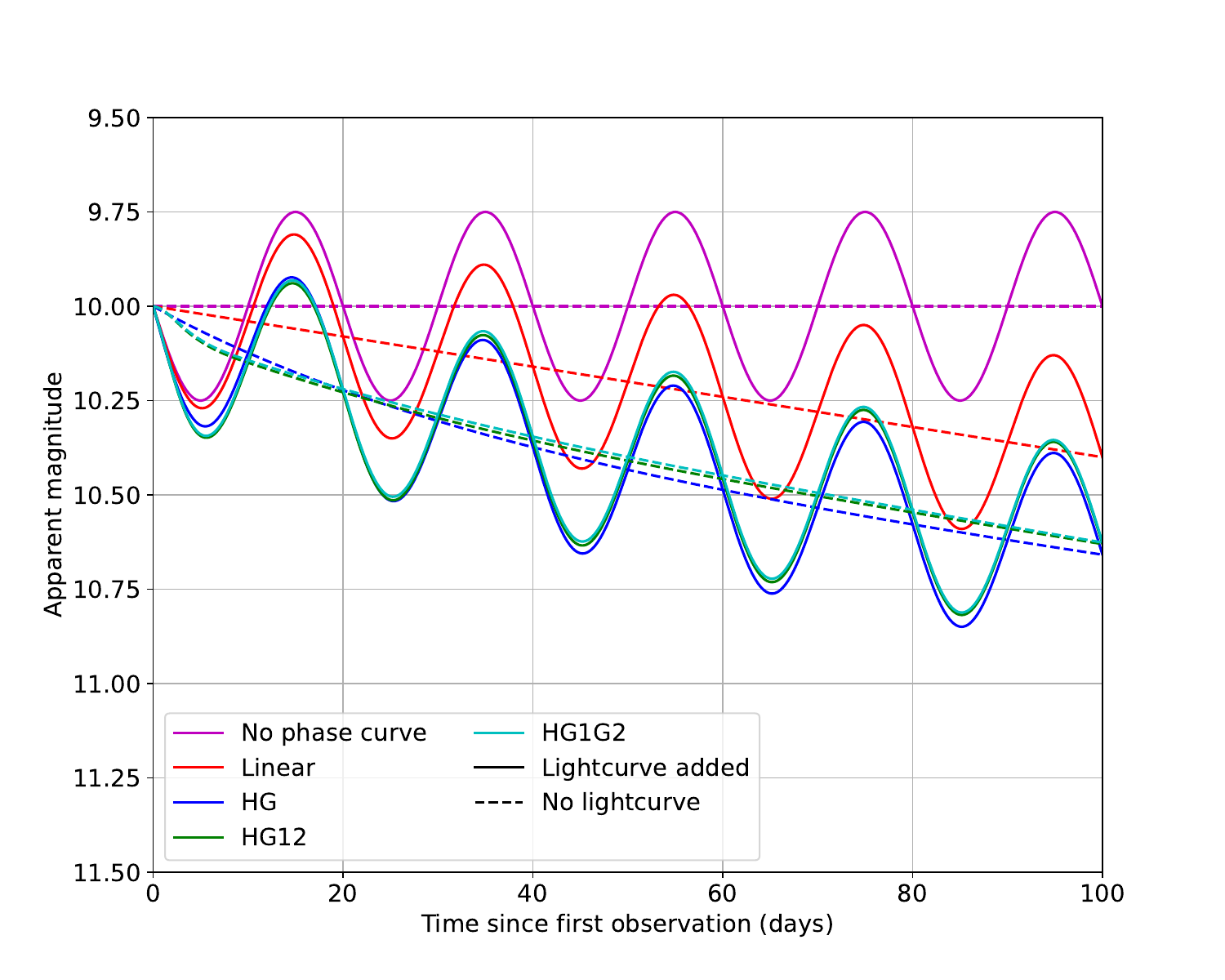}
    \caption{The effects of the simple sinusoidal light curve model included in the \sorcha \texttt{add-ons} package for different phase function models. The dashed lines show the unmodified apparent magnitude (trailed source magnitude); the solid lines show the apparent magnitude after the model has been applied.}
    \label{fig:simple_lightcurve}
\end{figure}

The light curves of small bodies are typically caused by rotation effects, where the variability is caused by a combination of effects such as the non-sphericity of the object, inhomogeneity in surface material, or the presence of an (unresolved) companion \citep[e.g. ][]{durech2010,lazzarin2010,showalter2021,chang2021,durech2022,ashton2023, hasegawa2024}. Further, with longer time scales (comparable to the expected 10 years of operations of LSST), the change in the aspect angle of the object can also lead to a different shape of the light curve. To be able to reproduce this wide variety of variability, and its effect on object discoverability, all the information associated with the object and the potential detections's survey observation are exposed to the usersupplied light curve calculation, which includes any input light curve parameters, as well as the additional information computed in \sorcha (such as the full state vector of the object during the visit). Currently, the \sorcha \texttt{add-ons} package includes only a simple sinusoidal change in magnitude (the \texttt{SinusoidalLightCurve} subclass): the effect of this model is shown in Figure~\ref{fig:simple_lightcurve}. However, the modular nature of the package and the access to the entire physical state of the object enable general calculations of the light curve in both magnitude and flux space.
\\
\\
\noindent \textbf{Relevant configuration file parameters: }  \texttt{lc$\_$model} 

\subsection{Calculating Trailing Losses and the PSF Magnitude} \label{sub:trailing_losses}

As mentioned in Section \ref{sec:rubin_apparent_mags}, depending on the camera pixel scale, object on-sky velocity and observation exposure time, solar system objects can produce elongated or trailed PSFs in survey images. In order to estimate the astrometric and photometric uncertainties and determine the PSF magnitude, \sorcha calculates for each potential detection the equivalent photometric losses caused by the elongated PSFs. \sorcha calculates these losses as magnitude offsets to the trailed source magnitude. There are two different trailing losses that must be calculated. The first is the trailing loss due to the smearing of the photometric signal over a larger number of pixels than for a point-source PSF, $\Delta m(\textrm{PSF}$) (the \emph{PSF trailing loss}). The second is the trailing loss due to the Rubin Data Management software \citep{juric2017} detection algorithm attempting to identify sources on the image using a stellar PSF-like matched filter. We refer to this as the \emph{detection trailing loss},  $\Delta m(\textrm{PSF + detection})$, as it accounts for both the matched filter excluding part of the trail and the SNR losses due to the object's flux being distributed differently than a point source for the pixels that are included by the detection algorithm. 

The  PSF magnitude $(m_{\textrm{PSF}})$  and the  trailed source magnitude ($m_{\textrm{trailed source}}$) are related by:

\begin{equation} \label{eq:trailed_source_PSF_mag_relation}
m_{\textrm{PSF}} = m_{\textrm{trailed source}}+\Delta m(\textrm{PSF + detection})
\end{equation}
The PSF trailing loss will be used later to calculate the uncertainty of the trailed source magnitude (detailed in Section \ref{sub:uncertainties}) as  $m_{\textrm{trailed source}}+\Delta m(\textrm{PSF})$ provides the apparent magnitude of a point-source with the equivalent SNR as what would be measured for the trailed PSF. In \sorcha, the PSF magnitude is calculated and stored. $\Delta m(\textrm{PSF}$) is then calculated separately and is temporarily stored as an auxiliary value. $\Delta m(\textrm{PSF}$) is later used to properly calculate the overall signal-to-noise ratio (SNR) of the potential detection for an assessment of astrometric and photometric errors of the trailed source magnitude downstream (Section \ref{sub:uncertainties}). 
 
\begin{figure}
    \centering
    \includegraphics[width=0.49\columnwidth]{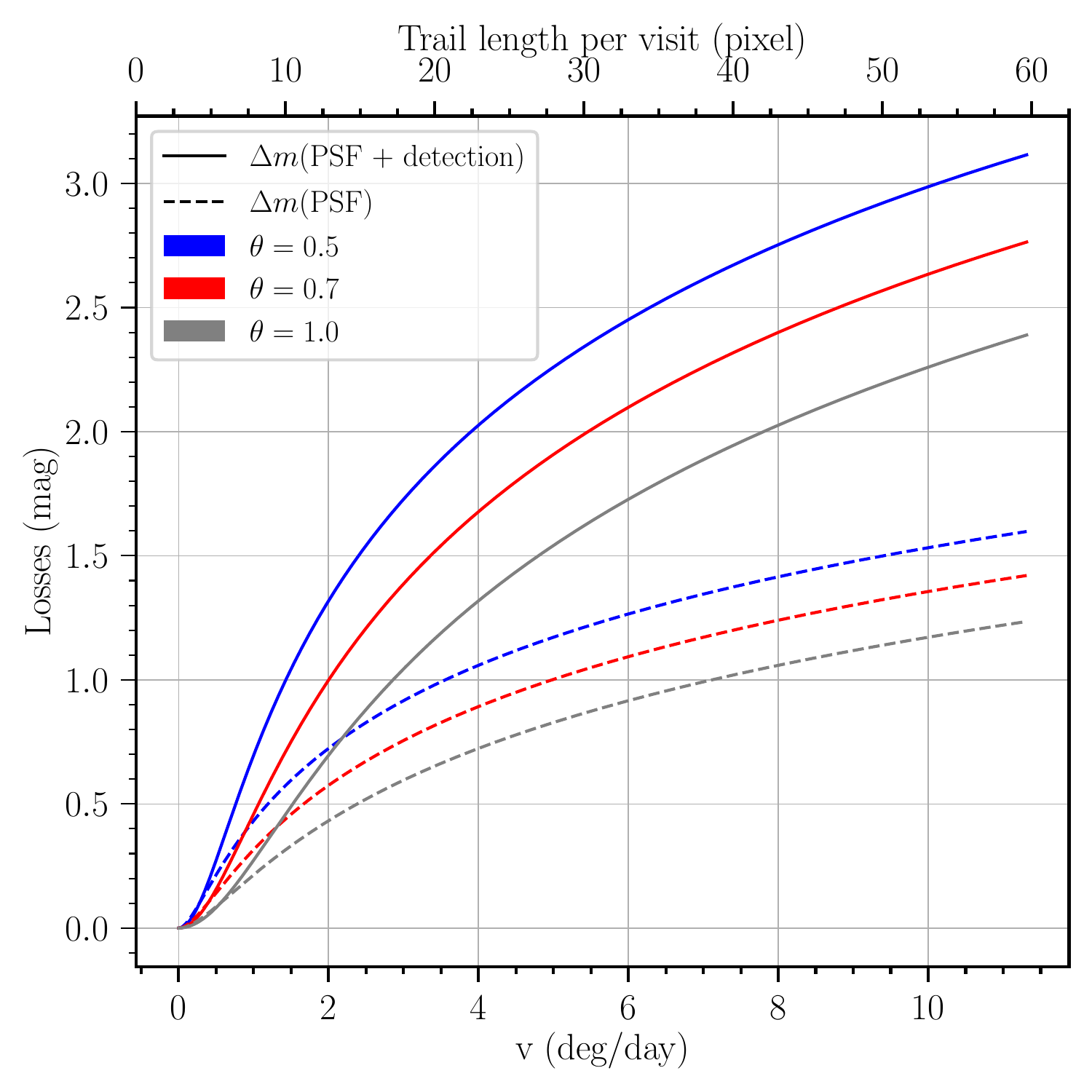}
        \includegraphics[width=0.49\columnwidth]{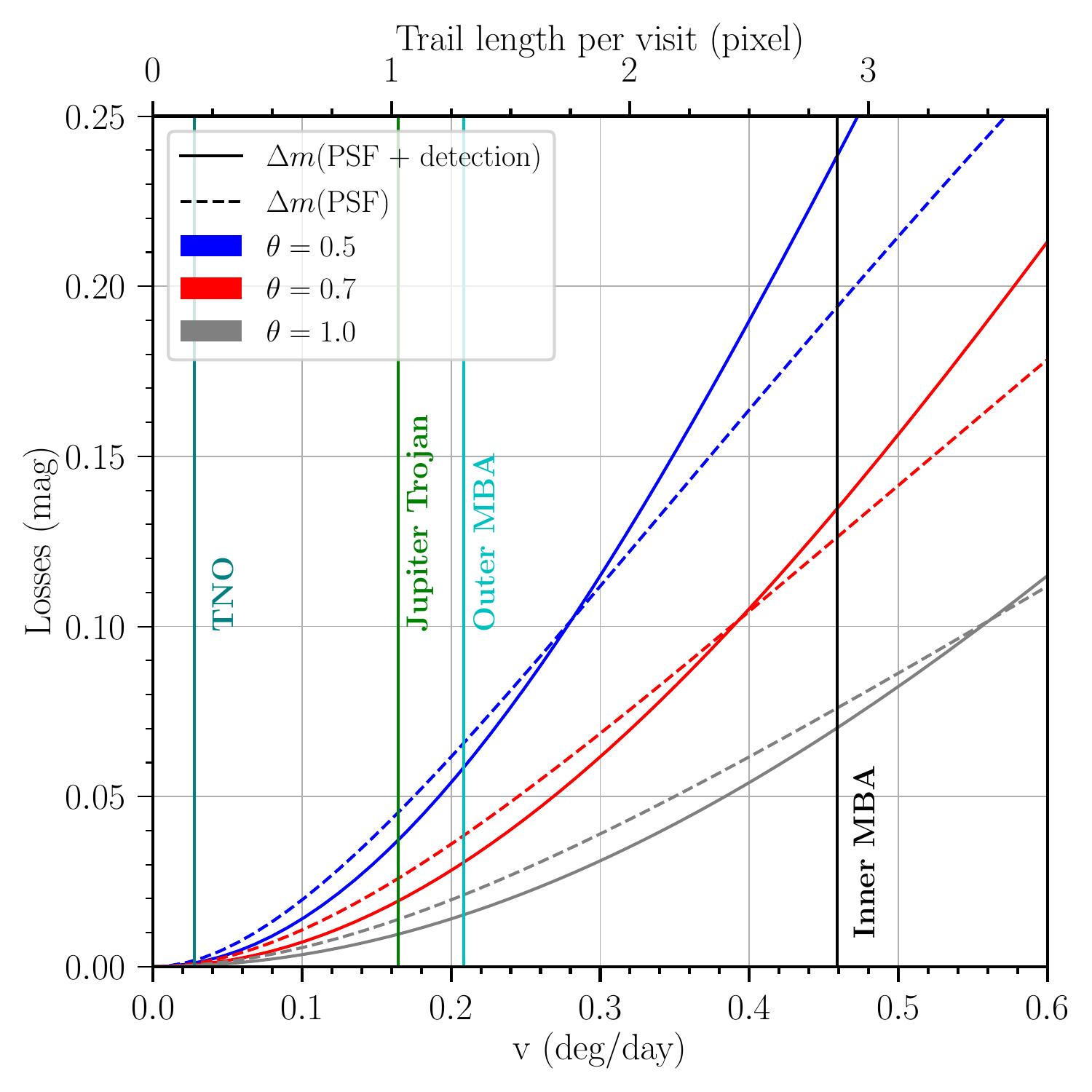}

    \caption{ \textbf{Left and Right}: The trailing losses for different values of the seeing $\theta$, shown as a function of the object's on-sky velocity $v$, given in degrees per day on the bottom axis and pixels per $30\, \mathrm{s}$ visit on the upper axis. \textbf{Right}: A zoomed in version of the figure on the left for low $v$.  Vertical lines represent the thresholds for typical on-sky motions of a TNO (Trans-Neptunian object), a Jupiter Trojan, and inner and outer MBAs (main-belt asteroids)\citep[][Equation 1]{luu1988}.}
    \label{fig:trailing_loss_2}
\end{figure}

We use the prescription developed by \citet{jones2018} to calculate both trailing losses for moving solar system objects based on the expected LSST photometric performance. The functions calculating these values in \sorcha were adapted from \texttt{rubin\_sim} \citep[][]{Connolly2014,yoachim2022}. Both components of the trailing loss can be described by the empirical formula:

\begin{equation}
\Delta m = -1.25 \log_{10} \left( 1 + \frac{ax^2}{1+bx} \right)
\end{equation}
where
\begin{equation}
x = \frac{v T_{exp}}{24 \theta}
\end{equation}
where $v$ is  the on-sky velocity (deg/day), $T_{exp}$ is the exposure time (seconds), and $\theta$ is the seeing full width at half-maximum (arcseconds). $a$ and $b$ are different combinations of trailing loss components. We use the values fit by \cite{jones2018} for  the predicted sensitivity and performance of the Rubin Observatory and the LSST, where $a=0.67$ and  $b=1.16$ for computing $\Delta m(\textrm{PSF}$), and $a=0.42$ and  $b=0$ is used to estimate $\Delta m(\textrm{PSF + detection})$. This is illustrated in Figure~\ref{fig:trailing_loss_2}. The calculated $\Delta m(\textrm{PSF})$ and $\Delta m(\textrm{PSF + detection})$ will always be greater than or equal to zero. 
\\
\\
\noindent \textbf{Relevant configuration file parameters:}  \texttt{trailing\_losses\_on}
\\ 
\\
\textbf{Code function:}  \texttt{PPTrailingLoss}

\subsection{Photometric and Astrometric Uncertainty Estimation and Randomization} \label{sub:uncertainties}
In the interest of estimating the quality of orbit fits and characterization of the objects LSST will find, and to model lost and opportunistic detections near the limiting magnitude of an image, the capability to inject the uncertainty on astrometric and photometric measurements by means of random draws is provided. We compute the uncertainties for each potential detection of the input population and use them to characterize a normal distribution with a mean equal to the true value. A random draw from this distribution is then taken to generate a measurement with simulated noise. True values are retained in the full \sorcha output for comparison to the resulting fits (see Section \ref{sec:outputs}). We note that the functions calculating these values in \sorcha were adapted from versions developed for \texttt{rubin\_sim} \citep[][]{Connolly2014,yoachim2022}.

The models for these uncertainties are primarily driven by the SNR for a point source in an image, following the methods in \cite{ivezic2019}. {Derivations of these relations are given in Appendix \ref{app:mag_uncertainty}. Uncertainties are computed individually for the PSF magnitude and the trailed source magnitude. The photometric error ($\sigma_{\textrm{mPSF}}$) for the PSF magnitude ($m_{\textrm{PSF}}$) is given in magnitudes by
\begin{equation}
\sigma_{\textrm{mPSF}}^2 = (0.04 - \gamma) \times 10^{0.4(m_{\textrm{PSF}}-m_{5\sigma})} + \gamma \times 10^{2*0.4(m_\textrm{PSF} -m_{5\sigma})}
\end{equation}
The photometric error ($\sigma_{\textrm{mtrailed}}$) for the trailed source magnitude ($m_{\textrm{trailed source}}$) is given in magnitudes by

\begin{equation}
\sigma_{\textrm{mtrailed}}^2 = (0.04 - \gamma) \times 10^{0.4(m_{\textrm{trailed source}}+\Delta m(\textrm{PSF}) -m_{5\sigma})} + \gamma \times 10^{2*0.4(m_\textrm{trailed source} + \Delta \update{m}(\textrm{PSF}) -m_{5\sigma})}
\end{equation}
$\Delta m(\textrm{PSF}$) is the PSF trailing loss (as discussed in Section \ref{sub:trailing_losses}) and  $m_{\textrm{trailed source}}+\Delta m(\textrm{PSF})$ provides the apparent magnitude of a point-source with the equivalent SNR as what would be measured from a moving object's extended PSF. $m_{5\sigma}$ is the observation's $5\sigma$ limiting magnitude at the object's location on the FOV and $\gamma=0.039$ is a blended single-value parameter gathering together, for example, the effects of sky background and readout noise. This is in principle a filter-specific value: however, analysis by \cite{ivezic2019} (Table 2)  indicates close agreement in all bands except $u$, where  $\gamma=0.038$. As $u$-filter observations are expected only for relatively bright solar system bodies where the photometric errors are small, we find the introduced error negligible, and for simplicity use $\gamma=0.039$ for all LSST bands. 

The single coordinate astrometric uncertainty is then given by: 

\begin{equation}
    \textrm{$\sigma_{\mathrm{astr}}$} = \textrm{$k$} \frac{\theta}{\textrm{SNR}}
\end{equation}

with the SNR is well approximated in the regime of SNR $\geq 2$ by 

\begin{equation} \label{eq:snr}
    \textrm{SNR} = \frac{1}{\sigma_{\textrm{mtrailed}}}
\end{equation}
where $\theta$ is the seeing and $k$ is a model dependent coefficient, which is 0.6 for a Gaussian PSF. The trailed source magnitude uncertainty is used here because the astrometic uncertainty is correlated with the trailed length. This error is then added in quadrature with 10 mas noise floor, the LSST design specification on astrometric precision \citep{lsst-SRD-2013}, such that the best positional accuracy achievable for the brightest image sources will never be better than the survey's reported precision. To avoid complicated conversion to uncertainty in right ascension and declination, the random perturbation applied to the astrometric measurement is computed in the local tangent plane, which is then translated into the reported values.

Since the magnitude error approximation is inversely proportional to SNR, at very low signal levels the photometric errors increase significantly. This means that a low-SNR source with a PSF magnitude significantly fainter than the observation's 5$\sigma$ limiting magnitude will have very large error bars.  If there are very high photometric uncertainties, applying a random perturbation on the computed apparent magnitudes will incorrectly generate a subset of ``bright sources" well above the observation limiting magnitude from objects that are too dim to be detected on the real survey images.  As a compromise between low-probability detections and unrealistic magnitude uncertainties producing ``fake detections", by default \sorcha removes all potential detections of the input population with SNR less than 2 after calculating the astrometric and photometric uncertainties.
\\
\\
\noindent \textbf{Relevant configuration file parameters:}  \texttt{randomization\_on}
\\ 
\\
\textbf{Code functions:}  \texttt{PPAddUncertainties, PPRandomizeMeasurements}

\begin{figure}[h]
    \centering
    \includegraphics[width=0.8\columnwidth]{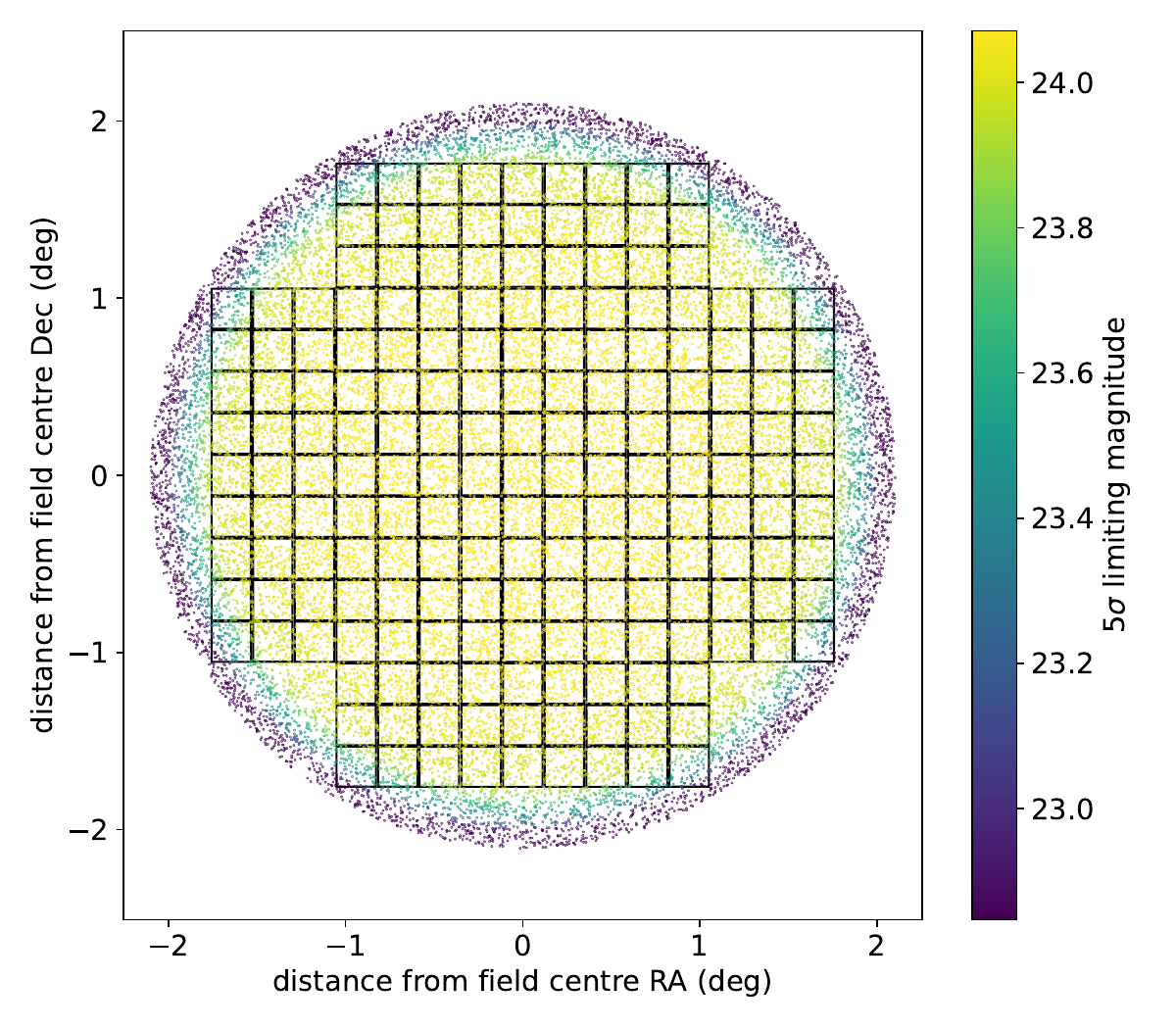}
    \caption{The effects of vignetting on the $5\sigma$ limiting magnitude for a randomized series of points on a circular FOV in the LSSTCam focal plane. \update{Locations further from the center of the FOV have shallower depths.} The LSSTCam detector footprint is also plotted.  The center of \update{this example} pointing has a limiting magnitude of 24.071.}
    \label{fig:vignetting}
\end{figure}
 
\subsection{Calculating the Limiting Magnitude at the Source FOV Location} \label{sub:vignetting}
As with all telescopes, the LSSTCam FOV will be unevenly illuminated due to vignetting with the edges of the focal plane receiving less light/photons than the center. The effect of this is to decrease the 5$\sigma$ limiting magnitude -- the apparent magnitude where a detected point source has exactly a 50\% probability of detection -- at the edges of the LSSTCam FOV. \sorcha accommodates this by calculating the effects of vignetting at the source's location on the focal plane and adjusting the $5\sigma$ limiting magnitude accordingly for each potential detection. This modified limiting magnitude will be used when applying the survey detection efficiency (described in Section \ref{sub:fading_function}). Around 7\% of LSSTCam's 2.06$^{\circ}$ radius circular focal plane is expected to be vignetted by more than a 0.1 magnitude \citep{veres2017a}, but only the outermost CCDs at the corners of the cross-shaped LSSTCam detector layout are far enough from center to be significantly impacted by vignetting, as shown in Figure~\ref{fig:vignetting}. \sorcha's built-in vignetting function is based on the functionality developed for \texttt{rubin\_sim} \citep{Connolly2014,yoachim2022} based on the parametrization from \cite{araujo-hauck2016}. It is tailored for the specific case of LSST and is therefore non-configurable.
\\
\\
\noindent \textbf{Relevant configuration file parameters:}  \texttt{vignetting\_on}
\\ 
\\
\textbf{Code function:}  \texttt{PPVignetting}

\subsection{Camera Footprint Filter} \label{sub:footprint}
The footprint filter models the effects of the camera detector layout within the footprint on discoverability. This is important for the LSST where the CCDs are arranged in a cross pattern across the circular focal plane, as shown in Figure~\ref{fig:LSSTCamfovs}.  189 CCDs are mounted in 3$\times$3 ``rafts'', with both the CCDs and the rafts are separated by narrow gaps \citep{lsst-sciencebook-ch5-2009,ivezic2019}. \sorcha sifts through the possible moving object detections and picks out those that actually land within the associated observation's camera footprint and therefore could be detectable if bright enough. Solar system objects may move into one of these chip gaps (or chip gaps and raft gaps in the case of the LSST) between nightly observations, affecting their detectability; this effect is most significant in the case of slow-moving objects such as TNOs which move only a small amount over the course of a night. However, the determination of which objects lie on the full camera footprint of the LSST can be computationally expensive, and is less concerning for faster-moving objects such as MBAs or NEOs. As a result, we provide the user with two mutually-exclusive methods of applying the camera footprint which can be selected using the configuration file: the circular footprint or the full camera footprint. 
\\
\\
\noindent \textbf{Relevant configuration file parameters:}  \texttt{camera\_model}
\\
\\
\textbf{Code function:}  \texttt{PPApplyFOVFilter}
\\

\begin{figure}
    \centering
    \includegraphics[width=0.6\columnwidth]{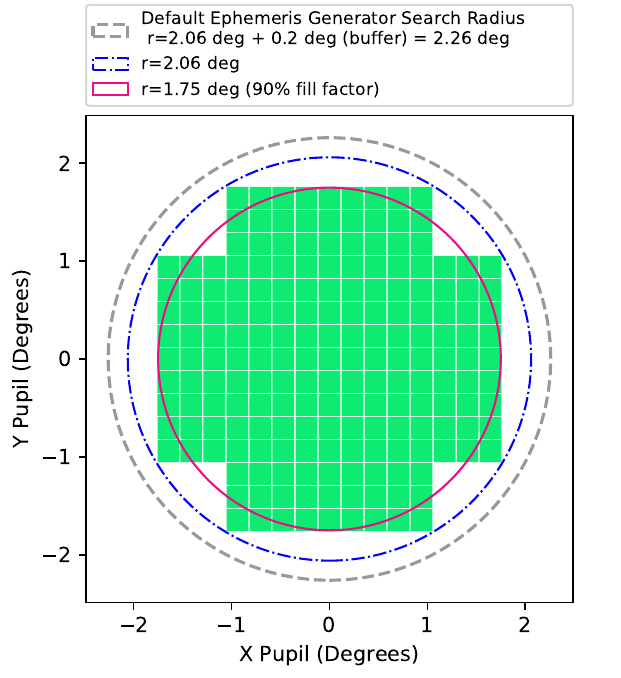}
    \caption{The LSSTCam field-of-view (FOV). The overplotted circles show the recommended default search area radius (including buffer) used in \sorcha's ephemeris generation stage (gray dashed line); the 2.06$^{\circ}$ radius circle encompasses all the CCD detectors (blue dash-dotted line); and the 1.75$^{\circ}$ radius circles contains 90\% of LSSTCam detector area (pink solid line).}
    \label{fig:LSSTCamfovs}
\end{figure}

\subsubsection{Circular Footprint}
\label{sec:circleFOV}
This filter applies a simple circular camera footprint of a user-determined radius. For the LSST, the minimal radius to include all observations (with a certain overlap) is 2.06$^{\circ}$, and the circle containing 90\% of observations has a radius of 1.75$^{\circ}$ \citep{jones2018}. These radii are illustrated in Figure~\ref{fig:LSSTCamfovs}. If using the circle footprint filter, \sorcha can also remove a user-defined fraction of random input objects that land within the circle footprint. This is intended to quickly mimic the effect of the gaps/areas without CCDs without performing the computationally-expensive procedure of explicitly detecting and removing specific detections which do not lie upon a detector. The ``fill factor'' (the fraction of observations to retain) is set in the configuration file. A circular radius of 1.7$^{\circ}$ and a fill factor of 90$\%$ mimics the LSSTCam detector area. Note that as described in Section~\ref{sec:ephemeris}, the in-built ephemeris generator already utilizes a circular `search field' analogous to this circular footprint with the radius set by the user in the configuration file. Thus, setting the radius of the circular footprint filter to be equal or larger than that set by the user for the ephemeris generation search radius (plus buffer) will have little effect. 
\\
\\
\noindent \textbf{Relevant configuration file parameters:}  \texttt{circle\_radius, fill\_factor}
\\ 
\\
\textbf{Code function:}  \texttt{PPCircleFootprint,PPSimpleSensorArea}

\subsubsection{Full Camera Footprint}
\label{sec:fullFOV}
The camera footprint filter is the most accurate and computationally most expensive version of the footprint filter. The function was adapted from a version utilized in \texttt{rubin\_sim} \citep[][]{Connolly2014,yoachim2022}. It checks every object against the full detector map of the instrument, taking into account the field rotation from the pointing database, and removes all objects which do not lie upon a detector. The effect of this is illustrated in Figure~\ref{fig:footprint_validation}. By default, the code will use an in-built detector map of LSSTCam. The user may also supply a detector map of their own devising, as described in Section~\ref{subsub:telescope_file}. Additionally, it is expected that sources which lie on the very edges of the detectors will not be correctly extracted. The user can, via the configuration file, parameterize the distance from the edge of a detector (in arcseconds on the focal plane) at which an object will not be detected. In \sorcha's example configuration files, this value is set to 2 arcseconds (10 pixels) following recommendations made by the Rubin Data Management Team.
\\
\\
\noindent \textbf{Relevant configuration file parameters:} \texttt{footprint\_path, footprint\_edge\_threshold}
\\ 
\\
\textbf{Code function:}  \texttt{footprint.applyFootprint}

\begin{figure}
    \centering
    \includegraphics[width=0.99\columnwidth]{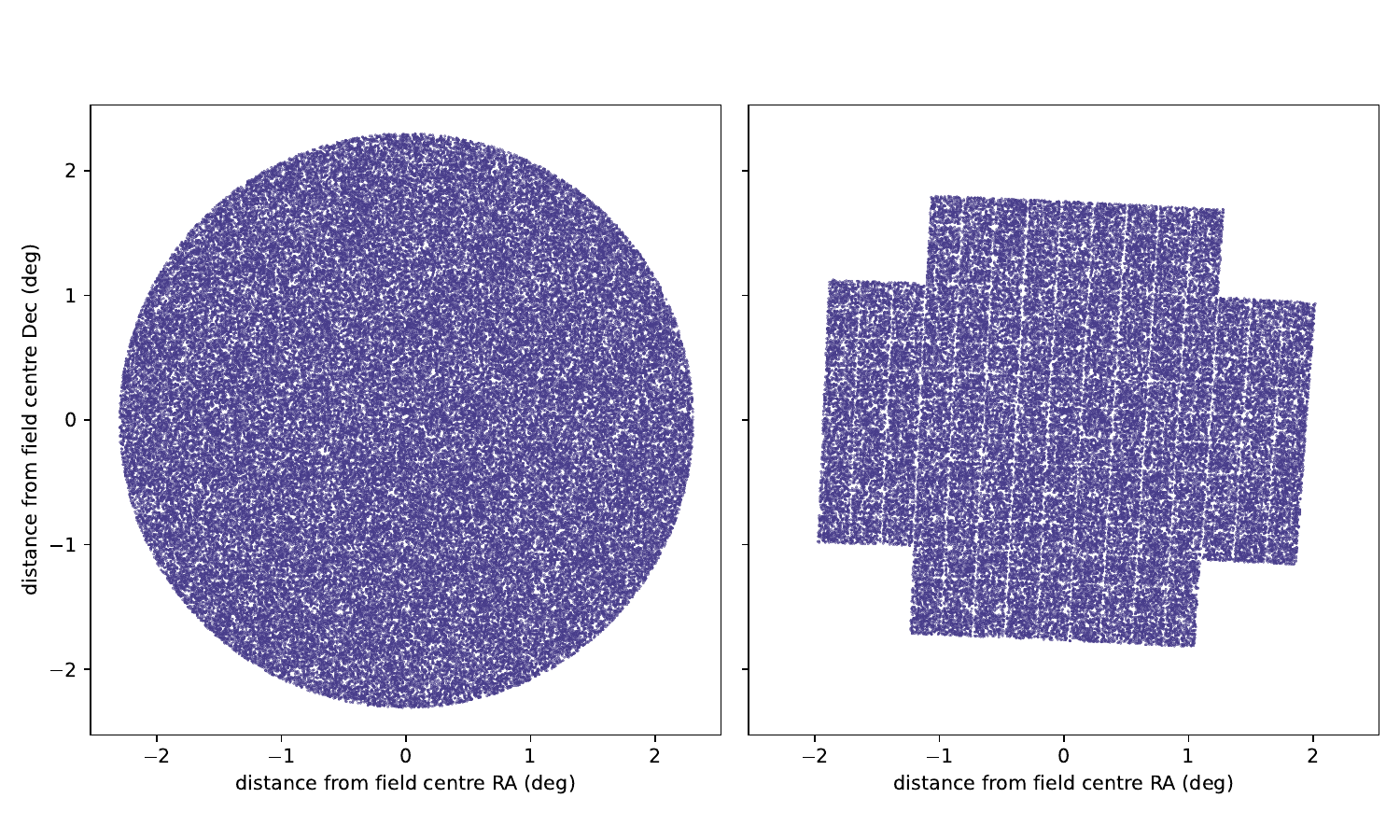}
    \caption{The effect of the full camera footprint filter on a selection of 100,000 random synthetic sources. \textbf{Left}: original sources, distributed over a circular FOV of radius 2.1$^{\circ}$. \textbf{Right}: the same sources after running \sorcha's full camera footprint filter. The shape of the LSSTCam detector footprint can be seen with the loss of detections in the raft and chip gaps.}
    \label{fig:footprint_validation}
\end{figure}

\begin{figure}
    \centering
    \includegraphics[width=0.7\columnwidth]{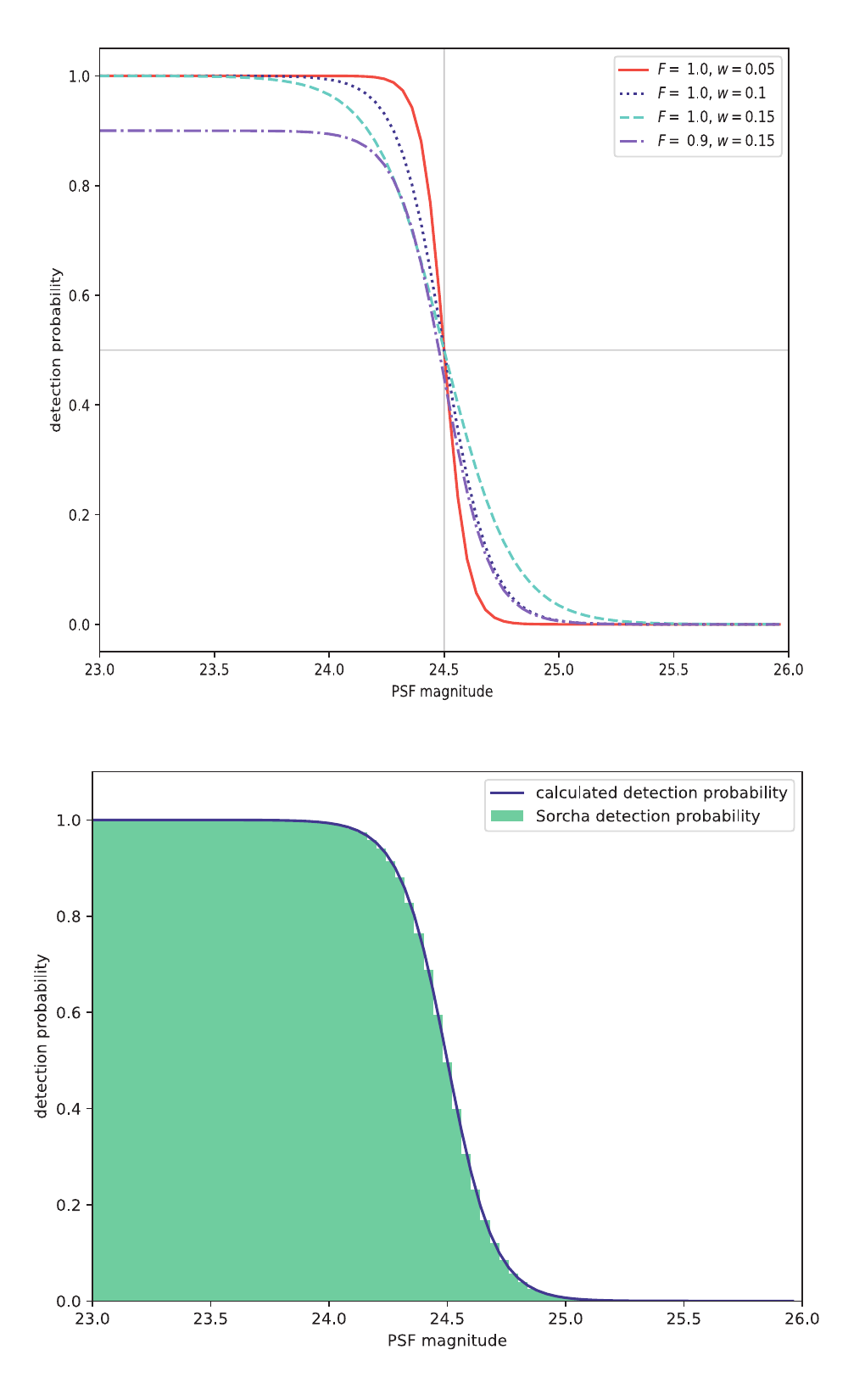}
    \caption{\textbf{Top}: the fading function as defined in Equation~\ref{eq:fadingfunction}, representing the fraction of observed point sources as a function of PSF magnitude. The different lines represent the effect of the variation of the peak detection efficiency $F$ and the width parameter $w$ on the shape of the function. The $5\sigma$ limiting magnitude at the source location is marked in gray ($m_{5\sigma}$=24.5). \textbf{Bottom}: a histogram showing detection probability of 10,000 point sources passed through \sorcha's fading function filter, with the actual calculated detection probability from Equation~\ref{eq:fadingfunction} overplotted as a solid line. Here, $F = 1.0$, $w = 0.1$, and $m_{5\sigma}$=24.5 and the binsize is  0.04 mag.}
    \label{fig:fadingfunction}
\end{figure}

\subsection{Source Detection Efficiency Filter} \label{sub:fading_function}
The detection efficiency filter serves to pick out which of the input small bodies that land within the FOV footprint of a given survey's observation would be detected by the survey's source extraction algorithm \citep[for the LSST, this is the Rubin Difference Imaging Analysis;][]{juric2021}. For each observation/pointing in the survey, we model the source detection efficiency as a fading function, the shape of which is two-fold: the detection efficiency remains constant when substantially above the 5$\sigma$ limiting magnitude of the observation, then drops swiftly when approaching the 5$\sigma$ limiting magnitude (shown in Figure \ref{fig:fadingfunction}). Following \citet{veres2017a}, \sorcha utilizes the following formulation of the fading function \citep{gladman1998, zavodny2008,gladman2009} to represent the per image source detection efficiency:
\begin{equation} \label{eq:fadingfunction}
\epsilon(m_{PSF}) = \frac{F}{1 + e^\frac{m_{PSF}-m_{5\sigma}}{w}}
\end{equation}
where $\epsilon(m_{PSF})$ is the probability of detection, $F$ is the peak detection efficiency, $m_{PSF}$ and $m_{5\sigma}$ are respectively the object's PSF magnitude and $5\sigma$ limiting magnitude of the observation at the source's location on the camera focal plane (see Section \ref{sub:vignetting}), and $w$ is the width of the fading function. In this formulation, by definition $\epsilon(m_{5\sigma}) = F/2$. The shape of the function and the variation of parameters $F$ and $w$ are visualized in Figure~\ref{fig:fadingfunction}. In our implementation, the detection efficiency $\epsilon(m_{PSF})$ is calculated at the PSF magnitude for each potential detection of an input synthetic small body, and compared to a random number selected for each detection opportunity from a uniform distribution. The PSF magnitude is used here because the Rubin DIA pipeline uses PSF filter matching as described in Section \ref{sec:rubin_apparent_mags}. Those potential detections whose drawn random number is less than or equal to $\epsilon(m_{PSF})$ will be deemed ``detected" as an astronomical source on the relevant survey observation/pointing and will continue to be passed on to later stages of post-processing. 

The values of $w$ and $F$ can be controlled via the user through the configuration file and are fixed for all pointings of the simulated survey. Both parameters are survey-specific, and the actual values for the LSST can only be estimated when on-sky data becomes available. For $w$, we adopt 0.1 as a baseline following the Sloan Digital Sky Survey (SDSS) \citep{annis2014}. For $F$, we adopt a default of 1: source detection for the LSST has been well-tuned for bright sources, and the true value will likely be very close to 1 \citep{juric2021}. 

\update{We acknowledge that the currently implemented method for accounting for the survey source detection efficiency within \sorcha uses median values measured/estimated for the entire survey being simulated. This strategy works well given the values provided within the pointing database generated by the \texttt{rubin\_sim} LSST cadence simulations and the information expected to be available early on at the start of the LSST.  Other representations, such as including per observation per CCD detector source detection efficiencies, may be a better fit once data is flowing from the LSST. We are planning for future releases of \sorcha to include additional options for representing realistic LSST source detection efficiency estimates. }
\\
\\
\noindent \textbf{Relevant configuration file parameters: }{fading\_function\_width, fading\_function\_peak\_efficiency}
\\ 
\\
\textbf{Code functions:} \texttt{PPFadingFunctionFilter}

\subsection{Saturation Limit Filter} \label{sub:saturation}
While saturated sources can often still be detected in astronomical images, their brightness cannot be reliably measured because either the pixels have such a high number of electrons that the response significantly \update{deviates} from linear or the pixel well has overflowed with electrons spilling into neighboring pixels. Moving solar system objects that saturate in LSSTCam exposures will not be identified. This is because the DIA algorithms employed by the Rubin Observatory Data Management Team for daily prompt processing and annual data releases will mask saturated pixels \citep{lsst-sciencebook-ch5-2009,ivezic2019}}.  \citet{ivezic2019} estimate that the saturation limit for LSST will be the 16th magnitude in the $r$-filter. \sorcha implements a simple magnitude cut-off. This cut-off threshold can be provided by the user in the configuration file as either a single value which will apply to all observational filters, or as a comma-separated list of filter-specific values.
\\
\\
\noindent \textbf{Relevant configuration file parameters:}   \texttt{bright\_limit}
\\ 
\\
\textbf{Code function:}  \texttt{PPBrightLimit}

\subsection{Linking Filter} \label{sub:linking_filter}
\begin{figure}
\begin{center}
\includegraphics[width=0.93\columnwidth]{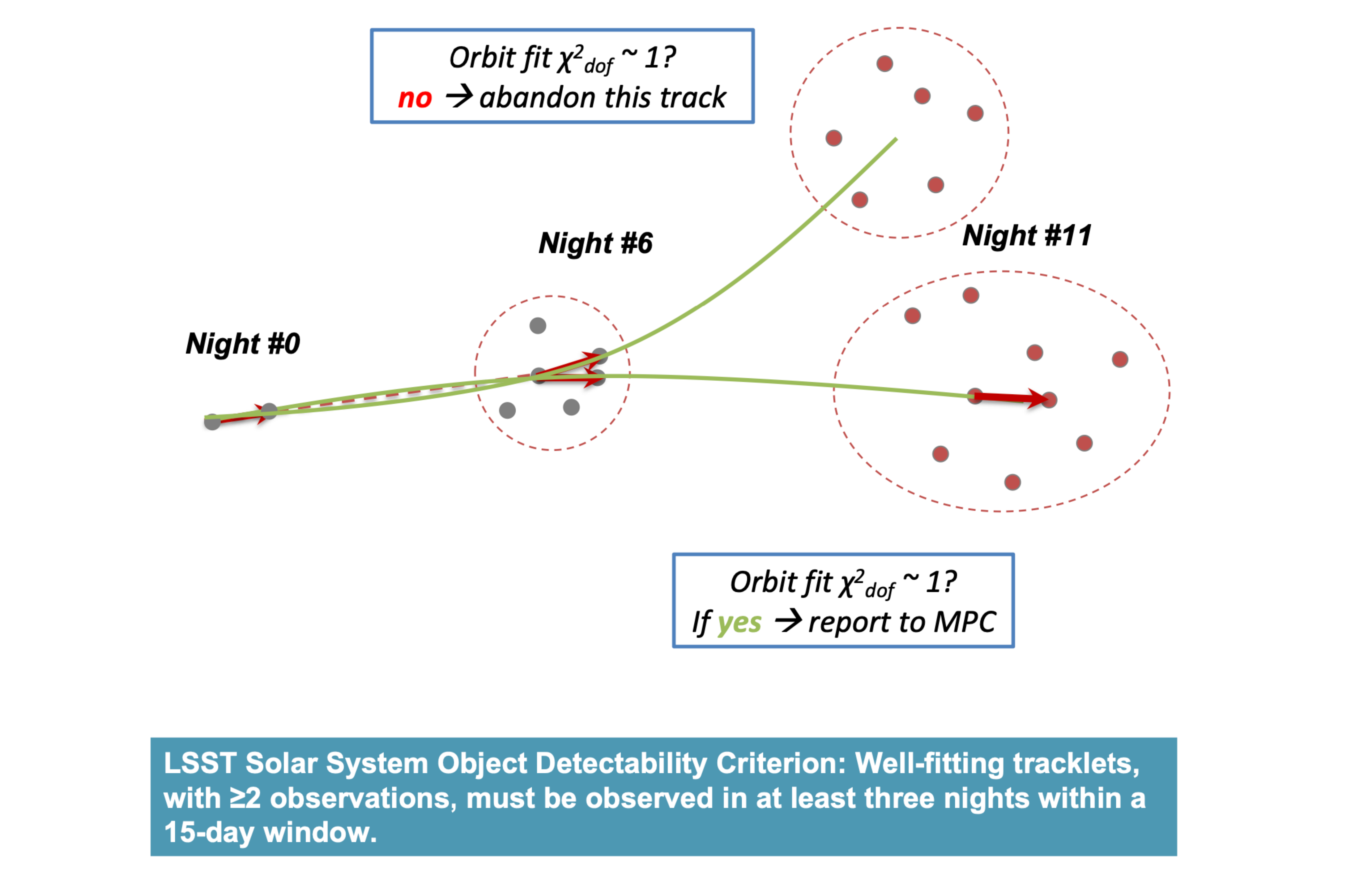}
\caption{A cartoon schematic of tracklet creation and the linking process for the Rubin Solar System Processing Pipeline (SSP). }
\label{fig:SSP}
\end{center}
\end{figure}

It is not enough for a moving object to be found by the source detection pipelines within a wide-field survey, it must also be identified as a moving solar system object. In practice automated algorithms sift through a survey's source catalogs ``linking'' transient detections together in order to identify potential tracks from objects on heliocentric orbits. This step oftentimes involves orbit fitting from the linked astrometric measurements. The linking filter within \sorcha mimics the selection function from automated discovery algorithms used to discover the solar system moving objects in wide-field surveys to pick out which members of the synthetic input population are discovered by the simulated survey and when the discovery happens. For \sorcha's v1.0 release, we primarily focused on replicating the Rubin SSP requirements \citep[see requirement {\tt OSS-REQ-0159} in \url{https://ls.st/oss}, and also][]{juric2020}, which builds tracklets (linkages of observations within a given night) and attempts to link three separate nights of tracklets over 14 days (see Figure \ref{fig:SSP}) onto heliocentric orbits. SSP relies on an upgraded version of the HelioLinC algorithm \citep{holman2018}, but the linking filter within \sorcha simply counts the number of detections per night to make tracklets, retaining detections from objects that satisfy a certain number of tracklet nights within a specific window while also applying a discovery efficiency. \update{For SSP, we set the discovery efficiency in our default configuration files to 95$\%$, the minimum efficiency required specified in the LSST Observatory System Specifications ({\tt OSS-REQ-0159} in \url{https://ls.st/oss}).} The software implementation of this algorithm -- named {\tt miniDifi} -- is highly vectorized for performance reasons and described in detail in the code comments.

 We assume perfect precovery and recovery from within the survey observation such that if the input object has sufficient observations to be linked at some point in the survey then the linking filter keeps all these observations associated with that object. Detections from objects not linked are removed by default and not passed to the next stages of post-processing. For the same \sorcha run, a user may want \update{to} compare the parts of the input population that would have been correctly linked and identified as a moving solar system object to the parts that were not linked but would be present in the survey source catalogs. We have made the linking filter configurable to either remove all observations from unlinked input objects  or keep all detections even if the object is not linked with a linking flag column to identify which input objects were linked. 

Although SSP will discover the majority of the solar system objects found during the LSST, SSP requires that there be motion between each observation that comprises a tracklet: this means that slow objects moving at distances beyond about 200 au will be undetectable by the SSP unless it is implemented in a way that exceeds Rubin requirements. Other bespoke algorithms targeting very slow/distant objects may have to be developed by the community, and applied later on during the lifetime of the LSST. We have made the linking filter flexible in that the minimum on-sky distance between nightly pairs within a tracklet, the number of observations required to make a tracklet, the number of tracklets that must be linked, and the linking window are all configurable by the user. 
\\
\\
\noindent \textbf{Relevant configuration file parameters:}  \texttt{SSP\_detection\_efficiency, SSP\_number\_observations, SSP\_separation\_threshold, SSP\_maximum\_time, SSP\_number\_tracklets, SSP\_track\_window, SSP\_night\_start\_utc, drop\_unlinked }
\\ 
\\
\textbf{Code function:}  \texttt{PPLinkingFilter}

\subsection{Additional Advanced Filters}
\label{sec:additional_advanced_filters}
We have also developed some additional filters to assist with specific science use cases and scenarios that could speed up \sorcha's performance. These filters are intended for the advanced user and are turned off by default within \sorcha. These filters can be activated by simply including the relevant keyword in the configuration file. 

\subsubsection{SNR/Magnitude Limit Filters} \label{sub:snr_mag_limits}
We have also included the option for the user to create a magnitude-limited or SNR-limited subsample of the synthetic survey discoveries through the magnitude limit and SNR filter functions.  The user may either implement the SNR limit, to remove all observations of objects below a user-defined SNR threshold; or the magnitude limit, to remove all observations of objects above a user-defined trailed source magnitude. This may be useful for estimating/predicting follow-up sample sizes, checking whether a known object is predicted to have been observed by the survey with a certain SNR or trailed source apparent magnitude, or analyzing population statistics related to follow-up observing programs. One such, use case we envision with \sorcha is for developing/testing the results from \update{an} observing program similar to the Colours of the Outer Solar System Origins Survey \citep[Col-OSSOS;][]{schwamb2019b,fraser2023}. Col-OSSOS observed 102 Centaurs and TNOs discovered by  Outer Solar System Origins Survey \citep[OSSOS;][]{bannister2016,bannister2018}, with discovery triplet brightness $r$ $<$ 23.6 mag. This brightness limited sample enabled cosmogonical composition studies of the Kuiper belt \citep{fraser2021,buchanan2022,pike2023} by combining the optical-Near infrared colors from Col-OSSOS  and the well characterized OSSOS detection efficiency and pointing history through the OSSOS survey simulator \citep{lawler2018} and other simulations. We note that using these filters is not the appropriate way of handling the source detection efficiency of the simulated survey (see Section~\ref{sub:fading_function}). This filter is applied before the source detection efficiency and linking effects are modeled within \sorcha. \update{Therefore, the user must take care  when combining the SNR/magnitude limit  filters with \sorcha's source detection efficiency and/or the linking filter to ensure that the output from the \sorcha simulation satisfies the user's science case as the generated \sorcha detections will not represent all the discoveries that the LSST should have found for the input population}. 
\\
\\
\noindent \textbf{Relevant configuration file parameters:}   \texttt{SNR\_limit, magnitude\_limit}
\\ 
\\
\textbf{Code functions:}   \texttt{PPSNRLimit, PPMagnitudeLimit}

\subsubsection{Faint Object Culling Filter} \label{sub:culling_filter}
By default \sorcha passes all input orbits to its built-in ephemeris generator as described in Section \ref{sec:ephemeris}. Applying the ephemeris generator's integrations and calculations to the entire input population can be costly given that small objects on moderately to highly eccentric objects will be preferentially discovered only near perihelion, but spend a large fraction of their time near aphelion undetectable by the simulated survey with apparent magnitudes well beyond the survey's brightness limit. In most cases with the power-law absolute magnitude/size distributions, the majority of the input population that \sorcha will process will not end up detected by the simulated survey. We provide the expert user the option to apply the faint object culling filter, which estimates the brightest trailed source magnitude any input object will potentially reach and removes those input objects that will never be bright enough to be detected by the simulated survey before moving on to the ephemeris generation stage. 

As a fast first pass, this filter takes the orbits of the input model and calculates a maximum trailed source magnitude in each filter for every object to be simulated as described in Section \ref{sub:magnitude_calc} (for phase angle $\alpha$ of 0) at the object's perihelion. The geocentric distances is approximated  as $(q-1)$ in order to calculate an estimate for the brightest possible  trailed source magnitude. The routines to calculate the perihelion if needed from the input parameters are described in Holman et al.(submitted). Modifications to this maximum brightness due to any activity or light curve models are incorporated at this stage according to the user's implementation of the \texttt{maxBrightness} method within each user supplied activity and light curve class (see Section \ref{sub:add-ons}). Finally, for each object a check is made if all of these calculated maximum trailed source magnitudes are fainter than 2 + the deepest survey observation per filter (as obtained from the survey pointing database) - if the object is fainter in all filters than this criteria, it is dropped and not sent on to the ephemeris generation stage. This approximation is only applied to those objects with a calculated perihelia greater than 2 au. 

This optional filter within \sorcha substantially decreases the compute time for a simulation. For the model of $\sim$6.9$\times$10$^6$ Centaurs from Murtagh et al. (submitted), split across 128 cores on Queen's University Belfast's HPC Kelvin2 with a chunk size of 5000, a full simulation of all objects takes $\sim$29 hours to run (or $\sim$3700 core-hours) - for the exact same model and simulation setup but with the filter applied this is reduced to only $\sim$1.25 hours (or $\sim$225 core-hours). 
\\
\\
\noindent \textbf{Relevant configuration file parameters:}   \texttt{brute\_force}
\\ 
\\
\textbf{Code functions:}   \texttt{PPFaintObjectCullingFilter, PPEstimatePerihelion}

\section{Outputs} \label{sec:outputs}
 The main output from sorcha is a file listing all predicted observations of objects from the input population.  Additionally, \sorcha can be configured to output the generated ASSIST+REBOUND ephemeris and a statistics file summarizing the main results. \sorcha also generates two log files (one with a  `.log' extension for general information, and one with a `.err' extension for error messages) each time \sorcha is run. These log files should contain helpful information for the user wishing to check if a \sorcha run has completed successfully and understand why a run has failed or produced unexpected results. We provide further details about the contents and formats of \sorcha's outputs below, but we recommend the reader review any updates to the output format by reading the \href{https://sorcha.readthedocs.io/}{online documentation}.

\subsection{Detections File}
 \sorcha's main output file is the detections file which contains all the predicted survey observations of input objects that are found by the survey. By default only the information associated with ``detected and linked'' objects from the input population are included in detections file. If the science cases requires it, a configuration file variable can be set to  allow successful detections of unlinked objects to be \update{included in the output}. The resulting file will contain an additional column, a flag identifying which detections are associated with input objects that have passed both the source detection and linking filters in post-processing (see Sections \ref{sub:fading_function} and \ref{sub:linking_filter} respectively)

The user also has complete control over how much information is provided in each row through the configuration file. By default, ``basic'' output includes column headers describing the pointing information for the observation and the object's predicted position and calculated brightness in the observational filter. Alternatively, the user may ask for ``all'' columns, which includes all columns from the input files and the full ephemeris information; this is useful in cases where the user wants all of the information accessible in a single file and does not need to worry about the larger file size. Finally, the advanced user can also specify a comma-separated list of column headings for their output file, choosing from the available headings to create a customized output. The potential column headings for the detections file are given in Table \ref{tab:output_cols}. The main output file can also be written out in multiple formats: comma-separated and whitespace separated text files, which are useful for readability when output file sizes are smaller; and HDF5 and SQLite3 databases, which perform better when output sizes become especially large and are generally recommended for HPC runs. The advanced user may also specify if they wish for the values to be rounded to a select number of decimals; by default, this option is off.
\\
\\
\noindent \textbf{Relevant configuration file parameters:} \texttt{output\_format}, \texttt{output\_columns}, \texttt{position\_decimals}, \texttt{magnitude\_decimals}

\startlongtable
\tabletypesize{\footnotesize}
\begin{deluxetable}{ll}
\tablecaption{Output Detection File Column Headings\label{tab:output_cols}}
\tablehead{
\colhead{Heading} & \colhead{Description}
}
\tablecolumns{2}
\startdata
\cutinhead{Default output columns}
\texttt{ObjID} & Unique object identifier of the simulated object (string) \\
\texttt{fieldMJD\_TAI}\tablenotemark{a} & Observation Mean Julian Date (MJD) in TAI ((International Atomic Time) \\
\texttt{fieldRA\_deg} & Right ascension (RA) of the center of the observation pointing (degrees) \\ 
\texttt{fieldDec\_deg} & Declination (Dec) of the center of the observation pointing (degrees) \\ 
\texttt{RA\_deg} & Object right ascension (RA) (degrees)\\ 
\texttt{Dec\_deg} & Object declination (Dec) (degrees) \\
\texttt{astrometricSigma\_deg} & Astrometric uncertainty in object RA and Dec position (degrees) \\ 
\texttt{optFilter} & Filter (band) for this observation ($ugrizy$) \\ 
\texttt{trailedSourceMag} &  Trailed Source Magnitude \\
\texttt{trailedSourceMagSigma} &  1$\sigma$ uncertainty on trailed source magnitude\\
\texttt{fiveSigmaDepth\_mag} &  5$\sigma$ limiting magnitude at the object's location on the camera focal plane \\
\texttt{phase\_deg} &  Sun-object-observer angle (phase angle) (degrees) \\
\texttt{Range\_LTC\_km} & Light-time-corrected object-observer distance (km) \\
\texttt{RangeRate\_LTC\_km\_s} & Light-time-corrected rate of change of the object-observer distance (km/s) \\
\texttt{Obj\_Sun\_LTC\_km} & Object-sun light-time-corrected distance (km) \\
\texttt{object\_linked}\tablenotemark{b} & Flag to identify where the synthetic object was linked by the linking filter \\
\cutinhead{All other output columns}
\texttt{FieldID} & Observation pointing field identifier\\
\texttt{RA\_true\_deg} &  Object right ascension unadjusted for astrometric uncertainty (degrees) \\
\texttt{Dec\_true\_deg} &  Object declination unadjusted for astrometric uncertainty (degrees) \\
\texttt{RARateCosDec\_deg\_day} & Object right ascension rate of motion (deg/day)(degrees) \\
\texttt{DecRate\_deg\_day} & Object right ascension rate of motion (deg/day)(degrees) \\
\texttt{visitTime} &  Total length of time for a visit (seconds) \\
\texttt{visitExposureTime} &  Total exposure time for a visit (seconds) \\
\texttt{fieldRotSkyPos\_deg} & Angle of the field y-axis and celestial north\\
\texttt{seeingFwhmGeom\_arcsec} & Geometric full-width half-maximum for the field (arcsec)\\
\texttt{seeingFwhmEff\_arcsec} & Effective full-width half-maximum for the field (arcsec)\\
\texttt{fieldFiveSigmaDepth\_mag} & 5-$\sigma$ limiting magnitude at the center of the field-of-view (FOV) \\
\texttt{H\_filter} & Absolute magnitude in the survey exposure's observing filter \\
\texttt{PSFMag} & Calculated PSF magnitude\\ 
\texttt{PSFMagSigma} & 1-$\sigma$ uncertainty on PSF magnitude \\
\texttt{trailedSourceMagTrue} & Calculated trailed apparent magnitude unadjusted for photometric uncertainty \\
\texttt{PSFMagTrue} & Calculated PSF magnitude unadjusted for photometric uncertainty \\
\texttt{SNR} & Predicted signal-to-noise (SNR) ratio of detection \\
\texttt{detectorID} &  Identifier of the camera detector covering the observation \\
\texttt{date\_linked\_MJD}\tablenotemark{c} & Date (MJD in TAI) on which the object was discovered by linking \\
Columns from orbits input & See Table~\ref{tab:orbits} \\
Columns from physical parameters input & See Table~\ref{tab:physical_parameters} \\
Columns from ephemeris generation & See Table~\ref{tab:ephemeris} \\
\enddata
\tablecomments{All positions and velocities are in respect to J2000.}
\tablenotetext{a}{Reported as midpoint of the simulated survey observation.}
\tablenotetext{b}{This column only appears if \sorcha has been instructed not to drop unlinked objects in the configuration file if the linking filter is on.}
\tablenotetext{c}{This column only appear if the linking filter is on.}

\end{deluxetable}

\subsection{Output Ephemeris File}
The user can output the ephemeris calculated by the ASSIST+REBOUND ephemeris generator (described in Section \ref{sec:ephemeris}), using a flag on the command line. The most common use-case for this ephemeris output is to create a file that can be read back in by \sorcha in subsequent runs, thus skipping the computationally-expensive process of ephemeris generation in cases where the user wishes to re-run \sorcha with the same orbital information but with changes to the simulation configuration or the physical parameters of the input population. The contents of this file are shown in Table~\ref{tab:ephemeris}. This ephemeris file can be output in several formats: comma-separated, whitespace-separated, or HDF5. Of the three, HDF5 format allows for the fastest read-in.
\\
\\
\noindent \textbf{Relevant configuration file parameters:} \texttt{eph\_format}

\subsection{Statistics (Tally) File}
\sorcha has the option to output a statistics or ``tally'' file, again using a flag on the command line. This file is designed to be a quick, easy-to-access overview of the output of a \sorcha run providing information per detected object per filter. The format and columns of the statistics file are listed in Table~\ref{tab:output_stats}. This file is useful in cases where the user wishes to know very simple information, such as which objects passed the linking filter and when, or the extent of the phase curves for the objects, without having to deal with a potentially large and unwieldy main output file.

\begin{deluxetable}{ll}
\tablecaption{ Output Statistics File Column Headings\label{tab:output_stats}}
\tablehead{
\colhead{Heading} & \colhead{Description}
}
\tablecolumns{2}
\startdata
\texttt{ObjID} & Unique object identifier for each input object (string) \\
\texttt{optFilter} & Relevant observing filter (band) ($ugrizy$) \\
\texttt{number\_obs} & Number of observations for this object in the given optFilter \\ 
\texttt{min\_apparent\_mag} & Minimum calculated apparent magnitude for this object in the given optFilter \\ 
\texttt{max\_apparent\_mag} & Maximum calculated apparent magnitude for this object in the given optFilter\\ 
\texttt{median\_apparent\_mag} & Median calculated apparent magnitude for this object in given optFilter \\
\texttt{min\_phase} & Minimum calculated phase angle for this object in the given optFilter (degrees) \\ 
\texttt{max\_phase} & Maximum calculated phase angle for this object for the given optFilter (degrees) \\ 
\texttt{object\_linked}\tablenotemark{$^*$}  & True/False whether the object passed the linking filter \\
\texttt{date\_linked\_MJD}\tablenotemark{$^{\dagger}$}&  Date the object was linked (if it was linked) in Mean Julian Date (MJD) in TAI \\
\enddata
\tablenotetext{^*}{This column only appears if \sorcha has been instructed not to drop unlinked objects in the configuration file and the linking filter is on.}
\tablenotetext{^{\dagger}}{This column only appear if the linking filter is on.}
\end{deluxetable}

\section{Validation} \label{sec:validation}
\begin{figure}
\begin{center}
\includegraphics[width=0.80\columnwidth]{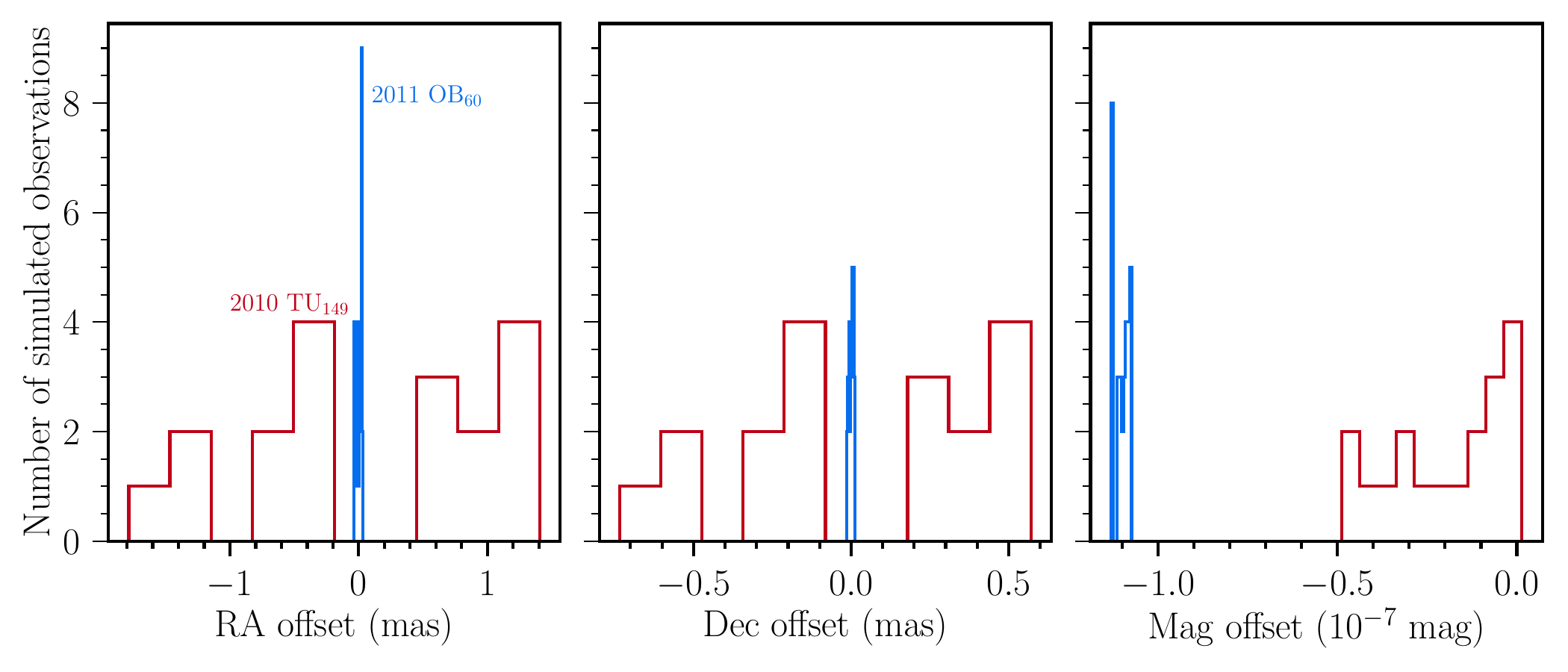}
\caption{Histogram of differences in right ascension (\textbf{left panel)}, declination (\textbf{center}) and trailed source magnitudes (\textbf{right}) between \sorcha derived values and an independent verification using JPL Horizons. The red curves show 2010 TU$_{149}$, a main belt asteroid, while the blue curves show 2011 OB$_{60}$, a TNO.}
\label{fig:verification}
\end{center}
\end{figure}

To verify whether the results from \sorcha are meaningful and properly reproduce what is expected from a survey, we have independently computed every step of the simulation for two objects (2011 OB$_{60}$, a TNO, and 2010 TU$_{149}$, a main belt asteroid), starting with the ephemerides as seen from the Rubin Observatory over the span of a month, derived using JPL Horizons\footnote{\url{https://ssd.jpl.nasa.gov/horizons/app.html\#/}}, and compared to the \sorcha outputs for one of the reference LSST baselines. Due to limitations in the Horizons system, the ephemerides can only be generated at a one point per minute cadence and, since the LSST is approximately twice as fast as that (one image every 30 seconds, plus a few seconds of overheads), we interpolated these positions to the exposure time midpoint. We verified (again using an independent implementation) whether the object was inside the simplified circular footprint for every one of the simulated $\approx 19$ thousand exposures in this period, and computed the \update{trailed} source apparent magnitude given similarly interpolated distances (topocentric and heliocentric), as well as including color terms. The \sorcha-derived and independently computed list of simulated observations agreed with each other (that is, the two sets of pointings that would have seen the object were identical), with the mean difference $\mu$ and standard deviation $\sigma$ between the \sorcha-derived right ascensions being $\mu = 8\times10^{-4}$ mas, $\sigma = 0.02$ mas for 2011 OB$_{60}$, and $\mu = 0.10$ mas, $\sigma = 1.02$ mas for 2010 TU$_{149}$. For declinations, we have $\mu = 9\times10^{-5}$ mas, $\sigma = 0.007$ mas for 2011 OB$_{60}$ and $\mu = 0.02$ mas, $\sigma = 0.42$ mas for 2010 TU$_{149}$. Both of these values show the high quality of the predictions expected from ASSIST \citep{holman2023}, and are below the LSST astrometric requirement of 10 mas precision \citep{ivezic2019}. The \sorcha-derived and independently computed trailed source magnitudes had a systematic offset of $-10^{-7} \, \mathrm{mag}$ for 2011 OB$_{60}$ and $-1.9 \times 10^{-8}$ mag for 2010 TU$_{149}$, several orders of magnitude below the LSST design goals of 10 mmag. We illustrate these results in Figure \ref{fig:verification}. Such a stringent test, then, demonstrates that \sorcha produces correct results. 

\section{Benchmarking and Performance} \label{sec:benchmarking}
A benchmarking script is included in the \href{https://github.com/dirac-institute/sorcha}{\sorcha repository}, which runs \sorcha on a test-set of 1000 MBAs for one year \update{of} the survey, using default \sorcha settings with all filters turned on. On a 2020 M1 MacBook Pro, this benchmark script takes around 115 seconds to run. Users may run this benchmarking script themselves if they wish to test their own machine or access more granular details of the benchmark, as described in the \href{https://sorcha.readthedocs.io/}{online documentation}. \sorcha has been tested extensively on HPC facilities. On the University of Rijeka's Bura supercomputer, 1 million model Centaurs with a chunk size of 5000, split across 100 cores, takes roughly 2 hours to run, or 200 core-hours. We note that the benchmark script may take longer to run on HPC facilities: HPC nodes typically have slower clocks than consumer-grade CPUs but many times the number of cores. These estimates are for \sorcha simulations that send all of the input population to ephemeris generator. Depending on the use case, it may be possible to further speed up \sorcha simulations by applying the faint object culling filter (as described in Section \ref{sub:culling_filter}) to estimate the maximum brightness of the input objects and remove those that will never get bright enough to be observed before the ephemeris generation stage. 

\section{Utility Scripts} \label{sec:utilities}
\sorcha has some useful utilities developed to aid the user, all of which are accessible via the CLI. For example, when \sorcha is installed, a demo can be run to check that the installation has proceeded correctly. The files used for the demo can be copied into a directory of the user's choice by running:

\begin{lstlisting}[language=bash, frame=single, caption={}, label=cmd:demo_files]
$ sorcha demo prepare
\end{lstlisting}
The demo command to run \sorcha using these files can be obtained through:

\begin{lstlisting}[language=bash, frame=single, caption={}, label=cmd:demo]
$ sorcha demo howto
\end{lstlisting}
The example configuration files given in Appendix~\ref{app:configs} can be copied into a directory of the user's choice with:

\begin{lstlisting}[language=bash, frame=single, caption={}, label=cmd:configs]
$ sorcha init 
\end{lstlisting}
The list of papers and software to cite for \sorcha can be found using:

\begin{lstlisting}[language=bash, frame=single, caption={}, label=cmd:cite]
$ sorcha cite
\end{lstlisting}
A utility also exists to collate the results of several \sorcha runs -- for example, those on HPC facilities -- into a single SQLite database, including the input files as separate tables on the database. This can be accessed via:

\begin{lstlisting}[language=bash, frame=single, caption={}, label=cmd:sqlite]
$ sorcha outputs create-sqlite
\end{lstlisting}
Finally, another utility for HPC users serves \update{to check all the log files from} many \sorcha runs and determine whether any of them failed, producing an output file which will list failed runs and give the reason for failure.

\begin{lstlisting}[language=bash, frame=single, caption={}, label=cmd:logs]
$ sorcha outputs check-logs
\end{lstlisting}
More detailed help and explanation is available for all of the command line utilities via the standard \texttt{--help} flag.

\section{Important Limitations and Caveats} \label{sec:caveats}

We built \sorcha to handle a wide variety of use cases, but there are some scenarios that the software is not designed to handle.  In addition, there are also some assumptions we have made. We list below a few noteworthy things that the user should be aware of and consider  before using \sorcha for science.

If using the built-in ephemeris generator (Section \ref{sec:ephemeris}), the orbits of the massive dwarf planet and asteroid perturbers within the main asteroid belt (listed in Table \ref{tab:perturber}) cannot not be directly simulated as they are already included as masses within the ASSIST+REBOUND integrations. Adding in a synthetic massless particle on the same orbit, would find that orbit significantly modified by the  interactions with the perturber.  Additionally, non-gravitational forces such as cometary outgassing,  Yarkovsky–O'Keefe–Radzievskii–Paddack \citep[YORP:][]{radzievskii1952,paddack1969, o'keefe1976} effect (see \cite{bottke2006} and \cite{vokrouhlicky2015} for a detailed review), Yarkovsky forces \citep{bottke2006},  or the impulse from a rocket engine (in the case of a moving artificial spacecraft) are not accounted for in the ASSIST+REBOUND N-body integrations used by \sorcha. At the time of writing, ASSIST is also unable to handle modeling small body collisions, breakup events, or accounting for planet impacts. If required, a user can provide separate ephemerides accounting for any of these situations as input into \sorcha rather than using the built-in ephemeris generator and its ASSIST+REBOUND integrations.

By default, \sorcha assumes that a synthetic planetoid's phase curve parameters are the same for each apparition. Over time as the object moves in its orbit, the viewing angle changes and more or less of a planetoid's surface may become visible or new regions with differing albedo may now be illuminated, changing the amount of sunlight reflected to the Earth and the observed phase curve. Distant objects like TNOs do not move very far in their orbits, so this effect is mostly negligible, but for some MBAs, Jupiter Trojans, and NEOs this effect can be significant \citep[e.g.][]{kwiatkowski1992, carvano2015, mahlke2021, schemel2021, jackson2022, robinson2024}. Bulk analysis of multi-year MBA phase curves from the Asteroid Terrestrial-impact Last Alert System (ATLAS) survey explored the impact of this effect on the MBAs. \cite{mahlke2021} measured a median value of 0.07 mags for the amplitude of the magnitude dispersion due to aspect angle changes between different apparitions. Future enhancements to \sorcha might include adding the recently proposed $sHG_1G_2$ phase curve model \citep{carry2024}, which accounts for the variable viewing geometry. 

Currently, we apply the same source detection and linking efficiencies across the sky regardless of the background stellar density. This may not necessarily be true in regions within or very near to the galactic plane, where the stellar crowding is significant. The LSST is using difference imaging to identify transient sources and  moving solar system objects, which can help mitigate some of the impact from high stellar crowding in the survey observations \citep{juric2021,ivezic2019}. Our assumption that  the efficiency is the same as in regions with no crowding is likely the best choice at this stage before LSSTCam commissioning and the start of Rubin Operations. It would be straightforward to modify \sorcha to include a variable detection efficiency as a function of galactic latitude if there is a significant difference in the galactic plane fields discovery rates compared to the rest of the LSST footprint. 

We also remind the reader that for the LSST, \sorcha computes two apparent magnitudes, the trailed source magnitude and the PSF magnitude. As introduced in Section \ref{sec:rubin_apparent_mags}, the PSF magnitude is used to assess whether a solar system object will be detected by the Rubin source detection algorithms. The trailed source magnitude is the resulting apparent magnitude from summing up all the photons hitting the detector, taking into account that the source may be trailed/streaked. For \emph{most} science cases, the user will need the trailed source magnitudes, and these values are \update{included} in the default output for \sorcha. If not using the default output option, we urge the user \emph{to take great care and ensure} that they are using the correct apparent magnitude column in \sorcha output files for their analysis.

\section{Summary and Conclusions} \label{sec:conclusions}
In this work, we have introduced the methodology and software design behind \sorcha, an open-source \python survey simulator that will take any defined \update{input} small body population and bias it to what a survey would have detected utilizing the survey's pointing history, observation metadata, camera footprint, search algorithm detection efficiency, and other information about the telescope and observatory. \sorcha is \texttt{pip} and \conda installable. The source code is hosted in an \href{https://github.com/dirac-institute/sorcha}{open repository}. Tutorials and \href{https://sorcha.readthedocs.io/}{online documentation} are available as well. We welcome the community to add enhancements to the simulator package through pull requests on the repository. Additional light curve or cometary activity classes developed by the community can be shared with a pull request to the \href{https://github.com/dirac-institute/sorcha-addons}{\texttt{sorcha add-ons} repository}.

The significant challenge in developing \sorcha has been to optimize for the scale of the LSST discoveries ($\sim$5 million+ detections/ $\sim$1 billion individual photometric and astrometric detections at the end of 10 years). We have prioritized making \sorcha fast and easy to use, whether the user intends to run on hundreds to thousands of cores on a HPC system or executing several processes on a laptop or desktop computer.  Our aim is for \sorcha to be a community resource that can accurately handle orbits from across the solar system, including hyperbolic and near-Earth orbits. One such example is how \sorcha will be used in the nightly Rubin SSP pipelines as part of the software to associate previously known solar system objects on secure orbits with real-time LSST transient detections within 60s of the LSST Camera (LSSTCam) reading out an exposure.

Currently, \sorcha is configured to use the outputs from the Rubin Observatory LSST survey strategy simulations. Once the Rubin Observatory is operational, we will add in the capability to ingest the LSST's observation metadata. Although \sorcha has been developed with LSST in mind, much of the code base is applicable to other wide-field and pencil-beam surveys. With some effort, it would be possible for \sorcha to be modified to simulate what these surveys should have detected for an input model small body population. Future versions of the software may  incorporate recent well characterized wide-field discovery surveys such as the Outer Solar System Origins Survey \citep[OSSOS; ][]{bannister2016, bannister2018}, and the Dark Energy Survey \citep[DES; ][]{bernardinelli2020, bernardinelli2022} or the Formation of the Outer Solar System: An Icy Legacy \citep[FOSSIL; ][]{chang2021, ashton2023} into \sorcha to enable joint constraints with the LSST Solar System discoveries.


\section*{Acknowledgments}

 This work was supported by a LSST Discovery Alliance LINCC Frameworks Incubator grant [2023-SFF-LFI-01-Schwamb]. Support was provided by Schmidt Sciences. S.R.M. and M.E.S. acknowledge support in part from UK Science and Technology Facilities Council (STFC) grants ST/V000691/1 and ST/X001253/1. G.F. acknowledges support in part from STFC grant ST/P000304/1. This project has received funding from the European Union’s Horizon 2020 research and innovation program under the Marie Sk\l{}odowska-Curie grant agreement No. 101032479. M.J. and P.H.B. acknowledge the support from the University of Washington College of Arts and Sciences, Department of Astronomy, and the DiRAC (Data-intensive Research in Astrophysics and Cosmology) Institute. The DiRAC Institute is supported through generous gifts from the Charles and Lisa Simonyi Fund for Arts and Sciences and the Washington Research Foundation. M.J. wishes to acknowledge the support of the Washington Research Foundation Data Science Term Chair fund, and the University of Washington Provost's Initiative in Data-Intensive Discovery. J. Murtagh acknowledges support from the Department for the Economy (DfE) Northern Ireland postgraduate studentship scheme and travel support from the STFC for UK participation in LSST through grant ST/S006206/1. J.A.K. and J. Murtagh thank the LSST-DA Data Science Fellowship Program, which is funded by LSST-DA, the Brinson Foundation, and the Moore Foundation; their participation in the program has benefited this work. S.E. and S.C. acknowledge support from the National Science Foundation through the following awards: Collaborative Research: SWIFT-SAT: Minimizing Science Impact on LSST and Observatories Worldwide through Accurate Predictions of Satellite Position and Optical Brightness NSF Award Number: 2332736 and Collaborative Research: Rubin Rocks: Enabling near-Earth asteroid science with LSST NSF Award Number: 2307570. R.R.L. was supported by the UK STFC grant ST/V506990/1. Any opinions, findings, and conclusions or recommendations expressed in this material are those of the authors and do not necessarily reflect the views of the National Science Foundation.
 
 This work was also supported via the Preparing for Astrophysics with LSST Program, funded by the Heising Simons Foundation through grant 2021-2975, and administered by Las Cumbres Observatory. This work was supported in part by the LSST Discovery Alliance Enabling Science grants program, the B612 Foundation, the University of Washington's DiRAC Institute, the Planetary Society, Karman+, Breakthrough Listen, and Adler Planetarium through generous support of the LSST Solar System Readiness Sprints. Breakthrough Listen is managed by the Breakthrough Initiatives, sponsored by the Breakthrough Prize Foundation (\url{http://www.breakthroughinitiatives.org}). 

 This research has made use of NASA’s Astrophysics Data System Bibliographic Services. This research has made use of data and/or services provided by the International Astronomical Union's Minor Planet Center. The SPICE Resource files used in this work are described in \citet{acton1996, acton2018}. Simulations in this paper made use of the REBOUND N-body code \citep{rein2012}. The simulations were integrated using IAS15, a 15th order Gauss-Radau integrator \citep{rein2015}. Some of the results in this paper have been derived using the \texttt{healpy} and HEALPix packages. This work made use of Astropy:\footnote{http://www.astropy.org} a community-developed core \python package and an ecosystem of tools and resources for astronomy \citep{astropy2013,astropy2018,astropy2022}. We thank the Vera C. Rubin Observatory Data Management Team and Scheduler Team for making their software open-source. We thank Dave Young and Conor MacBride for initial help setting up the \python project and repository. The authors also thank Michele Bannister and Rosemary Dorsey for conversations that helped improve the software's handling of interstellar objects. We also thank Aidan Berres, Ricardo B\'anffy, Richard Cannon, Konstantin Malanchev, Brian Rogers, and Richard Cannon,  for their contributions to documentation, discussions about feature implementations,and/or beta testing. We thank Max Mahlke for useful feedback on the manuscript. \update{We also thank Jean-Marc Petit for his detailed and thoughtful review of this manuscript.}  We are additionally grateful to the members of the Rubin Observatory LSST Solar System Science Collaboration for useful feedback at the LSST Solar System Readiness Sprints. We also thank the contributors to Stack Overflow for their examples and advice on common \python challenges that provided guidance on solving some of the programming challenges we have encountered. 

 This material or work is supported in part by the National Science Foundation through Cooperative Agreement AST-1258333 and Cooperative Support Agreement AST1836783 managed by the Association of Universities for Research in Astronomy (AURA), and the Department of Energy under Contract No. DE-AC02-76SF00515 with the SLAC National Accelerator Laboratory managed by Stanford University.  

We are grateful for the use of the computing resources from the Northern Ireland High Performance Computing (NI-HPC) service funded by EPSRC (EP/T022175). We gratefully acknowledge the support of the Center for Advanced Computing and Modelling, University of Rijeka (Croatia), for providing supercomputing resources at HPC (High Performance Computing) Bura.

The authors wish to acknowledge the researchers who worked tirelessly to rapidly develop COVID-19 vaccines and subsequent boosters. Without all their efforts, we would not have been able to pursue this work. 

We acknowledge the contribution of pets Isha Bernardinelli; Freddie and Millie Merritt, Stella Schwamb; Richard, Calcifer, and Buttons West; that, by keeping us awake at night, yowling during our meetings, or providing general emotional support, led to improvements in this software and manuscript. 

Data Access:  The software presented here is available open-source at \url{https://github.com/dirac-institute/sorcha}.

%

\vspace{5mm}
\facility{Rubin}


\software{sorcha, ASSIST \citep{holman2023,rein2023}, Astropy \citep{astropy2013,astropy2018,astropy2022}, Healpy \citep{gorski2005,zonca2019}, Matplotlib \citep{hunter2007}, Numba \citep{lam2015}, Numpy \citep{harris2020}, pandas \citep{mckinney2010, pandas2020}, Pooch \citep{uieda2020}, PyTables \citep{pytables2002}, REBOUND \citep{rein2012, rein2015}, rubin$\_$sim \citep{Connolly2014,yoachim2022},rubin$\_$scheduler \citep{naghib2019, yoachim2024}, sbpy \citep{mommert2019}, SciPy \citep{virtanen2020}, Spiceypy \citep{annex2020}, sqlite (\url{https://www.sqlite.org/index.html}), sqlite3 (\url{https://docs.python.org/3/library/sqlite3.html}), tqdm \citep{dacostaluis2023}, Black (\url{https://black.readthedocs.io/en/stable/faq.html}), Jupyter Notebooks \citep{kluyver2016}}.

\begin{contribution}
S.R.M. served as the lead developer of \sorcha from June 2021 onward, contributing a substantial portion of the current code base and the majority of unit tests. They also led new feature implementations, oversaw ongoing code maintenance and repository management, resolved software issues, conducted extensive testing on both HPC clusters and local systems, and contributed to the documentation. For this paper, they wrote the abstract and Sections~\ref{sec:inputs}, \ref{sec:outputs} and \ref{sec:benchmarking}; they wrote Sections~\ref{sec:intro} and \ref{sec:lsst} with reference to an earlier draft by G.F.; they wrote Appendix~\ref{app:running_sorcha} and the preambles to Section~\ref{sec:software_design} and \ref{sec:post-processing}; and they co-wrote Sections~\ref{sub:trailing_losses} to \ref{sub:saturation}. They created Figure~\ref{fig:fadingfunction} and generated Figures~\ref{fig:trailing_loss_2}, \ref{fig:vignetting} and \ref{fig:footprint_validation} based on Jupyter notebooks created by S.C. and G.F.. They made Tables~\ref{tab:output_cols} and \ref{tab:output_stats} and created Tables~\ref{tab:orbits}-\ref{tab:footprint} based on tables made by J.A.K.. They are also responsible for the overall layout and organization of the paper and for edits to the draft.

G.F. was the initial developer of \sorcha and contributed significantly to early drafts of the manuscript. Substrates of his text can be found in the article (in particular in Sections \ref{sec:intro}, \ref{sec:lsst}, \ref{sec:inputs} and \ref{sec:post-processing}), and in the code. He also produced Figure \ref{fig:trailing_loss} and contributed to the design of Figure \ref{fig:fadingfunction}. G.F. participated in the discussions and decisions about the overall design and implementation of \sorcha. He also provided feedback on the overall paper draft. 

M.E.S. served as principal investigator and project manager of the \sorcha team. She also served as PI of the \sorcha LINCC Frameworks Incubator Proposal. She contributed to the discussions and decisions about the overall design and implementation of \sorcha and the documentation: this included leading the team in-person meetings and calls. She helped beta test the software, contributed to the documentation, formatted docstrings, and also reviewed pull requests submitted to the \texttt{GitHub} repository. She also led the development of the built-in citation function. She wrote Sections~\ref{sub:documentation},  \ref{sec:ephemeris}, \ref{subsub:comet_activity},  \ref{sub:linking_filter}, \ref{sec:caveats}, and \ref{sec:conclusions}, and \ref{sub:random_numbers}.  She also contributed to Sections \ref{sub:magnitude_calc},  \ref{sub:trailing_losses},\ref{sub:footprint}, and  \ref{sub:linking_filter}. She made Figures~\ref{fig:main_workflow}, \ref{fig:overview}, \ref{fig:PSF_trailed_source_mags}, \ref{fig:simple_comet}, and \ref{fig:LSSTCamfovs}. She also provided significant feedback on the overall paper draft, assisted with incorporating co-author feedback into the manuscript, and made significant contributions to the \sorcha documentation. 

S.C. contributed to \sorcha software development by implementing several of the fundamental models of the telescope footprint and post-processing functions. S.C. also assisted in validation of intermediate calculations and final datasets, and created visualizations for the intermediate and final datasets. S.C. also contributed to this publication by authoring Section~\ref{sub:uncertainties}.

P.H.B. contributed to the development of several of the \sorcha routines for ephemerides generation and coordinate conversion described in Section \ref{sec:ephemeris}, as well as the design of the optional and user-defined cometary activity and light curve models of Section \ref{sub:add-ons}. He implemented the independent validation test described in Section \ref{sec:validation}. He also helped with the design of Figure \ref{fig:trailing_loss_2}. Besides these sections, he also wrote Appendix \ref{app:mag_uncertainty}, and provided feedback on the overall paper draft.

M.J. contributed the initial architecture design, the algorithm for the repeatable parallel random-number generation, the vectorized algorithm and implementation for the {\tt miniDifi} linking emulation package, and the design and implementation of UX (user experience) elements such as the fast command line activation and options. He also produced Figure \ref{fig:SSP}. M.J. also served as the liaison to the Rubin project and coordinated its application in Rubin Data Preview 0.3 (which was created using an early version of \sorcha).

M.J.H. led the development and refinement of Sorcha's ephemeris generator algorithm with ASSIST+REBOUND.  He also helped verify the output from the ephemeris generator components and improved the algorithm to better handle fast moving solar system objects. He also provided feedback on the paper manuscript. 

J.A.K. provided testing, user feedback, and bug-fixing throughout \sorcha's development. He also contributed to the \sorcha documentation. He also provided feedback on the paper manuscript. 

S.E. contributed to the conceptualization of \sorcha as well as providing feedback on the LINCC Frameworks Incubator proposal. He was involved in the discussions about the overall design and functionality of \sorcha. Moreover, S.E. contributed to the choice and implementation of the astrometric and photometric uncertainty models presented in Section~\ref{sub:magnitude_calc} as well as the editing of this article. He also provided feedback on the paper manuscript. 

D.O. applied the LF-PPT to the \sorcha code base, introduced several performance improvements, and implemented the plugin system allowing externally defined activity and light curve models to be used by \sorcha. D.O. also contributed text to Appendix~\ref{sub:testing_and_deployment}.

M.W. contributed to the \sorcha code base, including efficiency improvements in array handling and integrating the first draft of the ASSIST+REBOUND ephemeris generation.

J.K. contributed to the \sorcha code base, including improvements to input file reading and random number generation. J.K. also contributed text to Appendix \ref{sub:testing_and_deployment} and Section \ref{sub:readers}.

J. Murtagh contributed to the overall code base, in particular providing testing of the linking filter, cleaning up logging functionality, developing example Jupyter notebooks for estimating input colors, contributing to various sections of the documentation, and writing Section 6.12 detailing the operation of the faint object culling filter. He also provided feedback on the overall manuscript and figures, in particular Figure~\ref{fig:main_workflow} and Figure~\ref{fig:overview}.

R.L.J. developed \texttt{rubin\_sim} and \texttt{rubin\_scheduler}. Some of functions within \sorcha were adapted from \texttt{rubin\_sim}. R.L.J.  also contributed to useful discussion and provided feedback and guidance on how to use the simulated \texttt{rubin\_sim} survey cadence pointing databases and how to apply trailing losses and uncertainty calculations. She also produced a Jupyter notebook that was used as an example template to develop Figure \ref{fig:trailing_loss_2}. She also provided feedback on the paper manuscript. 

P.Y. developed  \texttt{rubin\_sim} and \texttt{rubin\_scheduler}. Some of functions within \sorcha were adapted from \texttt{rubin\_sim}. P.Y. also provided feedback and guidance on \texttt{rubin\_sim} cadence simulation pointing databases. 

R.R.L. implemented the data class configuration file reader. He also implemented the upgrades to enable the user to optionally specify the ephemeris generation auxiliary files and other additional variables in the configuration file. He also contributed updates to improve the overall user experience with the \sorcha CLI and other changes to improve the maintainability of the \sorcha code base. 

M.S.P.K. developed the initial version of the \texttt{LSSTCometActivity} class that was modified for inclusion in \texttt{Sorcha add-ons} and that led to the development of the abstract comet class implementation. 

J. Moeyens provided feedback and contributed to discussions during coding sprints at the University of Washington. He developed the simulated linking algorithm (\texttt{difi}) that was used in early iterations of \sorcha that led to the development of \texttt{miniDifi}. 

K.K. contributed to beta testing \sorcha for the edge case of simulated Earth impactors and very fast-moving potentially hazardous asteroids (PHAs). 

S.P.N. developed the \texttt{objectsInField} (\texttt{OIF}) ephemeris generator/simulator that enabled the development of \sorcha. The output from OIF was the input into early iterations of \sorcha as the source of ephemeris calculations.  

C.S. provided feedback and advice on incorporating cometary activity in to \sorcha's apparent magnitude calculations. 

S.M.M. made contributions to early drafts of \sorcha documentation and setting up early versions of sorcha to be pip installable.

C.O.C. contributed in the discussions about the development and enhancement of \sorcha during the LINCC Frameworks Incubator. 
\end{contribution}



\appendix

\section{Repository Setup, Software Testing, and Package Architecture and Deployment} \label{sub:testing_and_deployment}
\sorcha uses software engineering best practices, processes, and tools to help ensure the accuracy, usability, and long-term maintainability of the code. Many of these were configured through the use of the LINCC (LSST Interdisciplinary Collaboration for Computing)-Frameworks \python Project Template\footnote{\url{https://github.com/lincc-frameworks/python-project-template}} \citep[LF-PPT; ][]{oldag2024} which provides a suite of tools and conventions that are readily applied to new or existing projects. The LF-PPT automatically generates a familiar directory structure to organize code and provide the foundation for continuous integration (CI) with \texttt{GitHub} Actions\footnote{\url{https://docs.github.com/en/actions}} to exercise the code via unit tests and ensure adherence to user-defined coding style guides. These tools and practices ensure the accuracy of, and prevent regressions in, the software. To aid with long-term maintenance \update{of} the software, the LF-PPT provides a nightly CI smoke test that executes \sorcha's comprehensive suite of tests. This smoke test confirms that the package continues to build and run even as lower-level dependencies might change. This low-effort approach makes the code easier to maintain over many years. Additionally, to make documentation easy to produce, readily accessible, and maintainable, the LF-PPT provides a set of basic Sphinx documentation generator\footnote{\url{https://www.sphinx-doc.org/en/master/index.html}} configuration files needed to automatically publish documentation to ReadTheDocs\footnote{\url{https://readthedocs.org/}}. By default the LF-PPT documentation configuration automatically includes user generated pages, API (Application Programming Interface) docstrings, and rendered Jupyter notebooks.

\sorcha uses multiple well-adopted package repositories to provide a mechanism for users to install and update the package. \sorcha is both \conda/\mamba and \pip installable. The LF-PPT provides a workflow for automatically publishing a new release to \texttt{PyPI}\footnote{\url{https://pypi.org/project/sorcha/}} when a release is tagged in the \sorcha \href{https://github.com/dirac-institute/sorcha}{\texttt{Github} repository}. We also created additional recipes to automatically publish to \conda-forge\footnote{\url{https://github.com/conda-forge/sorcha-feedstock/}}. Minimal effort is required to deploy new versions of the code due to the automation. \texttt{PyPI} and \condaforge were selected as the release channels due to their popularity in the scientific community.

\subsection{Command Line Interface Implementation}
Importing the entire \sorcha \python package can take five seconds or more, and making the user wait that long just to print out an error message would be a poor user experience. To provide the user with a fast experience on the command line, \sorcha's command line interface (CLI) and parsing is separated from the core code. The large imports of \python packages, including the \sorcha package, are performed only in the execute function that runs the main \sorcha simulation. Importing \sorcha from the CLI execute function and not at the top-level of the module allows us to exit quickly and print the help information or print an error message (in case there was a mistake on the command line). 

\subsection{Random Number Generation} \label{sub:random_numbers}
\sorcha implements a per-module randomization approach that provides the ability to force deterministic behavior during testing regardless of the order in which modules are executed, as described in \cite{schwamb2024}. The \texttt{PerModuleRNG} class creates random number generators for each model using a global seed and a hash based on the module's name. By default \sorcha uses the 4 bytes returned from \texttt{os.urandom()}, which returns a string of random bytes from the operating system's randomness source, as the global seed to ensure random behavior for scientific runs. This ensures that the random seed is initialized uniquely when launching nearly simultaneous parallel \sorcha runs across many nodes in a HPC cluster. For reproducibility (mainly for testing, debugging purposes), a user can set a hidden environmental configuration parameter to override this setting and force deterministic behavior.

\subsection{Reader Classes} \label{sub:readers}
\sorcha reads data files via a series of reader classes that provide wrappers around common libraries for reading in data from such formats as CSV (comma-separated values), whitespace-separated text, HDF5, and SQLite databases. By using a common wrapper, derived from an abstract \texttt{ObjectDataReader} class, the individual reader classes present the user with a common interface for accessing data files with functions such as \texttt{read\_rows()}. Individual reader classes inherit from the \texttt{ObjectDataReader} class and are specialized for both the input format and category of data. For example, the \texttt{OrbitAuxReader} reads in data from comma or whitespace-separated files and contains logic for processing and standardizing different orbital specifications (e.g. Keplerian vs Cartesian parameters). The reader classes also incorporate additional error checking, such as required columns, and use \sorcha's logging framework.

\subsection{Ephemeris Generation Auxiliary Files} \label{sub:auxiliary}

\sorcha's ephemeris generator requires various SPICE (Spacecraft, Planet, Instrument, C-matrix, Events) kernels and resource files\footnote{\url{https://naif.jpl.nasa.gov/naif/data_generic.html}} \citep{acton1996, acton2018} from the Jet Propulsion Laboratory's Navigation and Ancillary Information Facility (NAIF) and the MPC's (Minor Planet Center's) observatory file (obscode\_extended.json.gz\footnote{\url{https://minorplanetcenter.net/data}}) to obtain the precise starting locations for the Moon, planets and 16 asteroid perturbers, as well as the observatory's location on the Earth to initialize the N-body simulations and then calculate the right ascension, declination, heliocentric state vectors, and light-time-corrected values. These auxiliary files present a challenge due to their size and frequency of updates.  Large files are impractical to include with packaged code and files that change frequently require additional effort to keep the distribution up to date. 
One alternative requires the end user to manually gather all the required files and then provide the paths at runtime. 
\sorcha addresses all of these issues by using the data retrieval tool \texttt{Pooch} \citep{uieda2020} which can be run as part of a stand alone bootstrap utility when \sorcha is first installed or automatically run within a \sorcha simulation if the auxiliary files are not found (see Section \ref{sec:utilities}). 
\sorcha defines a registry file with the URL (Uniform Resource Locator) locations of all required auxiliary files.
At runtime, \texttt{Pooch} uses the registry to download and cache the files in the default temporary directory of the operating system or on a directory defined by the user. 
During subsequent runs the files are retrieved from the local cache and only downloaded as needed. 
The user is always free to move or delete the files from the cache to trigger \texttt{Pooch} to download the latest copy of the default auxiliary files during the next \sorcha run. If there is a need to run \sorcha with a new or older version of the NAIF SPICE kernels or the MPC observatory file, the \sorcha configuration file can be used to override the default file names and URL addresses for any of the auxiliary files \sorcha uses.

\section{Quick-Start Guide}
\label{app:running_sorcha}
This brief quick-start guide shows the user how to set up and run a set of demonstration input files through \sorcha using the command line. For detailed installation instructions, we refer the user to our \href{https://sorcha.readthedocs.io/en/latest/installation.html}{documentation}.
\\
\\
\noindent \sorcha can be installed via \pip or \conda/\mamba from the \condaforge channel:

\begin{lstlisting}[language=bash, frame=single, caption={}, label=cmd:install_conda]
$ conda install -c conda-forge sorcha
\end{lstlisting}

or:

\begin{lstlisting}[language=bash, frame=single, caption={}, label=cmd:install_mamba]
$ mamba install -c conda-forge sorcha
\end{lstlisting}

or:

\begin{lstlisting}[language=bash, frame=single, caption={}, label=cmd:install_pip]
$ pip install sorcha
\end{lstlisting}
Once \sorcha is installed, the user must also download the auxiliary files used for ephemeris generation. This needs to be done only once. To install these files into the local system cache (there is an option to specify a directory for these files if running on an HPC setup):

\begin{lstlisting}[language=bash, frame=single, caption={}, label=cmd:bootstrap]
$ sorcha bootstrap
\end{lstlisting}
\sorcha comes packaged with a set of demo files for ten random solar system objects. This includes a small one-year version of the pointing database, a configuration file similar to those presented in Appendix~\ref{app:configs}, an orbits file and a physical parameters file. 
\\
\\
\noindent The contents of the input orbits file:
\begin{scriptsize}
\begin{verbatim}
ObjID		a		e		inc		node		argPeri		ma	  epochMJD_TDB	FORMAT
2010_TU149	2.2103867	0.8245336	1.9679		58.91911	92.60419	324.21145	60200	KEP
2010_TC209	2.6411033	0.2378397	17.51668	25.01678	18.48118	347.09673	60200	KEP
2011_OB60	103.711231	0.6461554	19.42877	142.66859	226.19366	358.36581	60200	KEP
2011_WJ157	99.2799891	0.6226317	27.06784	76.41243	58.43528	355.56379	60200	KEP
2011_YA42	3.1910723	0.1754333	17.90056	291.08053	99.36952	70.55556	60200	KEP
2012_AW12	1.6899997	0.6063531	51.51998	151.77178	246.62641	150.38051	60200	KEP
2012_BB14	1.0627063	0.098754	2.64121		316.78101	255.55235	135.8483	60200	KEP
2012_HW1	1.0607726	0.6615131	49.57978	207.81039	163.70136	63.97462	60200	KEP
2013_CC263	3.0000262	0.1126188	8.97766		336.5147	98.64025	69.02504	60200	KEP
2013_GD138	43.6879964	0.1086886	5.43381		57.00856	146.16056	24.30927	60200	KEP
\end{verbatim}
\end{scriptsize}
The contents of the input physical parameters file for the same ten synthetic objects with orbits described in the orbits file:

\begin{scriptsize}
\begin{verbatim}
ObjID		H_r	GS	u-r	g-r	i-r	z-r	y-r
2010_TU149	20.7	0.15	2.13	0.65	-0.19	 0.14	-0.14
2010_TC209	18.45	0.15	1.9	0.58	-0.21	-0.3	-0.39
2011_OB60	6.9	0.15	1.72	0.48	-0.11	-0.12	-0.12
2011_WJ157	4.92	0.15	2.55	0.92	-0.38	-0.59	-0.70
2011_YA42	16.36	0.15	2.13	0.65	-0.19	 0.14	-0.14
2012_AW12	19.79	0.15	1.9	0.58	-0.21	-0.3	-0.39
2012_BB14	24.99	0.15	1.72	0.48	-0.11	-0.12	-0.12
2012_HW1	19.99	0.15	2.13	0.65	-0.19	 0.14	-0.14
2013_CC263	18.46	0.15	1.72	0.48	-0.11	-0.12	-0.12
2013_GD138	8.14	0.15	2.55	0.92	-0.38	-0.59	-0.70
\end{verbatim}
\end{scriptsize}
 For more details on the contents of all the demonstration input files, we refer the user to Section~\ref{sec:inputs}. A single \sorcha command will copy these demo files into the local directory and print the command needed to run the \sorcha demo to the terminal:

\begin{lstlisting}[language=bash, frame=single, caption={}, label=cmd:demo_prepare_2]
$ sorcha demo prepare
\end{lstlisting}
We do not supply the main command to run sorcha here as this demo command may change in future versions of the code. If the user requires the current working demo command to be reprinted to the terminal, they can run:

\begin{lstlisting}[language=bash, frame=single, caption={}, label=cmd:demo_howto_2]
$ sorcha demo howto
\end{lstlisting}
The output \update{from the} \texttt{sorcha run} demo command can be simply copy-pasted into the terminal and run. Once \sorcha has finished running, it will have \update{produced} two log files and two output files: the main \sorcha output file, \texttt{testrun\_e2e.csv}, and a statistics file, \texttt{testrun\_stats.csv}. The contents of these files are described in Section~\ref{sec:outputs}. The top few lines of the main \sorcha output file will be a CSV file, the first few lines of which will appear similar to the following:

\begin{scriptsize}
\begin{verbatim}
ObjID,fieldMJD_TAI,fieldRA_deg,fieldDec_deg,RA_deg,Dec_deg,astrometricSigma_deg,optFilter,...
2011_OB60,60225.247167832895,2.8340797698367206,-12.19406486,1.9825549664962523,-11.89549965,7.4867112566597e-06,i,...
2011_OB60,60225.27094733885,2.8340797698367206,-12.19406486,1.9821061017355532,-11.89565259,1.970228013144157e-05,z,...
2011_OB60,60225.28270233429,2.8340797698367206,-12.19406486,1.981888824400869,-11.89577997,1.368449919373884e-05,z,...
2011_OB60,60227.24471872862,1.830365601304555,-10.65341974,1.9444094822390525,-11.91153435,1.2466492192790201e-05,r,...
2011_OB60,60227.26843110208,1.830365601304555,-10.65341974,1.9439523895215416,-11.91172213,1.0506087754733871e-05,i,...
\end{verbatim}
\end{scriptsize}
 Note that this output has been truncated in both directions to fit this paper and that the user's precise numbers will differ due to \sorcha's randomization of an object's astrometry and photometry about its uncertainties. For the full list of columns that should be available in this output, we refer the user to the basic output section of Table~\ref{tab:output_cols}.
\\
\\
\noindent The statistics file will look similar to this:

\begin{scriptsize}
\begin{verbatim}
ObjID,optFilter,number_obs,min_apparent_mag,max_apparent_mag,median_apparent_mag,min_phase,max_phase,date_linked_MJD
2010_TC209,g,1,21.042222892374717,21.042222892374717,21.042222892374717,4.822674862328868,4.822674862328868,60259.0
2010_TC209,i,2,20.263741481736623,20.39869118796203,20.331216334849326,4.902659658311991,7.356710425955291,60259.0
2010_TC209,r,2,20.444480018298275,20.617951470547577,20.531215744422926,4.821141532804455,7.345792889497799,60259.0
2010_TC209,u,0,,,,,,
2010_TC209,y,0,,,,,,
2010_TC209,z,1,20.19324699340425,20.19324699340425,20.19324699340425,4.899566415676175,4.899566415676175,60259.0
2011_OB60,g,7,22.946130649671932,23.19931119066643,23.037745264926205,0.378554335727471,1.2668113261408172,60228.0
2011_OB60,i,20,22.307405543931022,22.734689782707683,22.565578703812037,0.3324387390211166,1.5654399545866802,60228.0
2011_OB60,r,11,22.503812312947016,22.752436749239006,22.610941659961966,0.3325287235981864,1.50119506319347,60228.0
2011_OB60,u,1,24.5667047017521,24.5667047017521,24.5667047017521,1.3083690945964197,1.3083690945964197,60228.0
2011_OB60,y,2,22.51622985410243,22.578367849124156,22.54729885161329,0.4251751188145726,0.744770018195348,60228.0
2011_OB60,z,11,22.262555361397094,22.7722417002819,22.38455767193579,0.3323812401128109,1.0086113081302872,60228.0
2011_YA42,g,55,21.988268774434818,23.17042737356163,22.645034273517435,5.779172614054609,16.858086742640804,60294.0
2011_YA42,i,114,21.099983644512545,22.41865729383654,21.80604194913399,4.27818434969482,16.858826689815402,60294.0
2011_YA42,r,119,21.27632010408789,22.60535423791973,21.992096786094883,4.27905901648886,16.858790856474048,60294.0
2011_YA42,u,8,23.287772926582647,24.681129730878084,23.965335308765795,4.786855583707764,16.28483860182808,60294.0
2011_YA42,y,103,21.039217193403374,22.324927626389975,21.830816671100585,4.289386839279333,16.626087100847716,60294.0
2011_YA42,z,82,21.466672869498748,22.642891414440403,22.050717444544205,4.288267265701924,16.42954705183178,60294.0
2012_HW1,g,4,21.28135059249862,21.921530934981835,21.306082738885898,18.24670580343961,23.799694527700037,60401.0
2012_HW1,i,12,19.902649799186086,22.351360183436128,20.769568042204128,15.429492575574567,32.98342142925829,60401.0
2012_HW1,r,7,20.516025653001538,21.380105641386635,20.748282653915954,16.941927704030164,25.057711920957264,60401.0
2012_HW1,u,1,22.702522463962406,22.702522463962406,22.702522463962406,18.286703173487435,18.286703173487435,60401.0
2012_HW1,y,5,20.195605736059495,21.8182694730977,20.3557245332764,15.835262453124184,28.7883450541416,60401.0
2012_HW1,z,9,20.484442207810996,22.260849939372516,21.961964257261524,15.419928152211138,30.44395714693862,60401.0
2013_CC263,g,6,23.432040941615597,24.281615590762463,23.80201802687187,2.638120000168959,11.83521583002445,60375.0
2013_CC263,i,17,22.625645785273722,24.14087814493943,23.239324069683995,2.914130006135205,18.413844880602955,60375.0
2013_CC263,r,17,22.808332102472118,23.8611819839395,23.309735969610564,2.5782284903062447,15.665773562139645,60375.0
2013_CC263,u,0,,,,,,
2013_CC263,y,0,,,,,,
2013_CC263,z,3,22.86153172084036,23.016242989892422,22.92988785440862,3.0658304290927627,11.506994018448005,60375.0
\end{verbatim}
\end{scriptsize}
Once again, the user's precise numbers will differ due to \sorcha's randomization of the objects' photometry and astrometry. Further explanation of the output files can be found in Section~\ref{sec:outputs}.

\section{Example Configuration files}
\label{app:configs}

We provide three example configuration files that will initialize \sorcha for simulating what the LSST should discover and detect using a \texttt{rubin\_sim} cadence simulation as input for the survey pointing history. These example configuration files come installed with the \sorcha \python package and can be copied to any directory via the \update{\texttt{sorcha demo prepare}} command as outlined in Section~\ref{sec:utilities}. 

\subsection{LSST Full Camera Footprint}

This configuration file will force \sorcha to use the default model of the full LSST camera focal plane, accounting for detector and chip gaps. See Section \ref{sec:fullFOV} for further details. For most LSST science needs, we recommend using this setup for \sorcha unless there is a strong need to use the circular FOV approximation, such as speeding up the runtime.

\begin{scriptsize}
\begin{verbatim}
# Sorcha Configuration File
# This configuration file will set up Sorcha to simulate what the LSST would discover.
# This version uses the full LSSTCam detector footprint to detect whether objects are lost to chip gaps.
# This is slightly slower than using a circular approximation, but more accurate.


[INPUT]

# The simulation used for the ephemeris input. 
# ar=ASSIST+REBOUND interal ephemeris generation 
# external=providing an external input file from the command line
# Options: "ar", "external"
ephemerides_type = ar

# Format for ephemeris simulation input file if a file is specified at the command line. 
# This is also the format to which ephemeris files will be written out, if specified.
# Options: csv, whitespace, hdf5
eph_format = csv

# Sorcha chunk size: how many objects should be processed at once?
size_serial_chunk = 20000

# Format for the orbit, physical parameters, and complex physical parameters input files.
# Options: csv or whitespace
aux_format = csv

# SQL query for extracting data from the pointing database.
pointing_sql_query = SELECT observationId, observationStartMJD as observationStartMJD_TAI, visitTime, visitExposureTime, filter, seeingFwhmGeom as seeingFwhmGeom_arcsec, seeingFwhmEff as seeingFwhmEff_arcsec, fiveSigmaDepth as fieldFiveSigmaDepth_mag , fieldRA as fieldRA_deg, fieldDec as fieldDec_deg, rotSkyPos as fieldRotSkyPos_deg FROM observations order by observationId


[SIMULATION]
# Configuration for running the ASSIST+REBOUND ephemerides generator.

# the field of view of our search field, in degrees
ar_ang_fov = 2.06

# the buffer zone around the field of view we want to include, in degrees
ar_fov_buffer = 0.2

# the "picket" is our imprecise discretization of time that allows us to move progress
# our simulations forward without getting too granular when we don't have to.
# the unit is number of days.
ar_picket = 1

# the obscode is the MPC observatory code for the provided telescope.
ar_obs_code = X05

# the order of healpix which we will use for the healpy portions of the code.
# the nside is equivalent to 2**ar_healpix_order
ar_healpix_order = 6


[FILTERS]

# Filters of the observations you are interested in, comma-separated.
# Your physical parameters file must have H calculated in one of these filters
# and colour offset columns defined relative to that filter.
observing_filters = r,g,i,z,u,y


[SATURATION]

# Upper magnitude limit on sources that will overfill the detector pixels/have
# counts above the non-linearity regime of the pixels where one can’t do 
# photometry. Objects brighter than this limit (in magnitude) will be cut. 
# Comment out for no saturation limit.
# Two formats are accepted:
# Single float: applies same saturation limit to observations in all filters.
# Comma-separated list of floats: applies saturation limit per filter, in order as
# given in observing_filters keyword.
bright_limit = 16.0


[PHASECURVES]

# The phase function used to calculate apparent magnitude. The physical parameters input
# file must contain the columns needed to calculate the phase function.
# Options: HG, HG1G2, HG12, linear, none.
phase_function = HG


[FOV]

# Choose between circular or actual camera footprint, including chip gaps.
# Options: circle, footprint.
camera_model = footprint

# The distance from the edge of a detector (in arcseconds on the focal plane)
# at which we will not correctly extract an object. By default this is 10px or 2 arcseconds.
# Comment out or do not include if not using footprint camera model.
footprint_edge_threshold = 2.

# Path to camera footprint file. Uncomment to provide a path to the desired camera 
# detector configuration file if not using the default built-in LSSTCam detector 
# configuration for the actual camera footprint.
# footprint_path= ./data/detectors_corners.csv


[FADINGFUNCTION]

# Uses the fading function as outlined in
# Chesley and Vereš (2017) to remove observations.

# Width parameter for fading function. Should be greater than zero and less than 0.5.
# Suggested value is 0.1 after Chesley and Vereš (2017).
fading_function_width = 0.1

# Peak efficiency for the fading function, called the 'fill factor' in Chesley and Veres (2017).
# Suggested value is 1. Do not change this unless you are sure of what you are doing.
fading_function_peak_efficiency = 1.


[LINKINGFILTER]

# SSP detection efficiency. Which fraction of the objects will
# the automated Rubin Solar System Processing (SSP) pipeline successfully link? Float.
SSP_detection_efficiency = 0.95

# Length of tracklets. How many observations of an object during one night are
# required to produce a valid tracklet?
SSP_number_observations = 2

# Minimum separation (in arcsec) between two observations of an object required 
# for the linking software to distinguish them as separate and therefore as a valid tracklet.
SSP_separation_threshold = 0.5

# Maximum time separation (in days) between subsequent observations in a tracklet. 
# Default is 0.0625 days (90mins).
SSP_maximum_time = 0.0625

# Number of tracklets for detection. How many tracklets are required to classify
# an object as detected? 
SSP_number_tracklets = 3

# The number of tracklets defined above must occur in <= this number of days to 
# constitute a complete track/detection.
SSP_track_window = 15

# The time in UTC at which it is noon at the observatory location (in standard time).
# For the LSST, 12pm Chile Standard Time is 4pm UTC.
SSP_night_start_utc = 16.0


[OUTPUT]

# Output format of the output file[s]
# Options: csv, sqlite3, hdf5
output_format = sqlite3

# Controls which columns are in the output files.
# Options are "basic" and "all", which returns all columns.
output_columns = basic


[LIGHTCURVE]

# The unique name of the lightcurve model to use. Defined in the ``name_id`` method 
# of the subclasses of AbstractLightCurve. If not none, the complex physical parameters 
# file must be specified at the command line.lc_model = none
lc_model = none


[ACTIVITY]

# The unique name of the actvity model to use. Defined in the ``name_id`` method
#  of the subclasses of AbstractCometaryActivity.  If not none, a complex physical parameters 
# file must be specified at the command line.
comet_activity = none


\end{verbatim}
\end{scriptsize}

\subsection{LSST Circular Footprint Approximation}

This configuration file uses the circular footprint filter discussed in Section \ref{sec:circleFOV}. This setup is marginally faster than using the full LSSTCam footprint filter; for most science cases, however, we recommend the user use the configurations laid out in the previous file.

\begin{scriptsize}
\begin{verbatim}
# Sorcha Configuration File
# This configuration file will set up Sorcha to simulate what the LSST would discover.
# This version simulates objects lost to chip gaps between LSSTCam's detectors by 
# simulating a circular detector and approximating the chip gaps, removing all objects 
# outside of a 1.75deg radius and 10% of objects within. 
# This is slightly faster than using the full footprint.


[INPUT]

# The simulation used for the ephemeris input. 
# ar=ASSIST+REBOUND interal ephemeris generation 
# external=providing an external input file from the command line
# Options: "ar", "external"
ephemerides_type = ar

# Format for ephemeris simulation input file if a file is specified at the command line. 
# This is also the format to which ephemeris files will be written out, if specified.
# Options: csv, whitespace, hdf5
eph_format = csv

# Sorcha chunk size: how many objects should be processed at once?
size_serial_chunk = 20000

# Format for the orbit, physical parameters, and complex physical parameters input files.
# Options: csv or whitespace
aux_format = csv

# SQL query for extracting data from the pointing database.
pointing_sql_query = SELECT observationId, observationStartMJD as observationStartMJD_TAI, visitTime, visitExposureTime, filter, seeingFwhmGeom as seeingFwhmGeom_arcsec, seeingFwhmEff as seeingFwhmEff_arcsec, fiveSigmaDepth as fieldFiveSigmaDepth_mag , fieldRA as fieldRA_deg, fieldDec as fieldDec_deg, rotSkyPos as fieldRotSkyPos_deg FROM observations order by observationId


[SIMULATION]
# Configuration for running the ASSIST+REBOUND ephemerides generator.

# the field of view of our search field, in degrees
ar_ang_fov = 2.06

# the buffer zone around the field of view we want to include, in degrees
ar_fov_buffer = 0.2

# the "picket" is our imprecise discretization of time that allows us to move progress
# our simulations forward without getting too granular when we don't have to.
# the unit is number of days.
ar_picket = 1

# the obscode is the MPC observatory code for the provided telescope.
ar_obs_code = X05

# the order of healpix which we will use for the healpy portions of the code.
# the nside is equivalent to 2**ar_healpix_order
ar_healpix_order = 6


[FILTERS]

# Filters of the observations you are interested in, comma-separated.
# Your physical parameters file must have H calculated in one of these filters
# and colour offset columns defined relative to that filter.
observing_filters = r,g,i,z,u,y


[SATURATION]

# Upper magnitude limit on sources that will overfill the detector pixels/have
# counts above the non-linearity regime of the pixels where one can’t do 
# photometry. Objects brighter than this limit (in magnitude) will be cut. 
# Comment out for no saturation limit.
# Two formats are accepted:
# Single float: applies same saturation limit to observations in all filters.
# Comma-separated list of floats: applies saturation limit per filter, in order as
# given in observing_filters keyword.
bright_limit = 16.0


[PHASECURVES]

# The phase function used to calculate apparent magnitude. The physical parameters input
# file must contain the columns needed to calculate the phase function.
# Options: HG, HG1G2, HG12, linear, none.
phase_function = HG


[FOV]

# Choose between circular or actual camera footprint, including chip gaps.
# Options: circle, footprint.
camera_model = circle

# Fraction of detector surface area which contains CCD -- simulates chip gaps
# for OIF output. Comment out if using camera footprint.
fill_factor = 0.9

# Radius of the circle for a circular footprint (in degrees). Float.
# Comment out or do not include if using footprint camera model.
circle_radius = 1.75


[FADINGFUNCTION]

# Uses the fading function as outlined in
# Chesley and Vereš (2017) to remove observations.

# Width parameter for fading function. Should be greater than zero and less than 0.5.
# Suggested value is 0.1 after Chesley and Vereš (2017).
fading_function_width = 0.1

# Peak efficiency for the fading function, called the 'fill factor' in Chesley and Veres (2017).
# Suggested value is 1. Do not change this unless you are sure of what you are doing.
fading_function_peak_efficiency = 1.


[LINKINGFILTER]

# SSP detection efficiency. Which fraction of the objects will
# the automated Rubin Solar System Processing (SSP) pipeline successfully link? Float.
SSP_detection_efficiency = 0.95


# Length of tracklets. How many observations of an object during one night are
# required to produce a valid tracklet?
SSP_number_observations = 2

# Minimum separation (in arcsec) between two observations of an object required 
# for the linking software to distinguish them as separate and therefore as a valid tracklet.
SSP_separation_threshold = 0.5

# Maximum time separation (in days) between subsequent observations in a tracklet. 
# Default is 0.0625 days (90mins).
SSP_maximum_time = 0.0625

# Number of tracklets for detection. How many tracklets are required to classify
# an object as detected? 
SSP_number_tracklets = 3

# The number of tracklets defined above must occur in <= this number of days to 
# constitute a complete track/detection.
SSP_track_window = 15

# The time in UTC at which it is noon at the observatory location (in standard time).
# For the LSST, 12pm Chile Standard Time is 4pm UTC.
SSP_night_start_utc = 16.0


[OUTPUT]

# Output format of the output file[s]
# Options: csv, sqlite3, hdf5
output_format = sqlite3

# Controls which columns are in the output files.
# Options are "basic" and "all", which returns all columns.
output_columns = basic


[LIGHTCURVE]

# The unique name of the lightcurve model to use. Defined in the ``name_id`` method 
# of the subclasses of AbstractLightCurve. If not none, the complex physical parameters 
# file must be specified at the command line.
lc_model = none


[ACTIVITY]

# The unique name of the actvity model to use. Defined in the ``name_id`` method
#  of the subclasses of AbstractCometaryActivity.  If not none, a complex physical parameters 
# file must be specified at the command line.
comet_activity = none
\end{verbatim}
\end{scriptsize}

\subsection{LSST Known Object Predictions}
This configuration file will run \sorcha with default model of the full LSST camera focal plane, accounting for detector and chip gaps, but with randomization, fading function, vignetting, SSP linking, saturation limit and trailing losses turned off. This will result in output listing all detections which lie on the CCDs with unadulterated apparent magnitudes. This configuration file is suitable for instances where the user only cares about the locations of the population of interest; for example, if one wishes to predict where and when known objects appear in Rubin observations. As this configuration turns off almost all of \sorcha's post-processing steps, we do not recommend its use for science.

\begin{scriptsize}
\begin{verbatim}
# Sorcha Configuration File
# This configuration file is appropriate for running Sorcha using the full camera footprint 
# but with randomization, fading function, vignetting, SSP linking, saturation limit and 
# trailing losses off. This will output all detections which lie on the CCD with unadulterated 
# apparent magnitudes. This could thus be used to predict when and where known objects will appear 
# in Rubin observations. 
# WARNING: This configuration file turns off most of Sorcha's features,
# We do not recommend you use this unless you know what you are doing.


[INPUT]

# The simulation used for the ephemeris input.
# ar=ASSIST+REBOUND interal ephemeris generation
# external=providing an external input file from the command line
# Options: "ar", "external"
ephemerides_type = ar

# Format for ephemeris simulation input file if a file is specified at the command line. 
# This is also the format to which ephemeris files will be written out, if specified.
# Options: csv, whitespace, hdf5
eph_format = csv

# Sorcha chunk size: how many objects should be processed at once?
size_serial_chunk = 20000

# Format for the orbit, physical parameters, and complex physical parameters input files.
# Options: csv or whitespace
aux_format = csv

# SQL query for extracting data from the pointing database.
pointing_sql_query = SELECT observationId, observationStartMJD as observationStartMJD_TAI, visitTime, visitExposureTime, filter, seeingFwhmGeom as seeingFwhmGeom_arcsec, seeingFwhmEff as seeingFwhmEff_arcsec, fiveSigmaDepth as fieldFiveSigmaDepth_mag , fieldRA as fieldRA_deg, fieldDec as fieldDec_deg, rotSkyPos as fieldRotSkyPos_deg FROM observations order by observationId


[SIMULATION]
# Configs for running the ASSIST+REBOUND ephemerides generator.

# the field of view of our search field, in degrees
ar_ang_fov = 2.06

# the buffer zone around the field of view we want to include, in degrees
ar_fov_buffer = 0.2

# the "picket" is our imprecise discretization of time that allows us to move progress
# our simulations forward without getting too granular when we don't have to.
# the unit is number of days.
ar_picket = 1

# the obscode is the MPC observatory code for the provided telescope.
ar_obs_code = X05

# the order of healpix which we will use for the healpy portions of the code.
# the nside is equivalent to 2**ar_healpix_order
ar_healpix_order = 6


[FILTERS]

# Filters of the observations you are interested in, comma-separated.
# Your physical parameters file must have H calculated in one of these filters
# and colour offset columns defined relative to that filter.
observing_filters = r,g,i,z,u,y


[PHASECURVES]

# The phase function used to calculate apparent magnitude. The physical parameters input
# file must contain the columns needed to calculate the phase function.
# Options: HG, HG1G2, HG12, linear, none.
phase_function = HG


[FOV]

# Choose between circular or actual camera footprint, including chip gaps.
# Options: circle, footprint.
camera_model = footprint

# Path to camera footprint file. Uncomment to provide a path to the desired camera 
# detector configuration file if not using the default built-in LSSTCam detector 
# configuration for the actual camera footprint.
# footprint_path= ./data/detectors_corners.csv

# The distance from the edge of a detector (in arcseconds on the focal plane)
# at which we will not correctly extract an object. By default this is 10px or 2 arcseconds.
# Comment out or do not include if not using footprint camera model.
footprint_edge_threshold = 2.


[OUTPUT]

# Output format of the output file[s]
# Options: csv, sqlite3, hdf5
output_format = sqlite3

# Controls which columns are in the output files.
# Options are "basic" and "all", which returns all columns.
output_columns = basic


[LIGHTCURVE]

# The unique name of the lightcurve model to use. Defined in the ``name_id`` method 
# of the subclasses of AbstractLightCurve. If not none, the complex physical parameters 
# file must be specified at the command line.lc_model = none
lc_model = none


[ACTIVITY]

# The unique name of the actvity model to use. Defined in the ``name_id`` method
#  of the subclasses of AbstractCometaryActivity.  If not none, a complex physical parameters 
# file must be specified at the command line.
comet_activity = none


[EXPERT]
# Turning off all randomization, vignetting and trailing losses.
randomization_on = False
vignetting_on = False
trailing_losses_on = False
\end{verbatim}
\end{scriptsize}

\section{Magnitude Uncertainty Calculations}
\label{app:mag_uncertainty}
Throughout \sorcha, we use a few different approximations to the expected magnitude uncertainty at a given signal-to-noise. Here, we show that all of these formulations are equivalent, and any individual choice is purely a matter of convenience. By definition for a flux measurement $f$ with uncertainty $\sigma_f$, the signal-to-noise is $S/N \equiv f/\sigma_f$,

For a non-trailed source with flux $f$ and magnitude $m \equiv -2.5\log_{10} (f) + c$ (for some constant term $c$), we begin by directly deriving the magnitude $m'$ after applying a small perturbation $\pm \delta f$ \update{where $|\delta f < |f|$}:
\begin{align}
   m'  =   -2.5\log_{10}(f \pm \delta f) + c  = -2.5 \log_{10} f + c - 2.5\log_{10}(1 \pm \delta f/f) = m \mp 2.5 \log_{10}(1 + \delta f/f).
\end{align}
If we take $\delta f = \sigma_f$, then, we have that $m' = m \pm \sigma_m$, where 
\begin{equation}
    \sigma_m = 2.5\log_{10} \left(1 + (S/N)^{-1}\right). \label{eq:sndirect}
\end{equation}

We can also derive this uncertainty with the usual uncertainty propagation by the Jacobian of the transformation. In our case, we have that:
\begin{align} \sigma_m^2 & = \left(\frac{\partial f}{\partial m}\right)^2 \sigma_f^2  = \left(\frac{2.5}{\ln(10)} \times \frac {1}{f}\right)^2 \sigma_f^2 \\ \implies & \sigma_m = \left| \frac{2.5}{\ln(10)}  \frac{\sigma_f}{f} \right| \approx 1/(S/N).
\end{align}
To see that these two results are equivalent, we take the Taylor series of Equation \ref{eq:sndirect}, leading to
\begin{equation} -2.5 \log_{10}(1 \pm 1/x) \approx \frac{2.5}{x \ln(10)},  \end{equation}
where $x = S/N$. 

Finally, we derive the relationship between $m_{5\sigma}$ and $S/N$ of \cite{ivezic2019}. From Poisson statistics, we have that
\begin{equation}
    N^2 = N_0^2 + \alpha S,
\end{equation}
where $N_0$ is a constant noise floor \update{and $\alpha$ is an instrumental scaling factor dependent on the shot noise}. Then, 
\begin{equation}
    \sigma_m^2 \approx \frac{1}{(S/N)^2} = \frac{N_0^2}{S^2} + \frac{\alpha}{S}.
\end{equation}
By definition, $m_{5\sigma}$ is the magnitude where $S/N=5$, where we expect that $\sigma_m = 0.2$. Under this condition, and assuming that $S \propto 10^{-0.4 m}$, we define \update{$y \equiv 10^{0.4 (m-m_{5\sigma})}$}
\begin{equation}
     \sigma_m^2 = (0.2^2 - \gamma) y + \gamma y^2
\end{equation}
 \update{$\gamma$, similarly to $\alpha$, depends on image quality and instrumental properties \citep{ivezic2019}.} The term linear in \update{$y$} accounts for the uncertainty floor for bright sources (note \update{$y$} is small for $m \ll m_{5\sigma}$), whereas the quadratic term dominates in the regime $m \approx m_{5\sigma}$, where we expect the shot noise will dominate. \update{We note that, because of the typical steepness of Solar System absolute magnitude distributions, the majority of sources will be near $m_{5\sigma}$}


\bibliography{bibliography}{}

\begin{thebibliography}{}
\expandafter\ifx\csname natexlab\endcsname\relax\def\natexlab#1{#1}\fi
\providecommand{\url}[1]{\href{#1}{#1}}
\providecommand{\dodoi}[1]{doi:~\href{http://doi.org/#1}{\nolinkurl{#1}}}
\providecommand{\doeprint}[1]{\href{http://ascl.net/#1}{\nolinkurl{http://ascl.net/#1}}}
\providecommand{\doarXiv}[1]{\href{https://arxiv.org/abs/#1}{\nolinkurl{https://arxiv.org/abs/#1}}}

\bibitem[{T.~M.~C. {Abbott} {et~al.}(2021){Abbott}, {Adam{\'o}w}, {Aguena},
  {Allam}, {Amon}, {Annis}, {Avila}, {Bacon}, {Banerji}, {Bechtol}, {Becker},
  {Bernstein}, {Bertin}, {Bhargava}, {Bridle}, {Brooks}, {Burke}, {Carnero
  Rosell}, {Carrasco Kind}, {Carretero}, {Castander}, {Cawthon}, {Chang},
  {Choi}, {Conselice}, {Costanzi}, {Crocce}, {da Costa}, {Davis}, {De Vicente},
  {DeRose}, {Desai}, {Diehl}, {Dietrich}, {Drlica-Wagner}, {Eckert},
  {Elvin-Poole}, {Everett}, {Evrard}, {Ferrero}, {Fert{\'e}}, {Flaugher},
  {Fosalba}, {Friedel}, {Frieman}, {Garc{\'\i}a-Bellido}, {Gaztanaga},
  {Gelman}, {Gerdes}, {Giannantonio}, {Gill}, {Gruen}, {Gruendl}, {Gschwend},
  {Gutierrez}, {Hartley}, {Hinton}, {Hollowood}, {Honscheid}, {Huterer},
  {James}, {Jeltema}, {Johnson}, {Kent}, {Kron}, {Kuehn}, {Kuropatkin},
  {Lahav}, {Li}, {Lidman}, {Lin}, {MacCrann}, {Maia}, {Manning}, {Maloney},
  {March}, {Marshall}, {Martini}, {Melchior}, {Menanteau}, {Miquel}, {Morgan},
  {Myles}, {Neilsen}, {Ogando}, {Palmese}, {Paz-Chinch{\'o}n}, {Petravick},
  {Pieres}, {Plazas}, {Pond}, {Rodriguez-Monroy}, {Romer}, {Roodman}, {Rykoff},
  {Sako}, {Sanchez}, {Santiago}, {Scarpine}, {Serrano}, {Sevilla-Noarbe},
  {Smith}, {Smith}, {Soares-Santos}, {Suchyta}, {Swanson}, {Tarle}, {Thomas},
  {To}, {Tremblay}, {Troxel}, {Tucker}, {Turner}, {Varga}, {Walker},
  {Wechsler}, {Weller}, {Wester}, {Wilkinson}, {Yanny}, {Zhang}, {Nikutta},
  {Fitzpatrick}, {Jacques}, {Scott}, {Olsen}, {Huang}, {Herrera}, {Juneau},
  {Nidever}, {Weaver}, {Adean}, {Correia}, {de Freitas}, {Freitas},
  {Singulani}, {Vila-Verde}, \& {Linea Science Server}}]{abbott2021}
{Abbott}, T.~M.~C., {Adam{\'o}w}, M., {Aguena}, M., {et~al.} 2021,
  \bibinfo{title}{{The Dark Energy Survey Data Release 2},} \apjs, 255, 20,
  \dodoi{10.3847/1538-4365/ac00b3}

\bibitem[{C. {Acton} {et~al.}(2018){Acton}, {Bachman}, {Semenov}, \&
  {Wright}}]{acton2018}
{Acton}, C., {Bachman}, N., {Semenov}, B., \& {Wright}, E. 2018,
  \bibinfo{title}{{A look towards the future in the handling of space science
  mission geometry},} \planss, 150, 9, \dodoi{10.1016/j.pss.2017.02.013}

\bibitem[{C.~H. {Acton}(1996){Acton}}]{acton1996}
{Acton}, C.~H. 1996, \bibinfo{title}{Ancillary data services of NASA's
  Navigation and Ancillary Information Facility,} \planss, 44, 65,
  \dodoi{10.1016/0032-0633(95)00107-7}

\bibitem[{J. {Agarwal} {et~al.}(2024){Agarwal}, {Kim}, {Kelley}, \&
  {Marschall}}]{agarwal2023}
{Agarwal}, J., {Kim}, Y., {Kelley}, M. S.~P., \& {Marschall}, R. 2024, in
  Comets III, ed. K.~J. Meech, M.~R. Combi, D.~Bockel\'ee-Morvan, S.~N.
  Raymond, \& M.~Zolensky (University of Arizona Press)

\bibitem[{M.~F. {A'Hearn} {et~al.}(1984){A'Hearn}, {Schleicher}, {Millis},
  {Feldman}, \& {Thompson}}]{ahearn1984}
{A'Hearn}, M.~F., {Schleicher}, D.~G., {Millis}, R.~L., {Feldman}, P.~D., \&
  {Thompson}, D.~T. 1984, \bibinfo{title}{{Comet Bowell 1980b},} \aj, 89, 579,
  \dodoi{10.1086/113552}

\bibitem[{A. {Alvarez-Candal} {et~al.}(2022){Alvarez-Candal}, {Jimenez Corral},
  \& {Colazo}}]{alvarez-candal2022}
{Alvarez-Candal}, A., {Jimenez Corral}, S., \& {Colazo}, M. 2022,
  \bibinfo{title}{{Absolute colors and phase coefficients of asteroids},} \aap,
  667, A81, \dodoi{10.1051/0004-6361/202243479}

\bibitem[{A. {Annex} {et~al.}(2020){Annex}, {Pearson}, {Seignovert}, {Carcich},
  {Eichhorn}, {Mapel}, {von Forstner}, {McAuliffe}, {del Rio}, {Berry}, {Aye},
  {Stefko}, {de Val-Borro}, {Kulumani}, \& {Murakami}}]{annex2020}
{Annex}, A., {Pearson}, B., {Seignovert}, B., {et~al.} 2020,
  \bibinfo{title}{{SpiceyPy: a Pythonic Wrapper for the SPICE Toolkit},} The
  Journal of Open Source Software, 5, 2050, \dodoi{10.21105/joss.02050}

\bibitem[{J. Annis {et~al.}(2014)Annis, Soares-Santos, Strauss, Becker,
  Dodelson, Fan, Gunn, Hao, Ivezi{\'{c}}, Jester, Jiang, Johnston, Kubo,
  Lampeitl, Lin, Lupton, Miknaitis, Seo, Simet, \& Yanny}]{annis2014}
Annis, J., Soares-Santos, M., Strauss, M.~A., {et~al.} 2014,
  \bibinfo{title}{{THE} {SLOAN} {DIGITAL} {SKY} {SURVEY} {COADD}: 275
  deg$\less$sup$\greater$2$\less$/sup$\greater${OF} {DEEP} {SLOAN} {DIGITAL}
  {SKY} {SURVEY} {IMAGING} {ON} {STRIPE} 82,} The Astrophysical Journal, 794,
  120, \dodoi{10.1088/0004-637x/794/2/120}

\bibitem[{C. Araujo-Hauck {et~al.}(2016)Araujo-Hauck, Sebag, Liang, Neill,
  Muller, Thomas, Vucina, \& Gressler}]{araujo-hauck2016}
Araujo-Hauck, C., Sebag, J., Liang, M., {et~al.} 2016, in Ground-based and
  Airborne Telescopes VI, ed. H.~J. Hall, R.~Gilmozzi, \& H.~K. Marshall, Vol.
  9906, International Society for Optics and Photonics (SPIE), 202 -- 211,
  \dodoi{10.1117/12.2232923}

\bibitem[{E. {Ashton} {et~al.}(2023){Ashton}, {Chang}, {Chen}, {Lehner},
  {Wang}, {Alexandersen}, {Choi}, {Fraser}, {Granados Contreras}, {Ito},
  {Jeongahn}, {Ji}, {Kavelaars}, {Kim}, {Lawler}, {Li}, {Lin}, {Lykawka},
  {Moon}, {More}, {Mu{\~n}oz-Guti{\'e}rrez}, {Ohtsuki}, {Pike}, {Terai},
  {Urakawa}, {Yoshida}, {Zhang}, {Zhao}, {Zhou}, \& {Fossil
  Collaboration}}]{ashton2023}
{Ashton}, E., {Chang}, C.-K., {Chen}, Y.-T., {et~al.} 2023,
  \bibinfo{title}{{FOSSIL. III. Lightcurves of 371 Trans-Neptunian Objects},}
  \apjs, 267, 33, \dodoi{10.3847/1538-4365/acda1e}

\bibitem[{ {Astropy Collaboration} {et~al.}(2013){Astropy Collaboration},
  {Robitaille}, {Tollerud}, {Greenfield}, {Droettboom}, {Bray}, {Aldcroft},
  {Davis}, {Ginsburg}, {Price-Whelan}, {Kerzendorf}, {Conley}, {Crighton},
  {Barbary}, {Muna}, {Ferguson}, {Grollier}, {Parikh}, {Nair}, {Unther},
  {Deil}, {Woillez}, {Conseil}, {Kramer}, {Turner}, {Singer}, {Fox}, {Weaver},
  {Zabalza}, {Edwards}, {Azalee Bostroem}, {Burke}, {Casey}, {Crawford},
  {Dencheva}, {Ely}, {Jenness}, {Labrie}, {Lim}, {Pierfederici}, {Pontzen},
  {Ptak}, {Refsdal}, {Servillat}, \& {Streicher}}]{astropy2013}
{Astropy Collaboration}, {Robitaille}, T.~P., {Tollerud}, E.~J., {et~al.} 2013,
  \bibinfo{title}{{Astropy: A community Python package for astronomy},} \aap,
  558, A33, \dodoi{10.1051/0004-6361/201322068}

\bibitem[{ {Astropy Collaboration} {et~al.}(2018){Astropy Collaboration},
  {Price-Whelan}, {Sip{\H{o}}cz}, {G{\"u}nther}, {Lim}, {Crawford}, {Conseil},
  {Shupe}, {Craig}, {Dencheva}, {Ginsburg}, {Vand erPlas}, {Bradley},
  {P{\'e}rez-Su{\'a}rez}, {de Val-Borro}, {Aldcroft}, {Cruz}, {Robitaille},
  {Tollerud}, {Ardelean}, {Babej}, {Bach}, {Bachetti}, {Bakanov}, {Bamford},
  {Barentsen}, {Barmby}, {Baumbach}, {Berry}, {Biscani}, {Boquien}, {Bostroem},
  {Bouma}, {Brammer}, {Bray}, {Breytenbach}, {Buddelmeijer}, {Burke},
  {Calderone}, {Cano Rodr{\'\i}guez}, {Cara}, {Cardoso}, {Cheedella}, {Copin},
  {Corrales}, {Crichton}, {D'Avella}, {Deil}, {Depagne}, {Dietrich}, {Donath},
  {Droettboom}, {Earl}, {Erben}, {Fabbro}, {Ferreira}, {Finethy}, {Fox},
  {Garrison}, {Gibbons}, {Goldstein}, {Gommers}, {Greco}, {Greenfield},
  {Groener}, {Grollier}, {Hagen}, {Hirst}, {Homeier}, {Horton}, {Hosseinzadeh},
  {Hu}, {Hunkeler}, {Ivezi{\'c}}, {Jain}, {Jenness}, {Kanarek}, {Kendrew},
  {Kern}, {Kerzendorf}, {Khvalko}, {King}, {Kirkby}, {Kulkarni}, {Kumar},
  {Lee}, {Lenz}, {Littlefair}, {Ma}, {Macleod}, {Mastropietro}, {McCully},
  {Montagnac}, {Morris}, {Mueller}, {Mumford}, {Muna}, {Murphy}, {Nelson},
  {Nguyen}, {Ninan}, {N{\"o}the}, {Ogaz}, {Oh}, {Parejko}, {Parley}, {Pascual},
  {Patil}, {Patil}, {Plunkett}, {Prochaska}, {Rastogi}, {Reddy Janga},
  {Sabater}, {Sakurikar}, {Seifert}, {Sherbert}, {Sherwood-Taylor}, {Shih},
  {Sick}, {Silbiger}, {Singanamalla}, {Singer}, {Sladen}, {Sooley},
  {Sornarajah}, {Streicher}, {Teuben}, {Thomas}, {Tremblay}, {Turner},
  {Terr{\'o}n}, {van Kerkwijk}, {de la Vega}, {Watkins}, {Weaver}, {Whitmore},
  {Woillez}, {Zabalza}, \& {Astropy Contributors}}]{astropy2018}
{Astropy Collaboration}, {Price-Whelan}, A.~M., {Sip{\H{o}}cz}, B.~M., {et~al.}
  2018, \bibinfo{title}{{The Astropy Project: Building an Open-science Project
  and Status of the v2.0 Core Package},} \aj, 156, 123,
  \dodoi{10.3847/1538-3881/aabc4f}

\bibitem[{ {Astropy Collaboration} {et~al.}(2022){Astropy Collaboration},
  {Price-Whelan}, {Lim}, {Earl}, {Starkman}, {Bradley}, {Shupe}, {Patil},
  {Corrales}, {Brasseur}, {N{"o}the}, {Donath}, {Tollerud}, {Morris},
  {Ginsburg}, {Vaher}, {Weaver}, {Tocknell}, {Jamieson}, {van Kerkwijk},
  {Robitaille}, {Merry}, {Bachetti}, {G{"u}nther}, {Aldcroft},
  {Alvarado-Montes}, {Archibald}, {B{'o}di}, {Bapat}, {Barentsen}, {Baz{'a}n},
  {Biswas}, {Boquien}, {Burke}, {Cara}, {Cara}, {Conroy}, {Conseil}, {Craig},
  {Cross}, {Cruz}, {D'Eugenio}, {Dencheva}, {Devillepoix}, {Dietrich},
  {Eigenbrot}, {Erben}, {Ferreira}, {Foreman-Mackey}, {Fox}, {Freij}, {Garg},
  {Geda}, {Glattly}, {Gondhalekar}, {Gordon}, {Grant}, {Greenfield}, {Groener},
  {Guest}, {Gurovich}, {Handberg}, {Hart}, {Hatfield-Dodds}, {Homeier},
  {Hosseinzadeh}, {Jenness}, {Jones}, {Joseph}, {Kalmbach}, {Karamehmetoglu},
  {Ka{l}uszy{'n}ski}, {Kelley}, {Kern}, {Kerzendorf}, {Koch}, {Kulumani},
  {Lee}, {Ly}, {Ma}, {MacBride}, {Maljaars}, {Muna}, {Murphy}, {Norman},
  {O'Steen}, {Oman}, {Pacifici}, {Pascual}, {Pascual-Granado}, {Patil},
  {Perren}, {Pickering}, {Rastogi}, {Roulston}, {Ryan}, {Rykoff}, {Sabater},
  {Sakurikar}, {Salgado}, {Sanghi}, {Saunders}, {Savchenko}, {Schwardt},
  {Seifert-Eckert}, {Shih}, {Jain}, {Shukla}, {Sick}, {Simpson},
  {Singanamalla}, {Singer}, {Singhal}, {Sinha}, {Sip{H{o}}cz}, {Spitler},
  {Stansby}, {Streicher}, {{{S}}umak}, {Swinbank}, {Taranu}, {Tewary},
  {Tremblay}, {Val-Borro}, {Van Kooten}, {Vasovi{'c}}, {Verma}, {de Miranda
  Cardoso}, {Williams}, {Wilson}, {Winkel}, {Wood-Vasey}, {Xue}, {Yoachim},
  {Zhang}, {Zonca}, \& {Astropy Project Contributors}}]{astropy2022}
{Astropy Collaboration}, {Price-Whelan}, A.~M., {Lim}, P.~L., {et~al.} 2022,
  \bibinfo{title}{{The Astropy Project: Sustaining and Growing a
  Community-oriented Open-source Project and the Latest Major Release (v5.0) of
  the Core Package},} \apj, 935, 167, \dodoi{10.3847/1538-4357/ac7c74}

\bibitem[{M.~T. {Bannister} {et~al.}(2016){Bannister}, {Kavelaars}, {Petit},
  {Gladman}, {Gwyn}, {Chen}, {Volk}, {Alexandersen}, {Benecchi}, {Delsanti},
  {Fraser}, {Granvik}, {Grundy}, {Guilbert-Lepoutre}, {Hestroffer}, {Ip},
  {Jakubik}, {Jones}, {Kaib}, {Kavelaars}, {Lacerda}, {Lawler}, {Lehner},
  {Lin}, {Lister}, {Lykawka}, {Monty}, {Marsset}, {Murray-Clay}, {Noll},
  {Parker}, {Pike}, {Rousselot}, {Rusk}, {Schwamb}, {Shankman}, {Sicardy},
  {Vernazza}, \& {Wang}}]{bannister2016}
{Bannister}, M.~T., {Kavelaars}, J.~J., {Petit}, J.-M., {et~al.} 2016,
  \bibinfo{title}{{The Outer Solar System Origins Survey. I. Design and
  First-quarter Discoveries},} \aj, 152, 70, \dodoi{10.3847/0004-6256/152/3/70}

\bibitem[{M.~T. {Bannister} {et~al.}(2017){Bannister}, {Shankman}, {Volk},
  {Chen}, {Kaib}, {Gladman}, {Jakubik}, {Kavelaars}, {Fraser}, {Schwamb},
  {Petit}, {Wang}, {Gwyn}, {Alexandersen}, \& {Pike}}]{bannister2017}
{Bannister}, M.~T., {Shankman}, C., {Volk}, K., {et~al.} 2017,
  \bibinfo{title}{{OSSOS. V. Diffusion in the Orbit of a High-perihelion
  Distant Solar System Object},} \aj, 153, 262,
  \dodoi{10.3847/1538-3881/aa6db5}

\bibitem[{M.~T. {Bannister} {et~al.}(2018){Bannister}, {Gladman}, {Kavelaars},
  {Petit}, {Volk}, {Chen}, {Alexandersen}, {Gwyn}, {Schwamb}, {Ashton},
  {Benecchi}, {Cabral}, {Dawson}, {Delsanti}, {Fraser}, {Granvik},
  {Greenstreet}, {Guilbert-Lepoutre}, {Ip}, {Jakubik}, {Jones}, {Kaib},
  {Lacerda}, {Van Laerhoven}, {Lawler}, {Lehner}, {Lin}, {Lykawka}, {Marsset},
  {Murray-Clay}, {Pike}, {Rousselot}, {Shankman}, {Thirouin}, {Vernazza}, \&
  {Wang}}]{bannister2018}
{Bannister}, M.~T., {Gladman}, B.~J., {Kavelaars}, J.~J., {et~al.} 2018,
  \bibinfo{title}{{OSSOS. VII. 800+ Trans-Neptunian Objects{\textemdash}The
  Complete Data Release},} \apjs, 236, 18, \dodoi{10.3847/1538-4365/aab77a}

\bibitem[{K. {Batygin} {et~al.}(2019){Batygin}, {Adams}, {Brown}, \&
  {Becker}}]{batygin2019}
{Batygin}, K., {Adams}, F.~C., {Brown}, M.~E., \& {Becker}, J.~C. 2019,
  \bibinfo{title}{{The planet nine hypothesis},} \physrep, 805, 1,
  \dodoi{10.1016/j.physrep.2019.01.009}

\bibitem[{K. {Batygin} \& M.~E. {Brown}(2016){Batygin} \&
  {Brown}}]{batygin2016}
{Batygin}, K., \& {Brown}, M.~E. 2016, \bibinfo{title}{{Evidence for a Distant
  Giant Planet in the Solar System},} \aj, 151, 22,
  \dodoi{10.3847/0004-6256/151/2/22}

\bibitem[{M. Belyakov {et~al.}(2022)Belyakov, Bernardinelli, \&
  Brown}]{belyakov2022}
Belyakov, M., Bernardinelli, P.~H., \& Brown, M.~E. 2022,
  \bibinfo{title}{Limits on the Detection of Planet Nine in the Dark Energy
  Survey,} The Astronomical Journal, 163, 216, \dodoi{10.3847/1538-3881/ac5c56}

\bibitem[{P.~H. Bernardinelli {et~al.}(2020)Bernardinelli, Bernstein, Sako,
  Hamilton, Gerdes, Adams, Saunders, Aguena, Allam, Avila, Brooks, Diehl, Doel,
  Everett, García-Bellido, Gaztanaga, Gruendl, Honscheid, Ogando, Palmese,
  Tucker, Walker, \& Wester}]{bernardinelli2020}
Bernardinelli, P.~H., Bernstein, G.~M., Sako, M., {et~al.} 2020,
  \bibinfo{title}{Testing the isotropy of the {Dark} {Energy} {Survey}'s
  extreme trans-{Neptunian} objects,} The Planetary Science Journal, 1, 28,
  \dodoi{10.3847/PSJ/ab9d80}

\bibitem[{P.~H. Bernardinelli {et~al.}(2022)Bernardinelli, Bernstein, Sako,
  Yanny, Aguena, Allam, Andrade-Oliveira, Bertin, Brooks, Buckley-Geer, Burke,
  Rosell, Carrasco~Kind, Carretero, Conselice, Costanzi, da~Costa, De~Vicente,
  Desai, Diehl, Dietrich, Doel, Eckert, Everett, Ferrero, Flaugher, Fosalba,
  Frieman, García-Bellido, Gerdes, Gruen, Gruendl, Gschwend, Hinton,
  Hollowood, Honscheid, James, Kent, Kuehn, Kuropatkin, Lahav, Maia, March,
  Menanteau, Miquel, Morgan, Myles, Ogando, Palmese, Paz-Chinchón, Pieres,
  Malagón, Romer, Roodman, Sanchez, Scarpine, Schubnell, Serrano,
  Sevilla-Noarbe, Smith, Soares-Santos, Suchyta, Swanson, Tarle, To, Varga, \&
  Walker}]{bernardinelli2022}
Bernardinelli, P.~H., Bernstein, G.~M., Sako, M., {et~al.} 2022,
  \bibinfo{title}{A {Search} of the {Full} {Six} {Years} of the {Dark} {Energy}
  {Survey} for {Outer} {Solar} {System} {Objects},} The Astrophysical Journal
  Supplement Series, 258, 41, \dodoi{10.3847/1538-4365/ac3914}

\bibitem[{P.~H. {Bernardinelli} {et~al.}(2024){Bernardinelli}, {Smotherman},
  {Langford}, {Portillo}, {Connolly}, {Kalmbach}, {Stetzler}, {Juri{\'c}},
  {Oldroyd}, {Lin}, {Adams}, {Chandler}, {Fuentes}, {Gerdes}, {Holman},
  {Markwardt}, {McNeill}, {Mommert}, {Napier}, {Payne}, {Ragozzine}, {Rivkin},
  {Schlichting}, {Sheppard}, {Strauss}, {Trilling}, \&
  {Trujillo}}]{bernardinelli2024}
{Bernardinelli}, P.~H., {Smotherman}, H., {Langford}, Z., {et~al.} 2024,
  \bibinfo{title}{{The DECam Ecliptic Exploration Project (DEEP). III. Survey
  Characterization and Simulation Methods},} \aj, 167, 134,
  \dodoi{10.3847/1538-3881/ad1527}

\bibitem[{F. Bianco \&  the SCOC(2022)Bianco \& the SCOC}]{SCOC_Report_2}
Bianco, F., \& the SCOC. 2022, \bibinfo{title}{Survey Cadence Optimization
  Committee's Phase 2 Recommendations,} \url{https://pstn-055.lsst.io/}

\bibitem[{F. Bianco \&  the SCOC(2024)Bianco \& the SCOC}]{SCOC_Report_3}
Bianco, F., \& the SCOC. 2024, \bibinfo{title}{Survey Cadence Optimization
  Committee's Phase 3 Recommendations,} \url{https://pstn-056.lsst.io/}

\bibitem[{F.~B. {Bianco} {et~al.}(2022){Bianco}, {Ivezi{\'c}}, {Jones},
  {Graham}, {Marshall}, {Saha}, {Strauss}, {Yoachim}, {Ribeiro}, {Anguita},
  {Bauer}, {Bauer}, {Bellm}, {Blum}, {Brandt}, {Brough}, {Catelan}, {Clarkson},
  {Connolly}, {Gawiser}, {Gizis}, {Hlo{\v{z}}ek}, {Kaviraj}, {Liu}, {Lochner},
  {Mahabal}, {Mandelbaum}, {McGehee}, {Neilsen}, {Olsen}, {Peiris}, {Rhodes},
  {Richards}, {Ridgway}, {Schwamb}, {Scolnic}, {Shemmer}, {Slater}, {Slosar},
  {Smartt}, {Strader}, {Street}, {Trilling}, {Verma}, {Vivas}, {Wechsler}, \&
  {Willman}}]{bianco2022}
{Bianco}, F.~B., {Ivezi{\'c}}, {\v{Z}}., {Jones}, R.~L., {et~al.} 2022,
  \bibinfo{title}{{Optimization of the Observing Cadence for the Rubin
  Observatory Legacy Survey of Space and Time: A Pioneering Process of
  Community-focused Experimental Design},} \apjs, 258, 1,
  \dodoi{10.3847/1538-4365/ac3e72}

\bibitem[{W.~F. Bottke {et~al.}(2002)Bottke, Morbidelli, Jedicke, Petit,
  Levison, Michel, \& Metcalfe}]{bottke2002b}
Bottke, W.~F., Morbidelli, A., Jedicke, R., {et~al.} 2002,
  \bibinfo{title}{Debiased Orbital and Absolute Magnitude Distribution of the
  Near-Earth Objects,} Icarus, 156, 399,
  \dodoi{https://doi.org/10.1006/icar.2001.6788}

\bibitem[{W.~F. {Bottke} {et~al.}(2006){Bottke}, {Vokrouhlick{\'y}},
  {Rubincam}, \& {Nesvorn{\'y}}}]{bottke2006}
{Bottke}, Jr., W.~F., {Vokrouhlick{\'y}}, D., {Rubincam}, D.~P., \&
  {Nesvorn{\'y}}, D. 2006, \bibinfo{title}{{The Yarkovsky and Yorp Effects:
  Implications for Asteroid Dynamics},} Annual Review of Earth and Planetary
  Sciences, 34, 157, \dodoi{10.1146/annurev.earth.34.031405.125154}

\bibitem[{E. Bowell {et~al.}(1989)Bowell, Hapke, Domingue, Lumme, Peltoniemi,
  \& Harris}]{bowell1989}
Bowell, E., Hapke, B., Domingue, D., {et~al.} 1989, in {Asteroids II}, ed.
  R.~P. Binzel, T.~Gehrels, \& M.~S. Matthews ({University of Arizona Press}),
  524 -- 556

\bibitem[{M.~E. {Brown} \& K. {Batygin}(2019){Brown} \& {Batygin}}]{brown2019}
{Brown}, M.~E., \& {Batygin}, K. 2019, \bibinfo{title}{{Orbital Clustering in
  the Distant Solar System},} \aj, 157, 62, \dodoi{10.3847/1538-3881/aaf051}

\bibitem[{M.~E. {Brown} \& K. {Batygin}(2021){Brown} \& {Batygin}}]{brown2021}
{Brown}, M.~E., \& {Batygin}, K. 2021, \bibinfo{title}{{The Orbit of Planet
  Nine},} \aj, 162, 219, \dodoi{10.3847/1538-3881/ac2056}

\bibitem[{M.~E. {Brown} {et~al.}(2024){Brown}, {Holman}, \&
  {Batygin}}]{brown2024}
{Brown}, M.~E., {Holman}, M.~J., \& {Batygin}, K. 2024, \bibinfo{title}{{A
  Pan-STARRS1 Search for Planet Nine},} \aj, 167, 146,
  \dodoi{10.3847/1538-3881/ad24e9}

\bibitem[{L.~E. {Buchanan} {et~al.}(2022){Buchanan}, {Schwamb}, {Fraser},
  {Bannister}, {Marsset}, {Pike}, {Nesvorn{\'y}}, {Kavelaars}, {Benecchi},
  {Lehner}, {Wang}, {Peixinho}, {Volk}, {Alexandersen}, {Chen}, {Gladman},
  {Gwyn}, \& {Petit}}]{buchanan2022}
{Buchanan}, L.~E., {Schwamb}, M.~E., {Fraser}, W.~C., {et~al.} 2022,
  \bibinfo{title}{{Col-OSSOS: Probing Ice Line/Color Transitions within the
  Kuiper Belt's Progenitor Populations},} \psj, 3, 9,
  \dodoi{10.3847/PSJ/ac42c9}

\bibitem[{B. {Carry} {et~al.}(2024){Carry}, {Peloton}, {Le Montagner},
  {Mahlke}, \& {Berthier}}]{carry2024}
{Carry}, B., {Peloton}, J., {Le Montagner}, R., {Mahlke}, M., \& {Berthier}, J.
  2024, \bibinfo{title}{{Combined spin orientation and phase function of
  asteroids},} \aap, 687, A38, \dodoi{10.1051/0004-6361/202449789}

\bibitem[{J.~M. {Carvano} \& J.~A.~G. {Davalos}(2015){Carvano} \&
  {Davalos}}]{carvano2015}
{Carvano}, J.~M., \& {Davalos}, J.~A.~G. 2015, \bibinfo{title}{{Shape and solar
  phase angle effects on the taxonomic classification of asteroids},} \aap,
  580, A98, \dodoi{10.1051/0004-6361/201526268}

\bibitem[{C.-K. {Chang} {et~al.}(2021){Chang}, {Chen}, {Fraser}, {Yoshida},
  {Lehner}, {Wang}, {Kavelaars}, {Pike}, {Alexandersen}, {Ito}, {Choi},
  {Granados Contreras}, {Jeongahn}, {Ji}, {Kim}, {Lawler}, {Li}, {Lin}, {Sofia
  Lykawka}, {Moon}, {More}, {Mu{\~n}oz-Guti{\'e}rrez}, {Ohtsuki}, {Terai},
  {Urakawa}, {Zhang}, {Zhao}, {Zhou}, \& {Fossil Collaboration}}]{chang2021}
{Chang}, C.-K., {Chen}, Y.-T., {Fraser}, W.~C., {et~al.} 2021,
  \bibinfo{title}{{FOSSIL. I. The Spin Rate Limit of Jupiter Trojans},} \psj,
  2, 191, \dodoi{10.3847/PSJ/ac13a4}

\bibitem[{A.~L. {Cochran} {et~al.}(1989){Cochran}, {Green}, \&
  {Barker}}]{cochran1989}
{Cochran}, A.~L., {Green}, J.~R., \& {Barker}, E.~S. 1989, \bibinfo{title}{{Are
  low-activity comets intrinsically different from more active comets?},}
  \icarus, 79, 125, \dodoi{10.1016/0019-1035(89)90112-7}

\bibitem[{A.~J. {Connolly} {et~al.}(2014){Connolly}, {Angeli},
  {Chandrasekharan}, {Claver}, {Cook}, {Ivezic}, {Jones}, {Krughoff}, {Peng},
  {Peterson}, {Petry}, {Rasmussen}, {Ridgway}, {Saha}, {Sembroski},
  {vanderPlas}, \& {Yoachim}}]{Connolly2014}
{Connolly}, A.~J., {Angeli}, G.~Z., {Chandrasekharan}, S., {et~al.} 2014, in
  Society of Photo-Optical Instrumentation Engineers (SPIE) Conference Series,
  Vol. 9150, Modeling, Systems Engineering, and Project Management for
  Astronomy VI, ed. G.~Z. Angeli \& P.~Dierickx, 14, \dodoi{10.1117/12.2054953}

\bibitem[{N.~V. Cook {et~al.}(2016)Cook, Ragozzine, Granvik, \&
  Stephens}]{cook2016}
Cook, N.~V., Ragozzine, D., Granvik, M., \& Stephens, D.~C. 2016,
  \bibinfo{title}{REALISTIC DETECTABILITY OF CLOSE INTERSTELLAR COMETS,} The
  Astrophysical Journal, 825, 51, \dodoi{10.3847/0004-637X/825/1/51}

\bibitem[{C. da~Costa-Luis {et~al.}(2023)da~Costa-Luis, Larroque, Altendorf,
  Mary, richardsheridan, Korobov, Yorav-Raphael, Ivanov, Bargull, Rodrigues,
  Chen, Lee, Newey, CrazyPython, JC, Zugnoni, Pagel, mjstevens777, Dektyarev,
  Rothberg, Plavin, Dill, FichteFoll, Sturm, HeoHeo, van Kemenade, McCracken,
  MapleCCC, Nordlund, \& Boyle}]{dacostaluis2023}
da~Costa-Luis, C., Larroque, S.~K., Altendorf, K., {et~al.} 2023,
  \bibinfo{title}{{tqdm: A fast, Extensible Progress Bar for Python and CLI},},
  v4.66.1 Zenodo, \dodoi{10.5281/zenodo.8233425}

\bibitem[{F. {Delgado} {et~al.}(2014){Delgado}, {Saha}, {Chandrasekharan},
  {Cook}, {Petry}, \& {Ridgway}}]{delgado2014}
{Delgado}, F., {Saha}, A., {Chandrasekharan}, S., {et~al.} 2014, in Society of
  Photo-Optical Instrumentation Engineers (SPIE) Conference Series, Vol. 9150,
  Modeling, Systems Engineering, and Project Management for Astronomy VI, ed.
  G.~Z. {Angeli} \& P.~{Dierickx}, 915015, \dodoi{10.1117/12.2056898}

\bibitem[{L. Denneau {et~al.}(2015)Denneau, Jedicke, Fitzsimmons, Hsieh,
  Kleyna, Granvik, Micheli, Spahr, Vere\v{s}, Wainscoat, Burgett, Chambers,
  Draper, Flewelling, Huber, Kaiser, Morgan, \& Tonry}]{denneau2015}
Denneau, L., Jedicke, R., Fitzsimmons, A., {et~al.} 2015,
  \bibinfo{title}{Observational constraints on the catastrophic disruption rate
  of small main belt asteroids,} Icarus, 245, 1,
  \dodoi{https://doi.org/10.1016/j.icarus.2014.08.044}

\bibitem[{M.~M. {Dobson} {et~al.}(2021){Dobson}, {Schwamb}, {Fitzsimmons},
  {Kelley}, {Lister}, {Shingles}, {Denneau}, {Heinze}, {Smith}, {Tonry},
  {Weiland}, {Young}, {Benecchi}, \& {Verbiscer}}]{dobson2021}
{Dobson}, M.~M., {Schwamb}, M.~E., {Fitzsimmons}, A., {et~al.} 2021,
  \bibinfo{title}{{New or Increased Cometary Activity in (2060) 95P/Chiron},}
  Research Notes of the American Astronomical Society, 5, 211,
  \dodoi{10.3847/2515-5172/ac26c9}

\bibitem[{M.~M. {Dobson} {et~al.}(2023){Dobson}, {Schwamb}, {Benecchi},
  {Verbiscer}, {Fitzsimmons}, {Shingles}, {Denneau}, {Heinze}, {Smith},
  {Tonry}, {Weiland}, \& {Young}}]{dobson2023}
{Dobson}, M.~M., {Schwamb}, M.~E., {Benecchi}, S.~D., {et~al.} 2023,
  \bibinfo{title}{{Phase Curves of Kuiper Belt Objects, Centaurs, and
  Jupiter-family Comets from the ATLAS Survey},} \psj, 4, 75,
  \dodoi{10.3847/PSJ/acc463}

\bibitem[{T. {Engelhardt} {et~al.}(2017){Engelhardt}, {Jedicke}, {Vere{\v{s}}},
  {Fitzsimmons}, {Denneau}, {Beshore}, \& {Meinke}}]{engelhardt2017}
{Engelhardt}, T., {Jedicke}, R., {Vere{\v{s}}}, P., {et~al.} 2017,
  \bibinfo{title}{{An Observational Upper Limit on the Interstellar Number
  Density of Asteroids and Comets},} \aj, 153, 133,
  \dodoi{10.3847/1538-3881/aa5c8a}

\bibitem[{G. {Fedorets} {et~al.}(2020){Fedorets}, {Granvik}, {Jones},
  {Juri{\'c}}, \& {Jedicke}}]{fedorets2020}
{Fedorets}, G., {Granvik}, M., {Jones}, R.~L., {Juri{\'c}}, M., \& {Jedicke},
  R. 2020, \bibinfo{title}{{Discovering Earth's transient moons with the Large
  Synoptic Survey Telescope},} \icarus, 338, 113517,
  \dodoi{10.1016/j.icarus.2019.113517}

\bibitem[{B. {Flaugher} {et~al.}(2015){Flaugher}, {Diehl}, {Honscheid},
  {Abbott}, {Alvarez}, {Angstadt}, {Annis}, {Antonik}, {Ballester}, {Beaufore},
  {Bernstein}, {Bernstein}, {Bigelow}, {Bonati}, {Boprie}, {Brooks},
  {Buckley-Geer}, {Campa}, {Cardiel-Sas}, {Castander}, {Castilla}, {Cease},
  {Cela-Ruiz}, {Chappa}, {Chi}, {Cooper}, {da Costa}, {Dede}, {Derylo},
  {DePoy}, {de Vicente}, {Doel}, {Drlica-Wagner}, {Eiting}, {Elliott}, {Emes},
  {Estrada}, {Fausti Neto}, {Finley}, {Flores}, {Frieman}, {Gerdes},
  {Gladders}, {Gregory}, {Gutierrez}, {Hao}, {Holland}, {Holm}, {Huffman},
  {Jackson}, {James}, {Jonas}, {Karcher}, {Karliner}, {Kent}, {Kessler},
  {Kozlovsky}, {Kron}, {Kubik}, {Kuehn}, {Kuhlmann}, {Kuk}, {Lahav}, {Lathrop},
  {Lee}, {Levi}, {Lewis}, {Li}, {Mandrichenko}, {Marshall}, {Martinez},
  {Merritt}, {Miquel}, {Mu{\~n}oz}, {Neilsen}, {Nichol}, {Nord}, {Ogando},
  {Olsen}, {Palaio}, {Patton}, {Peoples}, {Plazas}, {Rauch}, {Reil}, {Rheault},
  {Roe}, {Rogers}, {Roodman}, {Sanchez}, {Scarpine}, {Schindler}, {Schmidt},
  {Schmitt}, {Schubnell}, {Schultz}, {Schurter}, {Scott}, {Serrano}, {Shaw},
  {Smith}, {Soares-Santos}, {Stefanik}, {Stuermer}, {Suchyta}, {Sypniewski},
  {Tarle}, {Thaler}, {Tighe}, {Tran}, {Tucker}, {Walker}, {Wang}, {Watson},
  {Weaverdyck}, {Wester}, {Woods}, {Yanny}, \& {DES
  Collaboration}}]{flaugher2015b}
{Flaugher}, B., {Diehl}, H.~T., {Honscheid}, K., {et~al.} 2015,
  \bibinfo{title}{{The Dark Energy Camera},} \aj, 150, 150,
  \dodoi{10.1088/0004-6256/150/5/150}

\bibitem[{W.~C. {Fraser} {et~al.}(2021){Fraser}, {Benecchi}, {Kavelaars},
  {Marsset}, {Pike}, {Bannister}, {Schwamb}, {Volk}, {Nesvorny},
  {Alexandersen}, {Chen}, {Gwyn}, {Lehner}, \& {Wang}}]{fraser2021}
{Fraser}, W.~C., {Benecchi}, S.~D., {Kavelaars}, J.~J., {et~al.} 2021,
  \bibinfo{title}{{Col-OSSOS: The Distinct Color Distribution of Single and
  Binary Cold Classical KBOs},} \psj, 2, 90, \dodoi{10.3847/PSJ/abf04a}

\bibitem[{W.~C. {Fraser} {et~al.}(2023){Fraser}, {Pike}, {Marsset}, {Schwamb},
  {Bannister}, {Buchanan}, {Kavelaars}, {Benecchi}, {Tan}, {Peixinho}, {Gwyn},
  {Alexandersen}, {Chen}, {Gladman}, \& {Volk}}]{fraser2023}
{Fraser}, W.~C., {Pike}, R.~E., {Marsset}, M., {et~al.} 2023,
  \bibinfo{title}{{Col-OSSOS: The Two Types of Kuiper Belt Surfaces},} \psj, 4,
  80, \dodoi{10.3847/PSJ/acc844}

\bibitem[{B. {Gladman} {et~al.}(1998){Gladman}, {Kavelaars}, {Nicholson},
  {Loredo}, \& {Burns}}]{gladman1998}
{Gladman}, B., {Kavelaars}, J.~J., {Nicholson}, P.~D., {Loredo}, T.~J., \&
  {Burns}, J.~A. 1998, \bibinfo{title}{{Pencil-Beam Surveys for Faint
  Trans-Neptunian Objects},} \aj, 116, 2042, \dodoi{10.1086/300573}

\bibitem[{B. Gladman {et~al.}(2012)Gladman, Lawler, Petit, Kavelaars, Jones,
  Parker, Laerhoven, Nicholson, Rousselot, Bieryla, \& Ashby}]{gladman2012}
Gladman, B., Lawler, S.~M., Petit, J.-M., {et~al.} 2012, \bibinfo{title}{{THE}
  {RESONANT} {TRANS}-{NEPTUNIAN} {POPULATIONS},} The Astronomical Journal, 144,
  23, \dodoi{10.1088/0004-6256/144/1/23}

\bibitem[{B.~J. Gladman {et~al.}(2009)Gladman, Davis, Neese, Jedicke, Williams,
  Kavelaars, Petit, Scholl, Holman, Warrington, Esquerdo, \&
  Tricarico}]{gladman2009}
Gladman, B.~J., Davis, D.~R., Neese, C., {et~al.} 2009, \bibinfo{title}{On the
  asteroid belt's orbital and size distribution,} Icarus, 202, 104,
  \dodoi{https://doi.org/10.1016/j.icarus.2009.02.012}

\bibitem[{K.~M. {G{'o}rski} {et~al.}(2005){G{'o}rski}, {Hivon}, {Banday},
  {Wandelt}, {Hansen}, {Reinecke}, \& {Bartelmann}}]{gorski2005}
{G{'o}rski}, K.~M., {Hivon}, E., {Banday}, A.~J., {et~al.} 2005,
  \bibinfo{title}{{HEALPix: A Framework for High-Resolution Discretization and
  Fast Analysis of Data Distributed on the Sphere},} \apj, 622, 759,
  \dodoi{10.1086/427976}

\bibitem[{T. {Grav} {et~al.}(2016){Grav}, {Mainzer}, \& {Spahr}}]{grav2016}
{Grav}, T., {Mainzer}, A.~K., \& {Spahr}, T. 2016, \bibinfo{title}{{Modeling
  the Performance of the LSST in Surveying the Near-Earth Object Population},}
  \aj, 151, 172, \dodoi{10.3847/0004-6256/151/6/172}

\bibitem[{C.~R. Harris {et~al.}(2020)Harris, Millman, van~der Walt, Gommers,
  Virtanen, Cournapeau, Wieser, Taylor, Berg, Smith, Kern, Picus, Hoyer, van
  Kerkwijk, Brett, Haldane, del R{\'{i}}o, Wiebe, Peterson,
  G{\'{e}}rard-Marchant, Sheppard, Reddy, Weckesser, Abbasi, Gohlke, \&
  Oliphant}]{harris2020}
Harris, C.~R., Millman, K.~J., van~der Walt, S.~J., {et~al.} 2020,
  \bibinfo{title}{Array programming with {NumPy},} Nature, 585, 357,
  \dodoi{10.1038/s41586-020-2649-2}

\bibitem[{S. {Hasegawa} {et~al.}(2024){Hasegawa}, {Marsset}, {DeMeo},
  {Hanu{\v{s}}}, {Binzel}, {Bus}, {Burt}, {Polishook}, {Thomas}, {Geem},
  {Ishiguro}, {Kuroda}, \& {Vernazza}}]{hasegawa2024}
{Hasegawa}, S., {Marsset}, M., {DeMeo}, F.~E., {et~al.} 2024,
  \bibinfo{title}{{Candidate Main-belt Asteroids for Surface Heterogeneity},}
  \aj, 167, 224, \dodoi{10.3847/1538-3881/ad3045}

\bibitem[{M.~J. {Holman} {et~al.}(2023){Holman}, {Akmal}, {Farnocchia}, {Rein},
  {Payne}, {Weryk}, {Tamayo}, \& {Hernandez}}]{holman2023}
{Holman}, M.~J., {Akmal}, A., {Farnocchia}, D., {et~al.} 2023,
  \bibinfo{title}{{ASSIST: An Ephemeris-quality Test-particle Integrator},}
  \psj, 4, 69, \dodoi{10.3847/PSJ/acc9a9}

\bibitem[{M.~J. {Holman} {et~al.}(2018){Holman}, {Payne}, {Blankley},
  {Janssen}, \& {Kuindersma}}]{holman2018}
{Holman}, M.~J., {Payne}, M.~J., {Blankley}, P., {Janssen}, R., \&
  {Kuindersma}, S. 2018, \bibinfo{title}{{HelioLinC: A Novel Approach to the
  Minor Planet Linking Problem},} \aj, 156, 135,
  \dodoi{10.3847/1538-3881/aad69a}

\bibitem[{D.~J. Hoover {et~al.}(2022)Hoover, Seligman, \& Payne}]{hoover2022}
Hoover, D.~J., Seligman, D.~Z., \& Payne, M.~J. 2022, \bibinfo{title}{The
  Population of Interstellar Objects Detectable with the {LSST} and Accessible
  for In Situ Rendezvous with Various Mission Designs,} The Planetary Science
  Journal, 3, 71, \dodoi{10.3847/psj/ac58fe}

\bibitem[{J.~D. Hunter(2007)Hunter}]{hunter2007}
Hunter, J.~D. 2007, \bibinfo{title}{Matplotlib: A 2D graphics environment,}
  Computing in Science \& Engineering, 9, 90, \dodoi{10.1109/MCSE.2007.55}

\bibitem[{{\v{Z}.}. Ivezi{\'c} \&  the LSST
  Science~Collaboration(2013)Ivezi{\'c} \& the LSST
  Science~Collaboration}]{lsst-SRD-2013}
Ivezi{\'c}, {\v{Z}.}., \& the LSST Science~Collaboration. 2013,
  \bibinfo{title}{LSST Science Requirements Document,}
  \url{http://ls.st/LPM-17}

\bibitem[{{\v{Z}.}. Ivezi{\'c} \&  the SCOC(2021)Ivezi{\'c} \& the
  SCOC}]{SCOC_Report_1}
Ivezi{\'c}, {\v{Z}.}., \& the SCOC. 2021, \bibinfo{title}{Survey Cadence
  Optimization Committee's Phase 1 Recommendation,}
  \url{https://pstn-053.lsst.io/}

\bibitem[{{\v{Z}}. {Ivezi{\'c}} {et~al.}(2019){Ivezi{\'c}}, {Kahn}, {Tyson},
  {Abel}, {Acosta}, {Allsman}, {Alonso}, {AlSayyad}, {Anderson}, {Andrew},
  {Angel}, {Angeli}, {Ansari}, {Antilogus}, {Araujo}, {Armstrong}, {Arndt},
  {Astier}, {Aubourg}, {Auza}, {Axelrod}, {Bard}, {Barr}, {Barrau}, {Bartlett},
  {Bauer}, {Bauman}, {Baumont}, {Bechtol}, {Bechtol}, {Becker}, {Becla},
  {Beldica}, {Bellavia}, {Bianco}, {Biswas}, {Blanc}, {Blazek}, {Blandford},
  {Bloom}, {Bogart}, {Bond}, {Booth}, {Borgland}, {Borne}, {Bosch}, {Boutigny},
  {Brackett}, {Bradshaw}, {Brandt}, {Brown}, {Bullock}, {Burchat}, {Burke},
  {Cagnoli}, {Calabrese}, {Callahan}, {Callen}, {Carlin}, {Carlson},
  {Chandrasekharan}, {Charles-Emerson}, {Chesley}, {Cheu}, {Chiang}, {Chiang},
  {Chirino}, {Chow}, {Ciardi}, {Claver}, {Cohen-Tanugi}, {Cockrum}, {Coles},
  {Connolly}, {Cook}, {Cooray}, {Covey}, {Cribbs}, {Cui}, {Cutri}, {Daly},
  {Daniel}, {Daruich}, {Daubard}, {Daues}, {Dawson}, {Delgado}, {Dellapenna},
  {de Peyster}, {de Val-Borro}, {Digel}, {Doherty}, {Dubois},
  {Dubois-Felsmann}, {Durech}, {Economou}, {Eifler}, {Eracleous}, {Emmons},
  {Fausti Neto}, {Ferguson}, {Figueroa}, {Fisher-Levine}, {Focke}, {Foss},
  {Frank}, {Freemon}, {Gangler}, {Gawiser}, {Geary}, {Gee}, {Geha}, {Gessner},
  {Gibson}, {Gilmore}, {Glanzman}, {Glick}, {Goldina}, {Goldstein}, {Goodenow},
  {Graham}, {Gressler}, {Gris}, {Guy}, {Guyonnet}, {Haller}, {Harris},
  {Hascall}, {Haupt}, {Hernandez}, {Herrmann}, {Hileman}, {Hoblitt}, {Hodgson},
  {Hogan}, {Howard}, {Huang}, {Huffer}, {Ingraham}, {Innes}, {Jacoby}, {Jain},
  {Jammes}, {Jee}, {Jenness}, {Jernigan}, {Jevremovi{\'c}}, {Johns}, {Johnson},
  {Johnson}, {Jones}, {Juramy-Gilles}, {Juri{\'c}}, {Kalirai}, {Kallivayalil},
  {Kalmbach}, {Kantor}, {Karst}, {Kasliwal}, {Kelly}, {Kessler}, {Kinnison},
  {Kirkby}, {Knox}, {Kotov}, {Krabbendam}, {Krughoff}, {Kub{\'a}nek},
  {Kuczewski}, {Kulkarni}, {Ku}, {Kurita}, {Lage}, {Lambert}, {Lange},
  {Langton}, {Le Guillou}, {Levine}, {Liang}, {Lim}, {Lintott}, {Long},
  {Lopez}, {Lotz}, {Lupton}, {Lust}, {MacArthur}, {Mahabal}, {Mandelbaum},
  {Markiewicz}, {Marsh}, {Marshall}, {Marshall}, {May}, {McKercher}, {McQueen},
  {Meyers}, {Migliore}, {Miller}, {Mills}, {Miraval}, {Moeyens}, {Moolekamp},
  {Monet}, {Moniez}, {Monkewitz}, {Montgomery}, {Morrison}, {Mueller},
  {Muller}, {Mu{\~n}oz Arancibia}, {Neill}, {Newbry}, {Nief}, {Nomerotski},
  {Nordby}, {O'Connor}, {Oliver}, {Olivier}, {Olsen}, {O'Mullane}, {Ortiz},
  {Osier}, {Owen}, {Pain}, {Palecek}, {Parejko}, {Parsons}, {Pease},
  {Peterson}, {Peterson}, {Petravick}, {Libby Petrick}, {Petry},
  {Pierfederici}, {Pietrowicz}, {Pike}, {Pinto}, {Plante}, {Plate}, {Plutchak},
  {Price}, {Prouza}, {Radeka}, {Rajagopal}, {Rasmussen}, {Regnault}, {Reil},
  {Reiss}, {Reuter}, {Ridgway}, {Riot}, {Ritz}, {Robinson}, {Roby}, {Roodman},
  {Rosing}, {Roucelle}, {Rumore}, {Russo}, {Saha}, {Sassolas}, {Schalk},
  {Schellart}, {Schindler}, {Schmidt}, {Schneider}, {Schneider}, {Schoening},
  {Schumacher}, {Schwamb}, {Sebag}, {Selvy}, {Sembroski}, {Seppala}, {Serio},
  {Serrano}, {Shaw}, {Shipsey}, {Sick}, {Silvestri}, {Slater}, {Smith},
  {Smith}, {Sobhani}, {Soldahl}, {Storrie-Lombardi}, {Stover}, {Strauss},
  {Street}, {Stubbs}, {Sullivan}, {Sweeney}, {Swinbank}, {Szalay}, {Takacs},
  {Tether}, {Thaler}, {Thayer}, {Thomas}, {Thornton}, {Thukral}, {Tice},
  {Trilling}, {Turri}, {Van Berg}, {Vanden Berk}, {Vetter}, {Virieux},
  {Vucina}, {Wahl}, {Walkowicz}, {Walsh}, {Walter}, {Wang}, {Wang}, {Warner},
  {Wiecha}, {Willman}, {Winters}, {Wittman}, {Wolff}, {Wood-Vasey}, {Wu},
  {Xin}, {Yoachim}, \& {Zhan}}]{ivezic2019}
{Ivezi{\'c}}, {\v{Z}}., {Kahn}, S.~M., {Tyson}, J.~A., {et~al.} 2019,
  \bibinfo{title}{{LSST: From Science Drivers to Reference Design and
  Anticipated Data Products},} \apj, 873, 111, \dodoi{10.3847/1538-4357/ab042c}

\bibitem[{S.~L. {Jackson} {et~al.}(2022){Jackson}, {Rozitis}, {Dover}, {Green},
  {Kolb}, {Andrews}, \& {Lowry}}]{jackson2022}
{Jackson}, S.~L., {Rozitis}, B., {Dover}, L.~R., {et~al.} 2022,
  \bibinfo{title}{{The effect of aspect changes on Near-Earth Asteroid phase
  curves},} \mnras, 513, 3076, \dodoi{10.1093/mnras/stac1053}

\bibitem[{D. {Jewitt} \& H.~H. {Hsieh}(2024){Jewitt} \& {Hsieh}}]{jewitt2022}
{Jewitt}, D., \& {Hsieh}, H.~H. 2024, in Comets III, ed. K.~J. Meech, M.~R.
  Combi, D.~Bockel\'ee-Morvan, S.~N. Raymond, \& M.~Zolensky (University of
  Arizona Press)

\bibitem[{R.~L. {Jones} {et~al.}(2006){Jones}, {Gladman}, {Petit}, {Rousselot},
  {Mousis}, {Kavelaars}, {Campo Bagatin}, {Bernabeu}, {Benavidez}, {Parker},
  {Nicholson}, {Holman}, {Grav}, {Doressoundiram}, {Veillet}, {Scholl}, \&
  {Mars}}]{jones2006}
{Jones}, R.~L., {Gladman}, B., {Petit}, J.~M., {et~al.} 2006,
  \bibinfo{title}{{The CFEPS Kuiper Belt Survey: Strategy and presurvey
  results},} \icarus, 185, 508, \dodoi{10.1016/j.icarus.2006.07.024}

\bibitem[{R.~L. {Jones} {et~al.}(2009){Jones}, {Chesley}, {Connolly}, {Harris},
  {Ivezic}, {Knezevic}, {Kubica}, {Milani}, \& {Trilling}}]{jones2009}
{Jones}, R.~L., {Chesley}, S.~R., {Connolly}, A.~J., {et~al.} 2009,
  \bibinfo{title}{{Solar System Science with LSST},} Earth Moon and Planets,
  105, 101, \dodoi{10.1007/s11038-009-9305-z}

\bibitem[{R.~L. {Jones} {et~al.}(2018){Jones}, {Slater}, {Moeyens}, {Allen},
  {Axelrod}, {Cook}, {Ivezi{\'c}}, {Juri{\'c}}, {Myers}, \&
  {Petry}}]{jones2018}
{Jones}, R.~L., {Slater}, C.~T., {Moeyens}, J., {et~al.} 2018,
  \bibinfo{title}{{The Large Synoptic Survey Telescope as a Near-Earth Object
  discovery machine},} \icarus, 303, 181, \dodoi{10.1016/j.icarus.2017.11.033}

\bibitem[{M. Juri\'{c} {et~al.}(2020)Juri\'{c}, Eggl, Moeyens, \&
  Jones}]{juric2020}
Juri\'{c}, M., Eggl, S., Moeyens, J., \& Jones, R.~L. 2020,
  \bibinfo{title}{Proposed Modifications to Solar System Processing and Data
  Products,} \url{https://dmtn-087.lsst.io}

\bibitem[{M. {Juri{\'c}} {et~al.}(2017){Juri{\'c}}, {Kantor}, {Lim}, {Lupton},
  {Dubois-Felsmann}, {Jenness}, {Axelrod}, {Aleksi{\'c}}, {Allsman},
  {AlSayyad}, {Alt}, {Armstrong}, {Basney}, {Becker}, {Becla}, {Biswas},
  {Bosch}, {Boutigny}, {Kind}, {Ciardi}, {Connolly}, {Daniel}, {Daues},
  {Economou}, {Chiang}, {Fausti}, {Fisher-Levine}, {Freemon}, {Gris},
  {Hernandez}, {Hoblitt}, {Ivezi{\'c}}, {Jammes}, {Jevremovi{\'c}}, {Jones},
  {Kalmbach}, {Kasliwal}, {Krughoff}, {Lurie}, {Lust}, {MacArthur}, {Melchior},
  {Moeyens}, {Nidever}, {Owen}, {Parejko}, {Peterson}, {Petravick},
  {Pietrowicz}, {Price}, {Reiss}, {Shaw}, {Sick}, {Slater}, {Strauss},
  {Sullivan}, {Swinbank}, {Van Dyk}, {Vuj{\v{c}}i{\'c}}, {Withers}, \&
  {Yoachim}}]{juric2017}
{Juri{\'c}}, M., {Kantor}, J., {Lim}, K.~T., {et~al.} 2017, in Astronomical
  Society of the Pacific Conference Series, Vol. 512, Astronomical Data
  Analysis Software and Systems XXV, ed. N.~P.~F. {Lorente}, K.~{Shortridge},
  \& R.~{Wayth}, 279.
\newblock \doarXiv{1512.07914}

\bibitem[{M. Juri\'{c} {et~al.}(2021)Juri\'{c}, Axelrod, Becker, Becla, Bellm,
  Bosch, Ciardi, Connolly, Dubois-Felsmann, Economou, Freemon, Gelman, Gill,
  Graham, Guy, Ivezi\'{c}, Jenness, Kantor, Krughoff, Lim, Lupton, Mueller,
  Nidever, O'Mullane, Patterson, Petravick, Shaw, Slater, Strauss, Swinbank,
  Tyson, Wood-Vasey, \& Wu}]{juric2021}
Juri\'{c}, M., Axelrod, T., Becker, A., {et~al.} 2021, \bibinfo{title}{{Data
  Products Definition Document},} \url{https://lse-163.lsst.io/}

\bibitem[{J.~J. {Kavelaars} {et~al.}(2021){Kavelaars}, {Petit}, {Gladman},
  {Bannister}, {Alexandersen}, {Chen}, {Gwyn}, \& {Volk}}]{kavelaars2021}
{Kavelaars}, J.~J., {Petit}, J.-M., {Gladman}, B., {et~al.} 2021,
  \bibinfo{title}{{OSSOS Finds an Exponential Cutoff in the Size Distribution
  of the Cold Classical Kuiper Belt},} \apjl, 920, L28,
  \dodoi{10.3847/2041-8213/ac2c72}

\bibitem[{J.~J. {Kavelaars} {et~al.}(2009){Kavelaars}, {Jones}, {Gladman},
  {Petit}, {Parker}, {Van Laerhoven}, {Nicholson}, {Rousselot}, {Scholl},
  {Mousis}, {Marsden}, {Benavidez}, {Bieryla}, {Campo Bagatin},
  {Doressoundiram}, {Margot}, {Murray}, \& {Veillet}}]{kavelaars2009}
{Kavelaars}, J.~J., {Jones}, R.~L., {Gladman}, B.~J., {et~al.} 2009,
  \bibinfo{title}{{The Canada-France Ecliptic Plane Survey{\textemdash}L3 Data
  Release: The Orbital Structure of the Kuiper Belt},} \aj, 137, 4917,
  \dodoi{10.1088/0004-6256/137/6/4917}

\bibitem[{T. Kluyver {et~al.}(2016)Kluyver, Ragan-Kelley, P{\'e}rez, Granger,
  Bussonnier, Frederic, Kelley, Hamrick, Grout, Corlay, Ivanov, Avila, Abdalla,
  Willing, \& development team}]{kluyver2016}
Kluyver, T., Ragan-Kelley, B., P{\'e}rez, F., {et~al.} 2016, in Positioning and
  Power in Academic Publishing: Players, Agents and Agendas (IOS Press),
  87--90.
\newblock \url{https://eprints.soton.ac.uk/403913/}

\bibitem[{J. Kubica {et~al.}(2007)Kubica, Denneau, Grav, Heasley, Jedicke,
  Masiero, Milani, Moore, Tholen, \& Wainscoat}]{kubica2007}
Kubica, J., Denneau, L., Grav, T., {et~al.} 2007, \bibinfo{title}{Efficient
  intra- and inter-night linking of asteroid detections using kd-trees,}
  Icarus, 189, 151, \dodoi{https://doi.org/10.1016/j.icarus.2007.01.008}

\bibitem[{J.~A. {Kurlander} {et~al.}(2025){Kurlander}, {Holman},
  {Bernardinelli}, {Juri{\'c}}, {Heinze}, \& {Payne}}]{kurlander2025}
{Kurlander}, J.~A., {Holman}, M.~J., {Bernardinelli}, P.~H., {et~al.} 2025,
  \bibinfo{title}{{A Well-Characterized Survey for Centaurs in Pan-STARRS1},}
  \aj, 169, 73, \dodoi{10.3847/1538-3881/ad9a58}

\bibitem[{T. {Kwiatkowski} \& A. {Kryszczynska}(1992){Kwiatkowski} \&
  {Kryszczynska}}]{kwiatkowski1992}
{Kwiatkowski}, T., \& {Kryszczynska}, A. 1992, in Liege International
  Astrophysical Colloquia, Vol.~30, Liege International Astrophysical
  Colloquia, ed. A.~{Brahic}, J.~C. {Gerard}, \& J.~{Surdej}, 353

\bibitem[{S.~K. {Lam} {et~al.}(2015){Lam}, {Pitrou}, \& {Seibert}}]{lam2015}
{Lam}, S.~K., {Pitrou}, A., \& {Seibert}, S. 2015, in Proc. Second Workshop on
  the LLVM Compiler Infrastructure in HPC, 1--6,
  \dodoi{10.1145/2833157.2833162}

\bibitem[{S.~M. Lawler {et~al.}(2018{\natexlab{a}})Lawler, Kavelaars,
  Alexandersen, Bannister, Gladman, Petit, \& Shankman}]{lawler2018}
Lawler, S.~M., Kavelaars, J.~J., Alexandersen, M., {et~al.} 2018{\natexlab{a}},
  \bibinfo{title}{OSSOS: X. How to Use a Survey Simulator: Statistical Testing
  of Dynamical Models Against the Real Kuiper Belt,} Frontiers in Astronomy and
  Space Sciences, 5, 14, \dodoi{10.3389/fspas.2018.00014}

\bibitem[{S.~M. Lawler {et~al.}(2018{\natexlab{b}})Lawler, Shankman, Kavelaars,
  Alexandersen, Bannister, Chen, Gladman, Fraser, Gwyn, Kaib, Petit, \&
  Volk}]{lawler2018b}
Lawler, S.~M., Shankman, C., Kavelaars, J.~J., {et~al.} 2018{\natexlab{b}},
  \bibinfo{title}{OSSOS. VIII. The Transition between Two Size Distribution
  Slopes in the Scattering Disk,} The Astronomical Journal, 155, 197,
  \dodoi{10.3847/1538-3881/aab8ff}

\bibitem[{O. {Lay} {et~al.}(2024){Lay}, {Masiero}, {Grav}, {Mainzer}, {Masci},
  \& {Wright}}]{lay2024}
{Lay}, O., {Masiero}, J., {Grav}, T., {et~al.} 2024, \bibinfo{title}{{Asteroid
  Impact Hazard Warning from the Near-Earth Object Surveyor Mission},} \psj, 5,
  149, \dodoi{10.3847/PSJ/ad4d9e}

\bibitem[{M. {Lazzarin} {et~al.}(2010){Lazzarin}, {Magrin}, {Marchi}, {Dotto},
  {Perna}, {Barbieri}, {Barucci}, \& {Fulchignoni}}]{lazzarin2010}
{Lazzarin}, M., {Magrin}, S., {Marchi}, S., {et~al.} 2010,
  \bibinfo{title}{{Rotational variation of the spectral slope of (21) Lutetia,
  the second asteroid target of ESA Rosetta mission},} \mnras, 408, 1433,
  \dodoi{10.1111/j.1365-2966.2010.17268.x}

\bibitem[{W.~G. {Levine} {et~al.}(2021){Levine}, {Cabot}, {Seligman}, \&
  {Laughlin}}]{levine2021}
{Levine}, W.~G., {Cabot}, S. H.~C., {Seligman}, D., \& {Laughlin}, G. 2021,
  \bibinfo{title}{{Constraints on the Occurrence of 'Oumuamua-Like Objects},}
  \apj, 922, 39, \dodoi{10.3847/1538-4357/ac1fe6}

\bibitem[{ {LSST Science Collaboration} {et~al.}(2009){LSST Science
  Collaboration}, {Abell}, {Allison}, {Anderson}, {Andrew}, {Angel}, {Armus},
  {Arnett}, {Asztalos}, {Axelrod}, \& et~al.}]{lsst-sciencebook-ch5-2009}
{LSST Science Collaboration}, {Abell}, P.~A., {Allison}, J., {et~al.} 2009,
  \bibinfo{title}{{LSST Science Book, Version 2.0},} ArXiv e-prints.
\newblock \doarXiv{0912.0201}

\bibitem[{J. Luu \& D. Jewitt(1989)Luu \& Jewitt}]{luu1989}
Luu, J., \& Jewitt, D. 1989, \bibinfo{title}{On the relative numbers of C types
  and S types among near-Earth asteroids,} Astronomical Journal (ISSN
  0004-6256), vol. 98, Nov. 1989, p. 1905-1911. Research supported by NSF., 98,
  1905

\bibitem[{J.~X. {Luu} \& D. {Jewitt}(1988){Luu} \& {Jewitt}}]{luu1988}
{Luu}, J.~X., \& {Jewitt}, D. 1988, \bibinfo{title}{{A Two-Part Search for
  Slow-Moving Objects},} \aj, 95, 1256, \dodoi{10.1086/114721}

\bibitem[{P.~S. {Lykawka} \& T. {Ito}(2023){Lykawka} \& {Ito}}]{lykawa2023}
{Lykawka}, P.~S., \& {Ito}, T. 2023, \bibinfo{title}{{Is There an Earth-like
  Planet in the Distant Kuiper Belt?},} \aj, 166, 118,
  \dodoi{10.3847/1538-3881/aceaf0}

\bibitem[{M. {Mahlke} {et~al.}(2021){Mahlke}, {Carry}, \&
  {Denneau}}]{mahlke2021}
{Mahlke}, M., {Carry}, B., \& {Denneau}, L. 2021, \bibinfo{title}{{Asteroid
  phase curves from ATLAS dual-band photometry},} \icarus, 354, 114094,
  \dodoi{10.1016/j.icarus.2020.114094}

\bibitem[{J.~R. {Masiero} {et~al.}(2023){Masiero}, {Dahlen}, {Mainzer},
  {Bottke}, {Bragg}, {Bauer}, \& {Grav}}]{masiero2023}
{Masiero}, J.~R., {Dahlen}, D.~W., {Mainzer}, A.~K., {et~al.} 2023,
  \bibinfo{title}{{Validation of the Survey Simulator Tool for the NEO Surveyor
  Mission Using NEOWISE Data},} \psj, 4, 225, \dodoi{10.3847/PSJ/ad00bb}

\bibitem[{J.~R. {Masiero} {et~al.}(2024){Masiero}, {Linder}, {Mainzer},
  {Dahlen}, \& {Kwon}}]{masiero2024}
{Masiero}, J.~R., {Linder}, T., {Mainzer}, A., {Dahlen}, D.~W., \& {Kwon},
  Y.~G. 2024, \bibinfo{title}{{Visual-band Brightnesses of Near-Earth Objects
  that will be Discovered in the Infrared by NEO Surveyor},} \psj, 5, 222,
  \dodoi{10.3847/PSJ/ad7859}

\bibitem[{W. {M}c{K}inney(2010){M}c{K}inney}]{mckinney2010}
{M}c{K}inney, W. 2010, in {P}roceedings of the 9th {P}ython in {S}cience
  {C}onference, ed. {S}t{\'{e}}fan van~der {W}alt \& {J}arrod {M}illman, 56 --
  61, \dodoi{10.25080/Majora-92bf1922-00a}

\bibitem[{E. McLoughlin {et~al.}(2015)McLoughlin, Fitzsimmons, \&
  McLoughlin}]{mcloughlin2015}
McLoughlin, E., Fitzsimmons, A., \& McLoughlin, A. 2015,
  \bibinfo{title}{Modelling the brightness increase signature due to asteroid
  collisions,} Icarus, 256, 37,
  \dodoi{https://doi.org/10.1016/j.icarus.2015.04.015}

\bibitem[{M. {Mommert} {et~al.}(2019){Mommert}, {Kelley}, {de Val-Borro}, {Li},
  {Guzman}, {Sip{\H{o}}cz}, {{\v{D}}urech}, {Granvik}, {Grundy}, {Moskovitz},
  {Penttil{\"a}}, \& {Samarasinha}}]{mommert2019}
{Mommert}, M., {Kelley}, M., {de Val-Borro}, M., {et~al.} 2019,
  \bibinfo{title}{{sbpy: A Python module for small-body planetary astronomy},}
  The Journal of Open Source Software, 4, 1426, \dodoi{10.21105/joss.01426}

\bibitem[{A. Moro-Mart{\'{\i}}n {et~al.}(2009)Moro-Mart{\'{\i}}n, Turner, \&
  Loeb}]{moro-martin2009}
Moro-Mart{\'{\i}}n, A., Turner, E.~L., \& Loeb, A. 2009, \bibinfo{title}{{WILL}
  {THE} {LARGE} {SYNOPTIC} {SURVEY} {TELESCOPE} {DETECT} {EXTRA}-{SOLAR}
  {PLANETESIMALS} {ENTERING} {THE} {SOLAR} {SYSTEM}?} The Astrophysical
  Journal, 704, 733, \dodoi{10.1088/0004-637x/704/1/733}

\bibitem[{K. Muinonen {et~al.}(2010)Muinonen, Belskaya, Cellino, Delb\`{o},
  Levasseur-Regourd, Penttil\"{a}, \& Tedesco}]{muinonen2010}
Muinonen, K., Belskaya, I.~N., Cellino, A., {et~al.} 2010, \bibinfo{title}{A
  three-parameter magnitude phase function for asteroids,} Icarus, 209, 542,
  \dodoi{https://doi.org/10.1016/j.icarus.2010.04.003}

\bibitem[{J. Myers {et~al.}(2013)Myers, Jones, \& Axelrod}]{myers2013}
Myers, J., Jones, R.~L., \& Axelrod, T. 2013, \bibinfo{title}{Moving Object
  Pipeline System Design,}
  \url{https://docushare.lsst.org/docushare/dsweb/Get/Version-24308/LDM-156.pdf}

\bibitem[{E. Naghib {et~al.}(2019)Naghib, Yoachim, Vanderbei, Connolly, \&
  Jones}]{naghib2019}
Naghib, E., Yoachim, P., Vanderbei, R.~J., Connolly, A.~J., \& Jones, R.~L.
  2019, \bibinfo{title}{A Framework for Telescope Schedulers: With Applications
  to the Large Synoptic Survey Telescope,} The Astronomical Journal, 157, 151,
  \dodoi{10.3847/1538-3881/aafece}

\bibitem[{K.~J. Napier {et~al.}(2021)Napier, Gerdes, Lin, Hamilton, Bernstein,
  Bernardinelli, Abbott, Aguena, Annis, Avila, Bacon, Bertin, Brooks, Burke,
  Rosell, Kind, Carretero, Costanzi, da~Costa, Vicente, Diehl, Doel, Everett,
  Ferrero, Fosalba, Garc{\'{\i}}a-Bellido, Gruen, Gruendl, Gutierrez,
  Hollowood, Honscheid, Hoyle, James, Kent, Kuehn, Kuropatkin, Maia, Menanteau,
  Miquel, Morgan, Palmese, Paz-Chinch{\'{o}}n, Plazas, Sanchez, Scarpine,
  Serrano, Sevilla-Noarbe, Smith, Suchyta, Swanson, To, Walker, \&
  Wilkinson}]{napier2021}
Napier, K.~J., Gerdes, D.~W., Lin, H.~W., {et~al.} 2021, \bibinfo{title}{No
  Evidence for Orbital Clustering in the Extreme Trans-Neptunian Objects,} The
  Planetary Science Journal, 2, 59, \dodoi{10.3847/psj/abe53e}

\bibitem[{J.~A. {O'Keefe}(1976){O'Keefe}}]{o'keefe1976}
{O'Keefe}, J.~A. 1976, \bibinfo{title}{{Tektites and their origin},} NASA
  STI/Recon Technical Report A, 77, 14534

\bibitem[{D. {Oldag} {et~al.}(2024){Oldag}, {DeLucchi}, {Beebe}, {Branton},
  {Campos}, {Chandler}, {Christofferson}, {Connolly}, {Kubica}, {Lynn},
  {Malanchev}, {Malz}, {Mandelbaum}, {McGuire}, \& {Wenneman}}]{oldag2024}
{Oldag}, D., {DeLucchi}, M., {Beebe}, W., {et~al.} 2024, \bibinfo{title}{{A
  Python Project Template for Healthy Scientific Software},} Research Notes of
  the American Astronomical Society, 8, 141, \dodoi{10.3847/2515-5172/ad4da1}

\bibitem[{W.~J. {Oldroyd} \& C.~A. {Trujillo}(2021){Oldroyd} \&
  {Trujillo}}]{oldroyd2021}
{Oldroyd}, W.~J., \& {Trujillo}, C.~A. 2021, \bibinfo{title}{{Outer Solar
  System Perihelion Gap Formation through Interactions with a Hypothetical
  Distant Giant Planet},} \aj, 162, 39, \dodoi{10.3847/1538-3881/abfb6f}

\bibitem[{S.~J. {Paddack}(1969){Paddack}}]{paddack1969}
{Paddack}, S.~J. 1969, \bibinfo{title}{{Rotational bursting of small celestial
  bodies: Effects of radiation pressure.},} \jgr, 74, 4379,
  \dodoi{10.1029/JB074i017p04379}

\bibitem[{A. {Penttil{\"a}} {et~al.}(2016){Penttil{\"a}}, {Shevchenko},
  {Wilkman}, \& {Muinonen}}]{penttila2016}
{Penttil{\"a}}, A., {Shevchenko}, V.~G., {Wilkman}, O., \& {Muinonen}, K. 2016,
  \bibinfo{title}{{H, G$_{1}$, G$_{2}$ photometric phase function extended to
  low-accuracy data},} Planetary and Space Science, 123, 117,
  \dodoi{10.1016/j.pss.2015.08.010}

\bibitem[{J.~M. {Petit} {et~al.}(2011){Petit}, {Kavelaars}, {Gladman}, {Jones},
  {Parker}, {Van Laerhoven}, {Nicholson}, {Mars}, {Rousselot}, {Mousis},
  {Marsden}, {Bieryla}, {Taylor}, {Ashby}, {Benavidez}, {Campo Bagatin}, \&
  {Bernabeu}}]{petit2011}
{Petit}, J.~M., {Kavelaars}, J.~J., {Gladman}, B.~J., {et~al.} 2011,
  \bibinfo{title}{{The Canada-France Ecliptic Plane Survey{\textemdash}Full
  Data Release: The Orbital Structure of the Kuiper Belt},} \aj, 142, 131,
  \dodoi{10.1088/0004-6256/142/4/131}

\bibitem[{R.~E. {Pike} {et~al.}(2023){Pike}, {Fraser}, {Volk}, {Kavelaars},
  {Marsset}, {Peixinho}, {Schwamb}, {Bannister}, {Peltier}, {Buchanan},
  {Benecchi}, \& {Tan}}]{pike2023}
{Pike}, R.~E., {Fraser}, W.~C., {Volk}, K., {et~al.} 2023,
  \bibinfo{title}{{Col-OSSOS: The Distribution of Surface Classes in Neptune's
  Resonances},} \psj, 4, 200, \dodoi{10.3847/PSJ/ace2c2}

\bibitem[{N. {Pinilla-Alonso} {et~al.}(2024){Pinilla-Alonso}, {Licandro},
  {Brunetto}, {Henault}, {Schambeau}, {Guilbert-Lepoutre}, {Stansberry},
  {Wong}, {Lunine}, {Holler}, {Emery}, {Protopapa}, {Cook}, {Hammel},
  {Villanueva}, {Milam}, {Cruikshank}, \& {de
  Souza-Feliciano}}]{pinillaalonso2024}
{Pinilla-Alonso}, N., {Licandro}, J., {Brunetto}, R., {et~al.} 2024,
  \bibinfo{title}{{Unveiling the ice and gas nature of active centaur (2060)
  Chiron using the James Webb Space Telescope},} arXiv e-prints,
  arXiv:2407.07761, \dodoi{10.48550/arXiv.2407.07761}

\bibitem[{P. {Pokorn{\'y}} {et~al.}(2020){Pokorn{\'y}}, {Kuchner}, \&
  {Sheppard}}]{pokorny2020}
{Pokorn{\'y}}, P., {Kuchner}, M.~J., \& {Sheppard}, S.~S. 2020,
  \bibinfo{title}{{A Deep Search for Stable Venus Co-orbital Asteroids: Limits
  on the Population},} \psj, 1, 47, \dodoi{10.3847/PSJ/abab9f}

\bibitem[{ {PyTables Developers Team}(2002){PyTables Developers
  Team}}]{pytables2002}
{PyTables Developers Team}. 2002, \bibinfo{title}{{PyTables}: Hierarchical
  Datasets in {Python},} \url{https://www.pytables.org/}

\bibitem[{V.~V. {Radzievskii}(1952){Radzievskii}}]{radzievskii1952}
{Radzievskii}, V.~V. 1952, \bibinfo{title}{{A mechanism for the disintegration
  of asteroids and meteorites},} \azh, 29, 162

\bibitem[{H. Rein {et~al.}(2023)Rein, Holman, \& Akmal}]{rein2023}
Rein, H., Holman, M., \& Akmal, A. 2023, \bibinfo{title}{matthewholman/assist:
  v1.1.1,}, v1.1.1 Zenodo, \dodoi{10.5281/zenodo.7778017}

\bibitem[{H. {Rein} \& S.~F. {Liu}(2012){Rein} \& {Liu}}]{rein2012}
{Rein}, H., \& {Liu}, S.~F. 2012, \bibinfo{title}{{REBOUND: an open-source
  multi-purpose N-body code for collisional dynamics},} \aap, 537, A128,
  \dodoi{10.1051/0004-6361/201118085}

\bibitem[{H. {Rein} \& D.~S. {Spiegel}(2015){Rein} \& {Spiegel}}]{rein2015}
{Rein}, H., \& {Spiegel}, D.~S. 2015, \bibinfo{title}{{IAS15: a fast, adaptive,
  high-order integrator for gravitational dynamics, accurate to machine
  precision over a billion orbits},} \mnras, 446, 1424,
  \dodoi{10.1093/mnras/stu2164}

\bibitem[{J.~E. {Robinson} {et~al.}(2024){Robinson}, {Fitzsimmons}, {Young},
  {Bannister}, {Denneau}, {Erasmus}, {Lawrence}, {Siverd}, \&
  {Tonry}}]{robinson2024}
{Robinson}, J.~E., {Fitzsimmons}, A., {Young}, D.~R., {et~al.} 2024,
  \bibinfo{title}{{Main-belt and Trojan asteroid phase curves from the ATLAS
  survey},} \mnras, 531, 304, \dodoi{10.1093/mnras/stae966}

\bibitem[{M. {Schemel} \& M.~E. {Brown}(2021){Schemel} \&
  {Brown}}]{schemel2021}
{Schemel}, M., \& {Brown}, M.~E. 2021, \bibinfo{title}{{Zwicky Transient
  Facility Observations of Trojan Asteroids: A Thousand Colors, Rotation
  Amplitudes, and Phase Functions},} \psj, 2, 40, \dodoi{10.3847/PSJ/abc752}

\bibitem[{M.~E. {Schwamb} {et~al.}(2010){Schwamb}, {Brown}, {Rabinowitz}, \&
  {Ragozzine}}]{schwamb2010}
{Schwamb}, M.~E., {Brown}, M.~E., {Rabinowitz}, D.~L., \& {Ragozzine}, D. 2010,
  \bibinfo{title}{{Properties of the Distant Kuiper Belt: Results from the
  Palomar Distant Solar System Survey},} \apj, 720, 1691,
  \dodoi{10.1088/0004-637X/720/2/1691}

\bibitem[{M.~E. Schwamb {et~al.}(2024)Schwamb, Kubica, Jurić, Oldag, West,
  DeLucchi, \& Holman}]{schwamb2024}
Schwamb, M.~E., Kubica, J., Jurić, M., {et~al.} 2024,
  \bibinfo{title}{Controlling Randomization in Astronomy Simulations,} Research
  Notes of the AAS, 8, 25, \dodoi{10.3847/2515-5172/ad1f6b}

\bibitem[{M.~E. Schwamb {et~al.}(2018)Schwamb, Levison, \& Buie}]{schwamb2018a}
Schwamb, M.~E., Levison, H.~F., \& Buie, M.~W. 2018,
  \bibinfo{title}{Opportunities for the Large Synoptic Survey Telescope to Find
  New L$\less$sub$\greater$5$\less$/sub$\greater$ Trojan and Hilda
  $\less$i$\greater$Lucy$\less$/i$\greater$ Encounter Targets,} Research Notes
  of the {AAS}, 2, 159, \dodoi{10.3847/2515-5172/aade00}

\bibitem[{M.~E. {Schwamb} {et~al.}(2018){Schwamb}, {Volk}, {Wen}, {Lin},
  {Kelley}, {Bannister}, {Hsieh}, {Jones}, {Mommert}, {Snodgrass}, {Ragozzine},
  {Chesley}, {Sheppard}, {Juric}, \& {Buie}}]{schwamb2018c}
{Schwamb}, M.~E., {Volk}, K., {Wen}, H., {et~al.} 2018, \bibinfo{title}{{A
  Northern Ecliptic Survey for Solar System Science},} arXiv e-prints,
  arXiv:1812.01149, \dodoi{10.48550/arXiv.1812.01149}

\bibitem[{M.~E. {Schwamb} {et~al.}(2019){Schwamb}, {Fraser}, {Bannister},
  {Marsset}, {Pike}, {Kavelaars}, {Benecchi}, {Lehner}, {Wang}, {Thirouin},
  {Delsanti}, {Peixinho}, {Volk}, {Alexandersen}, {Chen}, {Gladman}, {Gwyn}, \&
  {Petit}}]{schwamb2019b}
{Schwamb}, M.~E., {Fraser}, W.~C., {Bannister}, M.~T., {et~al.} 2019,
  \bibinfo{title}{{Col-OSSOS: The Colors of the Outer Solar System Origins
  Survey},} \apjs, 243, 12, \dodoi{10.3847/1538-4365/ab2194}

\bibitem[{M.~E. {Schwamb} {et~al.}(2021){Schwamb}, {Juri{\'c}}, {Bolin},
  {Dones}, {Greenstreet}, {Hsieh}, {Inno}, {Jones}, {Kelley}, {Knight},
  {Reach}, {Seccull}, {Snodgrass}, {Trilling}, \& {Vera C. Rubin Observatory
  LSST Solar System Science Collaboration}}]{schwamb2021}
{Schwamb}, M.~E., {Juri{\'c}}, M., {Bolin}, B.~T., {et~al.} 2021,
  \bibinfo{title}{{Year 1 of the Legacy Survey of Space and Time (LSST):
  Recommendations for Template Production to Enable Solar System Small Body
  Transient and Time Domain Science},} Research Notes of the American
  Astronomical Society, 5, 143, \dodoi{10.3847/2515-5172/ac090f}

\bibitem[{M.~E. {Schwamb} {et~al.}(2023){Schwamb}, {Jones}, {Yoachim}, {Volk},
  {Dorsey}, {Opitom}, {Greenstreet}, {Lister}, {Snodgrass}, {Bolin}, {Inno},
  {Bannister}, {Eggl}, {Solontoi}, {Kelley}, {Juri{\'c}}, {Lin}, {Ragozzine},
  {Bernardinelli}, {Chesley}, {Daylan}, {{\v{D}}urech}, {Fraser}, {Granvik},
  {Knight}, {Lisse}, {Malhotra}, {Oldroyd}, {Thirouin}, \& {Ye}}]{schwamb2023}
{Schwamb}, M.~E., {Jones}, R.~L., {Yoachim}, P., {et~al.} 2023,
  \bibinfo{title}{{Tuning the Legacy Survey of Space and Time (LSST) Observing
  Strategy for Solar System Science},} \apjs, 266, 22,
  \dodoi{10.3847/1538-4365/acc173}

\bibitem[{D. Seligman \& G. Laughlin(2018)Seligman \& Laughlin}]{seligman2018}
Seligman, D., \& Laughlin, G. 2018, \bibinfo{title}{The Feasibility and
  Benefits of In Situ Exploration of ‘Oumuamua-like Objects,} The
  Astronomical Journal, 155, 217, \dodoi{10.3847/1538-3881/aabd37}

\bibitem[{C. {Shankman} {et~al.}(2017){Shankman}, {Kavelaars}, {Bannister},
  {Gladman}, {Lawler}, {Chen}, {Jakubik}, {Kaib}, {Alexandersen}, {Gwyn},
  {Petit}, \& {Volk}}]{shankman2017}
{Shankman}, C., {Kavelaars}, J.~J., {Bannister}, M.~T., {et~al.} 2017,
  \bibinfo{title}{{OSSOS. VI. Striking Biases in the Detection of Large
  Semimajor Axis Trans-Neptunian Objects},} \aj, 154, 50,
  \dodoi{10.3847/1538-3881/aa7aed}

\bibitem[{A. {Shannon} {et~al.}(2015){Shannon}, {Jackson}, {Veras}, \&
  {Wyatt}}]{shannon2015}
{Shannon}, A., {Jackson}, A.~P., {Veras}, D., \& {Wyatt}, M. 2015,
  \bibinfo{title}{{Eight billion asteroids in the Oort cloud},} \mnras, 446,
  2059, \dodoi{10.1093/mnras/stu2267}

\bibitem[{S.~S. {Sheppard} \& C. {Trujillo}(2016){Sheppard} \&
  {Trujillo}}]{sheppard2016}
{Sheppard}, S.~S., \& {Trujillo}, C. 2016, \bibinfo{title}{{New Extreme
  Trans-Neptunian Objects: Toward a Super-Earth in the Outer Solar System},}
  \aj, 152, 221, \dodoi{10.3847/1538-3881/152/6/221}

\bibitem[{S.~S. {Sheppard} {et~al.}(2022){Sheppard}, {Tholen}, {Pokorn{\'y}},
  {Micheli}, {Dell'Antonio}, {Fu}, {Trujillo}, {Beaton}, {Carlsten},
  {Drlica-Wagner}, {Mart{\'\i}nez-V{\'a}zquez}, {Mau}, {Santana-Ros},
  {Santana-Silva}, {Sif{\'o}n}, {Simha}, {Thirouin}, {Trilling}, {Vivas}, \&
  {Zenteno}}]{sheppard2022}
{Sheppard}, S.~S., {Tholen}, D.~J., {Pokorn{\'y}}, P., {et~al.} 2022,
  \bibinfo{title}{{A Deep and Wide Twilight Survey for Asteroids Interior to
  Earth and Venus},} \aj, 164, 168, \dodoi{10.3847/1538-3881/ac8cff}

\bibitem[{M.~R. {Showalter} {et~al.}(2021){Showalter}, {Benecchi}, {Buie},
  {Grundy}, {Keane}, {Lisse}, {Olkin}, {Porter}, {Robbins}, {Singer},
  {Verbiscer}, {Weaver}, {Zangari}, {Hamilton}, {Kaufmann}, {Lauer}, {Mehoke},
  {Mehoke}, {Spencer}, {Throop}, {Parker}, {Stern}, {New Horizons Geology}, \&
  Team}]{showalter2021}
{Showalter}, M.~R., {Benecchi}, S.~D., {Buie}, M.~W., {et~al.} 2021,
  \bibinfo{title}{{A statistical review of light curves and the prevalence of
  contact binaries in the Kuiper Belt},} \icarus, 356, 114098,
  \dodoi{10.1016/j.icarus.2020.114098}

\bibitem[{K. {Silsbee} \& S. {Tremaine}(2016){Silsbee} \&
  {Tremaine}}]{silsbee2016}
{Silsbee}, K., \& {Tremaine}, S. 2016, \bibinfo{title}{{Modeling the Nearly
  Isotropic Comet Population in Anticipation of LSST Observations},} \aj, 152,
  103, \dodoi{10.3847/0004-6256/152/4/103}

\bibitem[{C. Snodgrass \& G.~H. Jones(2019)Snodgrass \& Jones}]{snodgrass2019}
Snodgrass, C., \& Jones, G.~H. 2019, \bibinfo{title}{The European Space
  Agency's Comet Interceptor lies in wait,} Nature Communications, 10, 5418,
  \dodoi{10.1038/s41467-019-13470-1}

\bibitem[{M. {Solontoi} {et~al.}(2010){Solontoi}, {Ivezi{\'c}}, {West},
  {Claire}, {Juri{\'c}}, {Becker}, {Jones}, {Hall}, {Kent}, {Lupton}, {Knapp},
  {Quinn}, {Gunn}, {Schneider}, \& {Loomis}}]{solontoi2010}
{Solontoi}, M., {Ivezi{\'c}}, {\v{Z}}., {West}, A.~A., {et~al.} 2010,
  \bibinfo{title}{{Detecting active comets in the SDSS},} \icarus, 205, 605,
  \dodoi{10.1016/j.icarus.2009.07.042}

\bibitem[{J.~S. Stuart \& R.~P. Binzel(2004)Stuart \& Binzel}]{stuart2004}
Stuart, J.~S., \& Binzel, R.~P. 2004, \bibinfo{title}{Bias-corrected
  population, size distribution, and impact hazard for the near-Earth objects,}
  Icarus, 170, 295, \dodoi{https://doi.org/10.1016/j.icarus.2004.03.018}

\bibitem[{ {The pandas development team}(2020){The pandas development
  team}}]{pandas2020}
{The pandas development team}. 2020, \bibinfo{title}{pandas-dev/pandas:
  Pandas,}, latest Zenodo, \dodoi{10.5281/zenodo.3509134}

\bibitem[{D.~E. {Trilling} {et~al.}(2018){Trilling}, {Bellm}, \&
  {Malhotra}}]{trilling2018}
{Trilling}, D.~E., {Bellm}, E.~C., \& {Malhotra}, R. 2018, \bibinfo{title}{{On
  the Detectability of Planet X with LSST},} \aj, 155, 243,
  \dodoi{10.3847/1538-3881/aabfc0}

\bibitem[{D.~E. Trilling {et~al.}(2017)Trilling, Valdes, Allen, James, Fuentes,
  Herrera, Axelrod, \& Rajagopal}]{trilling2017}
Trilling, D.~E., Valdes, F., Allen, L., {et~al.} 2017, \bibinfo{title}{The Size
  Distribution of Near-Earth Objects Larger Than 10 m,} The Astronomical
  Journal, 154, 170, \dodoi{10.3847/1538-3881/aa8036}

\bibitem[{D.~E. {Trilling} {et~al.}(2024){Trilling}, {Gerdes}, {Juri{\'c}},
  {Trujillo}, {Bernardinelli}, {Napier}, {Smotherman}, {Strauss}, {Fuentes},
  {Holman}, {Lin}, {Markwardt}, {McNeill}, {Mommert}, {Oldroyd}, {Payne},
  {Ragozzine}, {Rivkin}, {Schlichting}, {Sheppard}, {Adams}, \&
  {Chandler}}]{trilling2024}
{Trilling}, D.~E., {Gerdes}, D.~W., {Juri{\'c}}, M., {et~al.} 2024,
  \bibinfo{title}{{The DECam Ecliptic Exploration Project (DEEP). I. Survey
  Description, Science Questions, and Technical Demonstration},} \aj, 167, 132,
  \dodoi{10.3847/1538-3881/ad1529}

\bibitem[{C.~A. {Trujillo} \& S.~S. {Sheppard}(2014){Trujillo} \&
  {Sheppard}}]{trujillo2014}
{Trujillo}, C.~A., \& {Sheppard}, S.~S. 2014, \bibinfo{title}{{A Sedna-like
  body with a perihelion of 80 astronomical units},} \nat, 507, 471,
  \dodoi{10.1038/nature13156}

\bibitem[{L. Uieda {et~al.}(2020)Uieda, Soler, Rampin, van Kemenade, Turk,
  Shapero, Banihirwe, \& Leeman}]{uieda2020}
Uieda, L., Soler, S., Rampin, R., {et~al.} 2020, \bibinfo{title}{{Pooch}: {A}
  friend to fetch your data files,} Journal of Open Source Software, 5, 1943,
  \dodoi{10.21105/joss.01943}

\bibitem[{J. {{\v{D}}urech} {et~al.}(2010){{\v{D}}urech}, {Sidorin}, \&
  {Kaasalainen}}]{durech2010}
{{\v{D}}urech}, J., {Sidorin}, V., \& {Kaasalainen}, M. 2010,
  \bibinfo{title}{{DAMIT: a database of asteroid models},} \aap, 513, A46,
  \dodoi{10.1051/0004-6361/200912693}

\bibitem[{J. {{\v{D}}urech} {et~al.}(2022){{\v{D}}urech}, {V{\'a}vra},
  {Van{\v{c}}o}, \& {Erasmus}}]{durech2022}
{{\v{D}}urech}, J., {V{\'a}vra}, M., {Van{\v{c}}o}, R., \& {Erasmus}, N. 2022,
  \bibinfo{title}{{Rotation Periods of Asteroids Determined With Bootstrap
  Convex Inversion From ATLAS Photometry},} Frontiers in Astronomy and Space
  Sciences, 9, 809771, \dodoi{10.3389/fspas.2022.809771}

\bibitem[{P. Vere{\v{s}} \& S.~R. Chesley(2017{\natexlab{a}})Vere{\v{s}} \&
  Chesley}]{veres2017a}
Vere{\v{s}}, P., \& Chesley, S.~R. 2017{\natexlab{a}},
  \bibinfo{title}{High-fidelity Simulations of the Near-Earth Object Search
  Performance of the Large Synoptic Survey Telescope,} The Astronomical
  Journal, 154, 12, \dodoi{10.3847/1538-3881/aa73d1}

\bibitem[{P. Vere{\v{s}} \& S.~R. Chesley(2017{\natexlab{b}})Vere{\v{s}} \&
  Chesley}]{veres2017b}
Vere{\v{s}}, P., \& Chesley, S.~R. 2017{\natexlab{b}},
  \bibinfo{title}{Near-Earth Object Orbit Linking with the Large Synoptic
  Survey Telescope,} The Astronomical Journal, 154, 13,
  \dodoi{10.3847/1538-3881/aa73d0}

\bibitem[{P. Virtanen {et~al.}(2020)Virtanen, Gommers, Oliphant, Haberland,
  Reddy, Cournapeau, Burovski, Peterson, Weckesser, Bright, {van der Walt},
  Brett, Wilson, Millman, Mayorov, Nelson, Jones, Kern, Larson, Carey, Polat,
  Feng, Moore, {VanderPlas}, Laxalde, Perktold, Cimrman, Henriksen, Quintero,
  Harris, Archibald, Ribeiro, Pedregosa, {van Mulbregt}, \& {SciPy 1.0
  Contributors}}]{virtanen2020}
Virtanen, P., Gommers, R., Oliphant, T.~E., {et~al.} 2020,
  \bibinfo{title}{{{SciPy} 1.0: Fundamental Algorithms for Scientific Computing
  in Python},} Nature Methods, 17, 261, \dodoi{10.1038/s41592-019-0686-2}

\bibitem[{D. {Vokrouhlick{\'y}} {et~al.}(2015){Vokrouhlick{\'y}}, {Bottke},
  {Chesley}, {Scheeres}, \& {Statler}}]{vokrouhlicky2015}
{Vokrouhlick{\'y}}, D., {Bottke}, W.~F., {Chesley}, S.~R., {Scheeres}, D.~J.,
  \& {Statler}, T.~S. 2015, in Asteroids IV, ed. P.~Michel, F.~E. DeMeo, \&
  W.~F. Bottke ({University of Arizona Press}), 509--531,
  \dodoi{10.2458/azu_uapress_9780816532131-ch027}

\bibitem[{C.~N.~A. {Willmer}(2018){Willmer}}]{willmer2018}
{Willmer}, C. N.~A. 2018, \bibinfo{title}{{The Absolute Magnitude of the Sun in
  Several Filters},} \apjs, 236, 47, \dodoi{10.3847/1538-4365/aabfdf}

\bibitem[{P. Yoachim {et~al.}(2024)Yoachim, Jones, Eric H.~Neilsen, Bechtol,
  Becker, \& Ross}]{yoachim2024}
Yoachim, P., Jones, L., Eric H.~Neilsen, J., {et~al.} 2024,
  \bibinfo{title}{lsst/rubin\_scheduler: v3.4.0,}, v3.4.0 Zenodo,
  \dodoi{10.5281/zenodo.14232232}

\bibitem[{P. Yoachim {et~al.}(2023)Yoachim, Jones, Eric H.~Neilsen, Tiago,
  Parejko, Carlin, Becker, pgris, Prisinzano, Dennihy, Bellm, Sick, lmptc, LI,
  nsabrams, Guy, Bricman, Bregeon, Lim, Kelley, \& Andreoni}]{yoachim2022}
Yoachim, P., Jones, L., Eric H.~Neilsen, J., {et~al.} 2023,
  \bibinfo{title}{lsst/rubin\_sim: v2.0.0,}, v2.0.0 Zenodo,
  \dodoi{10.5281/zenodo.10215451}

\bibitem[{M. Zavodny {et~al.}(2008)Zavodny, Jedicke, Beshore, Bernardi, \&
  Larson}]{zavodny2008}
Zavodny, M., Jedicke, R., Beshore, E.~C., Bernardi, F., \& Larson, S. 2008,
  \bibinfo{title}{The orbit and size distribution of small Solar System objects
  orbiting the Sun interior to the Earth's orbit,} Icarus, 198, 284,
  \dodoi{https://doi.org/10.1016/j.icarus.2008.05.021}

\bibitem[{A. Zonca {et~al.}(2019)Zonca, Singer, Lenz, Reinecke, Rosset, Hivon,
  \& Gorski}]{zonca2019}
Zonca, A., Singer, L., Lenz, D., {et~al.} 2019, \bibinfo{title}{healpy: equal
  area pixelization and spherical harmonics transforms for data on the sphere
  in Python,} Journal of Open Source Software, 4, 1298,
  \dodoi{10.21105/joss.01298}

\end{thebibliography}
\bibliographystyle{aasjournalv7}



\end{document}